\documentclass[a4paper,12pt,onecolumn]{book}

\usepackage{graphicx}
\usepackage{bm}
\usepackage{amsmath}
\usepackage{exscale}
\usepackage{amssymb}
\usepackage{german}

\newcommand*{\MRM}[3]{Magn\ Reson\ Med #1;#2:#3}
\newcommand*{\MRI}[3]{Magn\ Reson\ Imaging #1;#2:#3}
\newcommand*{\JMRI}[3]{J\ Magn\ Reson\ Imaging #1;#2:#3}
\newcommand*{\JMR}[3]{J\ Magn\ Reson #1;#2:#3}

\newcommand*{\PR}[3]{Phys\ Rev\ #1;#2:#3}
\newcommand*{\PRL}[3]{Phys\ Rev\ Lett #1;#2:#3}
\newcommand*{\PRA}[3]{Phys\ Rev\ A #1;#2:#3}
\newcommand*{\PRB}[3]{Phys\ Rev\ B #1;#2:#3}
\newcommand*{\PRE}[3]{Phys\ Rev\ E #1;#2:#3}
\newcommand*{\JMP}[3]{J\ Math\ Phys #1;#2:#3}
\newcommand*{\JCP}[3]{J\ Chem\ Phys #1;#2:#3}
\newcommand*{\CPL}[3]{Chem\ Phys\ Lett #1;#2:#3}
\newcommand*{\JPC}[3]{J\ Phys\ Chem #1;#2:#3}

\newcommand{\I}{\mathrm{i}}
\newcommand{\E}{\mathrm{e}}

\begin{document}

\begin{titlepage}
\vfill
\vspace*{3cm}
\begin{minipage}{\textwidth}
\centering \Huge \bf
Suszeptibilit\"atseffekte in der Kernspinresonanzbildgebung\end{minipage}

\vspace{3cm}\begin{minipage}{\textwidth}
\centering{\Large{
										Dissertation zur Erlangung des\\
										naturwissenschaftlichen Doktorgrades\\
										der Julius-Maximilians-Universit\"at W\"urzburg\\[12ex]
										vorgelegt von\\[3ex]
										Christian H. Ziener\\[3ex]
										aus Weimar\\[12ex]
										W\"urzburg 2008
									}}
\end{minipage}

\vfill

\newpage

\thispagestyle{empty}

\phantom{Leerzeile}

\vfill

\flushleft
\begin{minipage}{\textwidth}
  \normalsize
  \begin{tabular}{ll}
    Eingereicht am: 04. 09. 2008\\
    \multicolumn{2}{l}{bei der Fakult\"at f\"ur Physik und Astronomie}\\
                               & \\
    1. Gutachter:              &\hspace{-2cm} {\sc Prof. Dr. rer. nat. Peter M. Jakob} \\
    2. Gutachter:              &\hspace{-2cm} {\sc Prof. Dr. med. Dr. rer. nat. Wolfgang R. Bauer} \\
    3. Gutachter:              &\hspace{-2cm} {\sc Prof. Dr. rer. nat. med. habil. J\"urgen R. Reichenbach} \\ 
    der Dissertation.\\
                               & \\
    1. Pr\"ufer:               &\hspace{-2cm} {\sc Prof. Dr. rer. nat. Peter M. Jakob} \\
    2. Pr\"ufer:               &\hspace{-2cm} {\sc Prof. Dr. med. Dr. rer. nat. Wolfgang R. Bauer} \\
    3. Pr\"ufer:               &\hspace{-2cm} {\sc Prof. Dr. rer. nat. Georg Reents} \\
    im Promotionskolloquium.\\
                                     & \\
    Tag des Promotionskolloquiums:   &09. 04. 2009\\
                                     & \\
    Doktorurkunde ausgeh\"andigt am: &20. 04. 2009
  \end{tabular}
\end{minipage}

\end{titlepage}


\pagestyle{empty}

\tableofcontents

\listoffigures
\addcontentsline{toc}{chapter}{Abbildungsverzeichnis}


\pagestyle{empty}

\chapter*{Vorwort}

\addcontentsline{toc}{chapter}{Vorwort}

\hfill\begin{minipage}{0.75\textwidth}\sl\vspace*{0.7cm} 
Aufgabe der Naturwissenschaft ist es nicht nur die Erfahrung zu erweitern,\\
sondern in diese Erfahrung eine Ordnung zu bringen.\\

{\sc Niels Bohr}
\end{minipage}
\\

\vspace{7ex}
Die Prinzipien der Kernspinresonanz haben sich in den letzten Jahrzehnten zu einem wichtigen Werkzeug f\"ur verschiedene Naturwissenschaften entwickelt. Die Anwendungsgebiete reichen heute von der Physik \"uber die Chemie und die Biologie bis hin zur Medizin. Grundlage zum Verst\"andnis der NMR-Techniken sind physikalische Modelle, die den relevanten Fragestellungen angepasst und gegebenenfalls erweitert werden m\"ussen. Daher ist es unverzichtbar, dass sowohl numerische Simulationen und analytische Rechnungen als auch experimentelle Untersuchungen das entwickelte physikalische Modell best\"atigen und neue Aspekte er\"offnen. So konnten in dieser Arbeit interdisziplin\"are Fragestellungen der Biophysik durch Analogieschl\"usse zu anderen Gebieten der Physik bearbeitet und schlie{\ss}lich in einem allgemeinen Kontext dargestellt werden.

An dieser Stelle m\"ochte ich mich bei denen bedanken, die direkt oder indirekt zum Gelingen dieser Arbeit beigetragen haben. Insbesondere danke ich meinem Betreuer Prof. Dr. Dr. Wolfgang Bauer f\"ur die sehr gute Betreuung. Die Diskussionen mit ihm, seine Anregungen und Ideen waren mir eine gro{\ss}e Hilfe. Ganz besonders m\"ochte ich mich bei Prof. Dr. Peter Jakob daf\"ur bedanken, dass er meine Arbeit stetig unterst\"utzt hat, viele Ideen und Anregungen geliefert hat und jederzeit f\"ur Diskussionen zur Verf\"ugung stand. F\"ur viele Hinweise und Ratschl\"age bez\"uglich einiger mathematisch interessanter Fragestellungen m\"ochte ich Prof. Dr. Georg Reents vom Institut f\"ur Theoretische Physik und Astrophysik danken.

Besonderer Dank geb\"uhrt der Schering-Stiftung, die einen wesentlichen Teil zum Gelingen dieser Arbeit beitrug. Als Stipendiat der Schering-Stiftung bot sich mir die M\"oglichkeit, mit anderen Wissenschaftlern in Kontakt zu treten, Erfahrungen auszutauschen und gemeinsam \"uber Probleme zu diskutieren. Erforderliches interdisziplin\"ares und vernetztes Denken und Arbeiten wurde dadurch nachhaltig gest\"arkt. Besonders m\"ochte ich ganz herzlich Frau Dr. Monika Lessl und Frau Ines St\"ohr von der Schering-Stiftung danken.

Dem Berufsverband Deutscher Internisten, insbesondere Herrn Dr. Wolfgang Wesiack danke ich f\"ur die Aufnahme in das studentische F\"orderprogramm.

Im Rahmen des an der Universit\"at W\"urzburg neu etablierten Sonderforschungsbereiches 688 \glqq Mechanismen und Bildgebung von Zell-Zell-Wechselwirkungen im kardiovaskul\"aren System\grqq\ war es m\"oglich, die Ergebnisse in ein umfassenderes Konzept zu integrieren und so die Grenzen herk\"ommlicher Disziplinen und Traditionen aufzubrechen.

Der Arbeitsgruppe Medizinische Physik des Instituts f\"ur Diagnostische und Interventionelle Radiologie der Fiedrich-Schiller-Universit\"at Jena, insbesondere Jan Sedlacik, Dr. Alexander Rauscher und Prof. Dr. J\"urgen Reichenbach danke ich f\"ur die Zusammenarbeit und die wertvollen Diskussionen bez\"uglich Fragen der Signalentstehung und Frequenzverteilung im Kapillarnetzwerk.

F\"ur die weitreichende Unterst\"utzung bei der Behandlung vieler numerischer Probleme danke ich Dr. Stephan Glutsch, der schon als Betreuer meiner Diplomarbeit viel Hilfestellung gab, Zusammenh\"ange erkl\"arte und wesentlich zu meinem bisherigen wissenschaftlichen Werdegang beitrug.

Weiterhin bedanke ich mich bei meinen Eltern f\"ur Ihre Unterst\"utzung, bei meinen Gro{\ss}eltern, die mir die Ferienkurse am Forschungszentrum J\"ulich erm\"oglichten, sowie meinen Arbeitskollegen vom Lehrstuhl f\"ur Experimentelle Physik 5 der Universit\"at W\"urzburg f\"ur die stetige Motivation und Hilfe.\\

\vspace*{2cm}
\noindent
W\"urzburg, im September 2008
\hfill
{\sc Christian H. Ziener}

\chapter{\label{Kap.Einleitung}Einleitung}
\pagestyle{headings}

\noindent

Methoden, die unter Verwendung von NMR-Signalen r\"aumliche Abbildungen oder Zeugmatogramme (griechisch: Zeugma = das Joch, die Zusammenf\"uhrung, die Verbindung) der Verteilung von Kernspinmomenten in einer makroskopischen Probe erstellen, werden mit dem Begriff Zeugmatographie bezeichnet \cite{Lauterbur72,Lauterbur73}. Speziell bei medizinischen Anwendungen wurde der Begriff Kernspinresonanztomographie eingef\"uhrt. Dabei besitzt die zeugmatographische Abbildung \"Ahnlichkeiten mit optischen Experimenten, bei denen Phasendifferenzen infolge unterschiedlicher Wegl\"angen einer Abbildung der Objekte zugrunde liegen. Im zeugmatographischen Experiment entstehen die Phasendifferenzen durch die unterschiedlichen Pr\"azessionsfrequenzen, welche die Kernmomente an verschiedenen Stellen des \"au{\ss}eren Magnetfeldes mit einem definierten Feldgradienten besitzen. 

Einen wichtigen Einfluss auf die Abbildung biologischen Gewebes haben mikroskopische Magnetfeldinhomogenit\"aten, die z. B. durch endogene oder exogene Kontrastmittel erzeugt werden. Ausschlaggebend f\"ur die intensive Befassung mit der Auswirkung von Magnetfeldinhomogenit\"aten auf das NMR-Signal war die Entdeckung des BOLD-Effekts (BOLD=Blood Oxygenation Level Dependent) von Ogawa im Jahre 1990 \cite{Ogawa90}. Ogawa untersuchte die Auswirkungen eines im Jahre 1936 von Pauling studierten interessanten Ph\"anomens der unterschiedlichen Eigenschaften von oxygenierten und desoxygenierten H\"amoglobin \cite{Pauling36}. Der BOLD-Effekt basiert auf diesen unterschiedlichen magnetischen Eigenschaften von oxygenierten und desoxygenierten H\"amoglobin und erlaubt es, Blut selbst als endogenes Kontrastmittel f\"ur die funktionelle Bildgebung zu nutzen. Blutgef\"ullte Kapillaren k\"onnen demnach als mikroskopische magnetisierte K\"orper aufgefasst werden, deren Suszeptibilit\"atskontrast zum umgebenden Gewebe vom Oxygenierungsgrad des enthaltenen Blutes abh\"angt. Diese Suszeptibilit\"atsdifferenz erzeugt die charakteristischen Eigenschaften des NMR-Signals. Nun besteht die Aufgabe darin, aus dem gemessenen NMR-Signal R\"uckschl\"usse auf den Oxygenierungsgrad des Blutes oder noch allgemeiner auf die geometrischen Eigenschaften des untersuchten Gewebes zu ziehen. Dazu m\"ussen jedoch zuerst physikalische Modelle entwickelt werden, welche die Signaleigenschaften des untersuchten Gewebes und die darin enthaltenen Feldinhomogenit\"aten ausreichend beschreiben. So wird z. B. als Gewebeeigenschaft die Frequenzverteilung um eine Kapillare in der vorliegenden Arbeit untersucht. Aus der mit NMR-Methoden gemessenen Frequenzverteilung kann dann die Suszeptibilit\"atsdifferenz zwischen Kapillare und umgebenden Medium sowie die Kapillardichte ermittelt werden. Anordnung der Kapillaren und Oxygenierungsgrad des Blutes bestimmen auch die Relaxationseigenschaften des untersuchten Gewebes. Dadurch ist es beispielsweise m\"oglich, aus der messbaren transversalen Relaxationszeit das regionale Blutvolumen des Herzmuskelgewebes zu ermitteln und so die eventuellen Auswirkungen eines Myokardinfarktes zu quantifizieren \cite{Wacker03}.

In Analogie zum endogenen Kontrastmittel Blut stellen sich die Verh\"altnisse bei exogenen Kontrastmitteln dar, die von Zellen aufgenommen werden und einen Suszeptibilit\"atskontrast zum umgebenden Gewebe erzeugen. Bei den exogenen Kontrastmitteln handelt es sich um sehr kleine paramagnetische Partikel (USPIO=Ultra Small Paramagnetic Iron Oxide), bestehend aus einem paramagnetischen Kern (Durchmesser $\approx 5\, \text{nm}$) und einer Dextranh\"ulle (Durchmesser $\approx 50\, \text{nm}$). Durch Phagozytose nehmen Zellen diese Partikel auf und werden danach als magnetisch markierte Zellen bezeichnet. Diese magnetisch markierten Zellen erzeugen eine charakteristische Frequenzverteilung im Voxel, die von der Konzentration des Kontrastmittels und weiteren Parametern, wie z. B. dem Diffusionskoeffizienten des umgebenden Gewebes, abh\"angt. Diese Parameter haben auch Einfluss auf die transversale Relaxationszeit. Die physikalischen Modelle sollen nun helfen, von der Relaxationszeit auf die Parameter des Gewebes, wie z. B. die Kontrastmittelkonzentration (oder besser die Konzentration der Zellen) zu schlie{\ss}en. 

Das Fundament zur L\"osung dieser Aufgaben ist ein tiefgreifendes Verst\"andnis des Einflusses von Magnetfeldinhomogenit\"aten auf die Entstehung des NMR-Signals. Im Verlauf dieser Arbeit wird genutzt werden, dass sich der aus der statistischen Physik bekannte Formalismus der Zustandsdichten auf das von den magnetisierten K\"orpern erzeugte lokale Magnetfeld \"ubertragen l\"asst, und dass dieser Formalismus ein wichtiges Werkzeug zur Beschreibung der Suszeptibilit\"atseffekte darstellt.

Eine grundlegende Arbeit zur Beschreibung von Magnetfeldinhomogenit\"aten stellt die Arbeit \glqq Theory of NMR signal behavior in magnetically inhomogeneous tissues: the static dephasing regime\grqq\  von Yablonskiy und Haacke aus dem Jahre 1994 dar \cite{Yablonskiy94}, in der die Signaleigenschaften untersucht werden, die von lokalen inhomogenen Feldern hervorgerufen werden. Allerdings wird in der Arbeit der Einfluss der Diffusion im umgebenden Medium nicht ber\"ucksichtigt. Solange es sich um Objekte mit gro{\ss}en Abmessungen (mm-Bereich) bzw. um starke Suszeptibilit\"atsspr\"unge zwischen Objekt und umgebenden Medium handelt, ist die Vernachl\"assigung der Diffusionseffekte gerechtfertigt. In dem Fall spricht man vom Static-Dephasing-Regime, d. h. die signalgebenden Spins des umgebenden Mediums k\"onnen als unbeweglich angesehen werden. Sobald es sich jedoch um relativ kleine Objekte wie Kontrastmittelteilchen oder um geringe Suszeptibilit\"atsdifferenzen -- wie sie z. B. durch die magnetischen Eigenschaften des H\"amoglobins hervorgerufen werden -- handelt, haben Diffusionseffekte einen wesentlichen Einfluss auf das NMR-Signal und k\"onnen nicht mehr vernachl\"assigt werden. Der andere Grenzfall, in dem die Suszeptibilit\"atseffekte vernachl\"assigt werden k\"onnen und in dem haupts\"achlich Diffusionseffekte die NMR-Signalentstehung bestimmen, ist das Motional-Narrowing-Regime, das bereits in den fr\"uhen Jahren der Kernspinresonanzforschung ausf\"uhrlich untersucht wurde \cite{Callaghan}. Bauer et al. gelang es 1999, die aus der Theorie der Linienform bekannte Strong-Collision-Approximation auf Fragestellungen der Signalentstehung in magnetisch inhomogenen Gewebe anzuwenden \cite{Bauer99}. Mit diesem Formalismus ist es m\"oglich, den gesamten Dynamikbereich vom Static-Dephasing-Regime bis zum Motional-Narrowing-Regime zu beschreiben, wobei die beiden Grenzf\"alle selber enthalten sind. Mit Hilfe dieser N\"aherung konnte die Relaxationszeit des Myokards in Abh\"angigkeiten vom Oxygenierungsgrad des Blutes, vom regionalen Blutvolumen und von der Diffusion des umgebenden Mediums bestimmt werden \cite{Bauer99PRL}. In der vorliegenden Arbeit wird die Strong-Collision-Approximation angewandt, um das Relaxationsverhalten von Gewebe zu untersuchen, in dem sich magnetisch markierte Zellen befinden, die lokale Magnetfeldinhomogenit\"aten erzeugen. Des Weiteren wird der Formalismus der Zustandsdichten angewandt, um die Frequenzverteilungen um magnetisierte Objekte zu beschreiben. Mittels der Strong-Collision-Approximation kann auch deren Diffusionsabh\"angigkeit untersucht werden. 

Eine \"Ubersicht \"uber die physikalischen Grundlagen der Beschreibung von Kernspins im Magnetfeld, die biologischen Grundlagen zur Charakterisierung von Geweben sowie die mathematischen Grundlagen der Strong-Collision-N\"aherung werden in Kapitel \ref{Kap.Grundlagen} gegeben. 

Basierend auf einfachen Annahmen zur Anordnung von magnetisierten Objekten in einem \"au{\ss}eren Magnetfeld wird in Kapitel \ref{Kap.Korr} die Korrelationszeit $\tau$, die den Diffusionsprozess um diese Objekte beschreibt, untersucht. Die Korrelationszeit ist abh\"angig vom charakteristischen Durchmesser des Objektes, vom Diffusionskoeffizienten des umgebenden Mediums, vom Volumenanteil des Objektes am gesamten Voxel und von der Permeabilit\"at der Oberfl\"ache. Ein einfaches Verfahren zur Ermittlung der Korrelationszeit wird vorgestellt. Aufgrund seiner mathematischen Einfachheit kann dieses Verfahren f\"ur weitere Anwendungen genutzt werden. F\"ur den Spezialfall von Kugeln und Zylindern werden analytische Ausdr\"ucke angegeben, die zur Charakterisierung des Diffusionsprozesses oder zur Beschreibung von Relaxationsraten genutzt werden k\"onnen.

Zur allgemeinen Beschreibung von NMR-Signalen, die durch die Anwesenheit von magnetisierten Objekten beeinflusst werden, wird in Kapitel \ref{Kap:Frequenz} ein Formalismus zur Beschreibung der Frequenzverteilung in einem Voxel entwickelt. Dabei wird davon ausgegangen, dass das Voxel ein magnetisiertes Objekt enth\"alt, welches ein lokales inhomogenes Magnetfeld erzeugt. Das Objekt, welches das inhomogene Magnetfeld erzeugt, wird vom Dephasierungsvolumen umgeben, in dem sich die diffundierenden Spins befinden. Der Diffusionsprozess wird durch den Diffusionskoeffizienten $D$ charakterisiert. Es kann gezeigt werden, dass allein die Form des magnetischen K\"orpers, die St\"arke der Suszeptibilit\"atsdifferenz zwischen K\"orper und umgebenden Medium und der Diffusionskoeffizient die Frequenzverteilung festlegen.

Weitere Verfahren zur Untersuchung der Diffusionseffekte auf das NMR-Signal wurden in der letzten Zeit von verschiedenen Autoren entwickelt. Kiselev und Posse erarbeiteten eine Erweiterung des Static-Dephasing-Regimes, basierend auf einem St\"orungsansatz im lokalen Magnetfeld \cite{Kiselev99}. Sukstanskii und Yablonskiy nutzten die Gau{\ss}sche N\"aherung zur Beschreibung der Signaldephasierung \cite{Sukstanskii02,Sukstanskii03,Sukstanskii04}. Jedoch beruht diese Gau{\ss}sche N\"aherung auf dem aus der Festk\"orperphysik bekannten Anderson-Weiss-Modell \cite{Anderson53}, dessen Anwendbarkeitskriterien f\"ur die NMR-Relaxationstheo\-rie erst in dieser Arbeit genauer untersucht werden. Die oben beschriebenen Anwendbarkeitskriterien des Anderson-Weiss-Modells f\"uhrten auf die Problematik der Nicht-Gau{\ss}schen Dephasierung und werden in Kapitel \ref{Chap:Dephas} erarbeitet.

Zur Charakterisierung von Geweben werden oft die transversalen Relaxationszeiten genutzt, die sich experimentell bestimmen lassen. Aufgrund der wachsenden Bedeutung von Stammzellen stellte sich die Frage, wie diese Zellen mit Hilfe der Kernspinresonanzbildgebung sichtbar gemacht werden k\"onnen. In Analogie zum oben beschriebenen BOLD-Effekt, der die Eigenschaften des endogenen Kontrastmittels Blut ausnutzt, werden die zu untersuchenden Zellen vor der Injektion in das zu untersuchende Gewebe mit kleinen paramagnetischen Kontrastmittelteichen (USPIOs) markiert. Abh\"angig von den Parametern dieser Kontrastmittelteilchen und abh\"angig von den Eigenschaften des umgebenden Mediums \"andern sich die Dephasierungseigenschaften der die Zellen umgebenden Wassermolek\"ule, was zu einer Ver\"anderung der Relaxationseigenschaften f\"uhrt. In Kapitel \ref{Kap:Relaxation} wird dieser Zusammenhang zwischen der Relaxationszeit, den Parametern der Kontrastmittelteilchen und dem Diffusionseffekt der umgebenden Wassermolek\"ule untersucht.

Ausgehend von einfachen Annahmen \"uber die Verteilung magnetischen Materials innerhalb eines Voxels werden in Kapitel \ref{Kap:Skalierung} einfache Skalierungsgesetze f\"ur die transversalen Relaxationszeiten $T_2^*$ und $T_2$ abgeleitet. Dabei werden schon bekannte Ergebnisse verallgemeinert. Mit den hergeleiteten Gesetzen kann man die Auswirkung der Ver\"anderung von Suszeptibilit\"atsdifferenz, \"au{\ss}erem Magnetfeld, charakteristischer Gr\"o{\ss}e des Objekts oder Diffusionseigenschaften des umgebenden Mediums auf die transversalen Relaxationszeiten vorhersagen. Von Bedeutung ist dabei, dass das Skalierungsverhalten der Relaxationszeiten f\"ur jede beliebige Form einer Feldinhomogenit\"at gilt. Das hei{\ss}t, im Gegensatz zu fr\"uheren Ergebnissen k\"onnen diese Skalierungsgesetze auf beliebige Geometrien unabh\"angig von der Form der Magnetfeldinhomogenit\"at angewandt werden. Damit k\"onnen verschiedene Parameter eines Experiments in einer einfachen Weise miteinander verkn\"upft werden, ohne die gesamte Komplexit\"at des Systems betrachten zu m\"ussen.

\chapter{\label{Kap.Grundlagen}Grundlagen}

\section{Kernspin im Magnetfeld} \label{Spin}
Betrachtet wird ein System von $N$ isolierten Spins in einem \"au{\ss}eren Magnetfeld mit der Flussdichte $B_0$ in $z$-Richtung, die das Volumen $V$ einnehmen. Das magnetische Moment $\bm{\mu}$ eines Kernes ist proportional zum Drehimpuls $\mathbf{I}$ des Kerns:
\begin{equation}
\bm{\mu}=\gamma\mathbf{I} \,,
\end{equation}
wobei $\gamma$ das f\"ur den betrachteten Kern spezifische gyromagnetische Verh\"altnis darstellt:
\begin{equation}
\gamma = g \frac{\mu_N}{\hbar} \,.
\end{equation}
Der g-Faktor $g$ ist vom betrachteten Kern abh\"angig und das Kernmagneton ist definiert durch
\begin{equation}
\mu_N = \frac{e \hbar}{2 m_K} \,,
\end{equation}
wobei $e$ die Elementarladung, $\hbar$ die Plancksche Konstante und $m_K$ die Masse des Kerns darstellt. F\"ur die Komponenten des Drehimpulses $\mathbf{I}$ gelten die Vertauschungsregeln $\left[ I_i , I_j \right]=\text{i} \hbar \varepsilon_{ijk} I_k$ sowie $\forall i$: $\left[ \mathbf{I}^2 , I_i \right]=0$. Seien $|Im\rangle$ die gemeinsamen Eigenfunktionen der Operatoren $\mathbf{I}^2$ und $I_z$ mit den Quantenzahlen $I$ und $m$ ($-I \leq m \leq +I$ ), so gilt
\begin{align}
\mathbf{I}^2 |Im\rangle &= \hbar^2 I(I+1)|Im\rangle \;\;\text{und}\\[2ex]
I_z |Im\rangle &= \hbar m|Im\rangle \,.
\end{align}
Der Hamilton-Operator f\"ur die Zeeman-Energie
\begin{align}
\mathcal{H}_z=-\mu_z B_0
\end{align}
erf\"ullt demnach die Eigenwertgleichung
\begin{align} \label{Ham-Zee}
\mathcal{H}_z |Im\rangle = E_m |Im\rangle
\end{align}
mit den Eigenwerten
\begin{align}
E_m = - \gamma \hbar B_0 \cdot m \,,
\end{align}
wobei $m$ die $2I+1$ Werte $-I,-I+1,...,I-1,I$ annehmen kann. Diese Gleichung beschreibt die Energieaufspaltung aufgrund der Richtungsquantelung des magnetischen Moments. Beim \"Ubergang zwischen zwei benachbarten Zust\"anden wird ein Photon mit der Energie $\hbar \omega_0 = |E_{m \pm 1}-E_m|$ ausgesandt. Damit ergibt sich die als Larmor-Relation bekannte Resonanzbedingung
\begin{equation} \label{Larmor}
\omega_0 = \gamma B_0 \,.
\end{equation}
In dieser Arbeit werden nur Protonen mit einem g-Faktor von $g \approx 5,59$ untersucht. Damit ergibt sich ein gyromagnetisches Verh\"altnis von $\gamma = 2,677 \cdot 10^8 \, \text{C} / \text{kg}$. 

Die Wechselwirkung zwischen den Spinmomenten f\"uhrt zum Aufbau einer makroskopischen Magnetisierung, deren Gleichgewichtswert durch
\begin{align} \label{Mnull}
M_0 = \frac{N}{V} \langle \mu_z \rangle
\end{align}
gegeben ist. Der Erwartungswert f\"ur die Komponente des magnetischen Moments in Magnetfeldrichtung ergibt sich zu
\begin{align}
\langle \mu_z \rangle = \gamma \hbar I B_I(x) \,,
\end{align}
mit der Brillouin-Funktion
\begin{align}
B_I(x)=\left[ \frac{2I+1}{2I} \coth \left( \frac{2I+1}{2I}x \right) - \frac{1}{2I} \coth \left( \frac{x}{2I} \right) \right]
\end{align}
und dem dimensionslosen Argument
\begin{align}
x=\frac{\gamma \hbar B_0}{kT} \,.
\end{align}

Die Wechselwirkung zwischen dem Spinsystem und dem \"au{\ss}eren Hochfrequenzfeld $B_1(t)$, wird durch den Hamilton-Operator
\begin{align}
\mathcal{H}_1(t) = - 2 B_1 \cos \omega t \cdot \gamma \hbar I_x
\end{align}
beschrieben, der sich als kleine St\"orung zum Hamilton-Operator der Zeeman-Energie (\ref{Ham-Zee}) addiert. Um den Einfluss dieser St\"orung zu beschreiben, werden in Analogie zu Gleichung (\ref{Mnull}) die Erwartungswerte der Komponenten $M_k$ der Magnetisierung ($k=x,y,z$) mit Hilfe des Dichteoperators $\varrho$ durch die Gleichung
\begin{align}
M_k(t)=\frac{N}{V} \gamma \hbar \, \text{Sp}\{ \varrho(t)I_k \}
\end{align}
beschrieben. Die Bewegungsgleichung des Dichteoperators 
\begin{align}
\frac{\partial \varrho}{\partial t} = - \frac{\text{i}}{\hbar} [\varrho, \mathcal{H}]
\end{align}
enth\"alt den Hamilton-Operator $\mathcal{H}(t)=\mathcal{H}_z + \mathcal{H}_1(t)$, der die Wechselwirkung mit dem konstanten \"au{\ss}eren $B_0$-Feld und dem hochfrequenten Wechselfeld $B_1(t)$ beschreibt. Die Zeitentwicklung des Dichteoperators kann nun z. B. im Dirac-Bild geschehen. Allerdings lassen sich f\"ur die betrachteten Systeme mit hinreichend hoher thermischer Beweglichkeit, f\"ur die sich die Umorientierungen und Translationen der spintragenden Teilchen in Zeitr\"aumen abspielen, die sehr kurz gegen\"uber der Beobachtungszeit im Kernresonanzexperiment sind, die magnetischen Eigenschaften eines Ensembles von Kernspins in einem \"au{\ss}eren Magnetfeld gut durch die ph\"anomenologischen Bloch-Gleichungen beschreiben:
\begin{align}
\frac{\text{d}\mathbf{M}}{\text{d}t} = \gamma [\mathbf{M}(t) \times \mathbf{B}] - \frac{M_x \mathbf{e}_x + M_y \mathbf{e}_y}{T_2} - \frac{M_z-M_0}{T_1} \mathbf{e}_z \,.
\end{align}
Dabei charakterisieren die Relaxationszeiten $T_1$ und $T_2$ den \"Ubergang des Spinsystems vom angeregten Zustand in das Gleichgewicht. Hierbei beschreibt die longitudinale Relaxationszeit $T_1$ die Wechselwirkung der Spins mit der Umgebung (Spin-Gitter-Relaxation oder Energierelaxation), und die transversale Relaxationszeit $T_2$ die Wechselwirkung der Spins untereinander (Spin-Spin-Relaxation oder Entropierelaxation).
Nach einem Anregungspuls in Form eines hochfrequenten $B_1$-Felds ergibt sich als L\"osung dieser Gleichung der freie Induktionsabfall (FID) der transversalen Komponente, sowie ein exponentieller Anstieg der longitudinalen Komponente:
\begin{align}
M_k(t) & = M_k(0) \text{e}^{-\frac{t}{T_2}}, \;\;\; k=x,y  \;\;\; \text{und} \\[2ex]
M_z(t) & = M_0-(M_0-M_z(0)) \text{e}^{-\frac{t}{T_1}}\,,
\end{align}
wobei $M_0$ die Gleichgewichtsmagnetisierung nach Gleichung (\ref{Mnull}) darstellt. Bisher wurden allerdings nur isochromatische Spins betrachtet, d. h. es wurde angenommen, dass alle Spins bei der Frequenz $\omega_0 = \gamma B_0$ pr\"azedieren. Werden jedoch magnetisierte K\"orper ber\"ucksichtigt, die eine lokale Resonanzfrequenz $\omega(\mathbf{r})$ in der Probe erzeugen, f\"uhrt dies zu einer zus\"atzlichen Dephasierung der Spins. Die lokale Resonanzfrequenz erzeugt die Frequenzverteilung $p(\omega)$, die angibt, wie oft jede Resonanzfrequenz anzutreffen ist. Dies f\"uhrt zu einem zus\"atzlichen Relaxationsterm f\"ur die transversale Komponente der Magnetisierung:
\begin{align} \label{Mk}
M_k(t) = M_k(0) \, \text{e}^{-\frac{t}{T_2}} \int_{-\infty}^{+\infty} \text{d} \omega \, p(\omega) \text{e}^{\text{i}\omega t} \;\;\;k=x,y \,.
\end{align}
F\"ur diese Frequenzverteilung wird oft ein lorentzf\"ormiges Profil von der Form 
\begin{align} \label{Lorentz}
p(\omega)=\frac{T_{2}^{'}}{\pi} \frac{1}{1+(\omega T_{2}^{'})^2}
\end{align}
angenommen, wobei $T_{2}^{'}$ die St\"arke der Magnetfeldinhomogenit\"at charakterisiert \cite{Kennan94}, was sich in der Breite der Lorentz-Linie niederschl\"agt. Unter der Annahme dieser lorentzf\"ormigen Frequenzverteilung l\"asst sich das Integral in obiger Gleichung auswerten. F\"ur den freien Induktionszerfall ergibt sich der Zusammenhang
\begin{align}
M_k(t) = M_k(0) \text{e}^{-\frac{t}{T_{2}^{*}}} \;\;\;\text{und}\;\;\;k=x,y \;\;\; \text{mit} \;\;\; \frac{1}{T_{2}^{*}}=\frac{1}{T_2}+\frac{1}{T_{2}^{'}} \,.
\end{align}
Im Fall einer gau{\ss}f\"ormigen Frequenzverteilung der Form $p(\omega) \propto \exp(- \omega^2 /2 \sigma^2)$ ergibt sich f\"ur den Signal-Zeit-Verlauf eine Mischung aus exponentiellem und gau{\ss}f\"ormigem Abfall der Form $M_k(t) = M_k(0) \exp(-t/T_2) \exp(-\sigma^2 t^2 /2)$. In Kapitel \ref{Chap:Dephas} wird der zugrunde liegenden Dephasierungsmechanismus betrachtet und die Form des freien Induktionszerfalls genauer untersucht. 

Die Frequenzverteilung wurde bisher als lorentzf\"ormig angenommen, damit sich ein exponentieller Signalzerfall ergab. Die genaue Form dieser Frequenzverteilung wird in Kapitel \ref{Kap:Frequenz} untersucht. Dazu wird der aus der statistischen Physik bekannte Formalismus der Zustandsdichten auf das von den magnetisierten Objekten erzeugte Frequenzfeld angewandt. 

\begin{figure}
\begin{center}
\includegraphics[width=12cm]{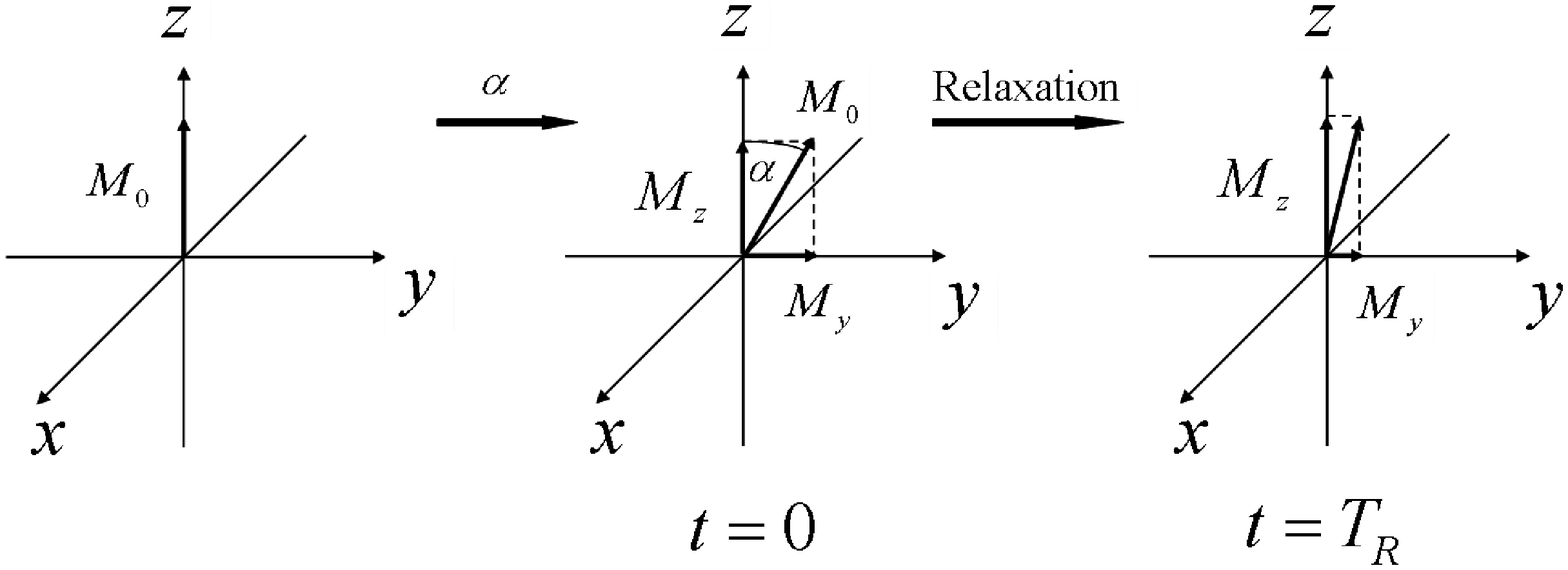}
\caption[Anregung und Relaxation]{\label{Fig:FLASH}{\footnotesize Anregung und Relaxation. In einem \"au{\ss}eren Magnetfeld stellt sich die Magnetisierung parallel zur Magnetfeldrichtung ein. Ein zum Zeitpunkt $t=0$ eingestrahlter Hochfrequenzpuls lenkt die Magnetisierung um den Winkel $\alpha$ aus. Die jetzt stattfindende Relaxation versucht die Magnetisierung wieder in das Gleichgewicht zu bringen, bis nach der Repetitionszeit $T_R$ der n\"achste Hochfrequenzpuls eingestrahlt wird. Die L\"ange des Magnetisierungsvektors bleibt w\"ahrend der Relaxation nicht konstant, da longitudinale und transversale Relaxation mit unterschiedlichen Relaxationszeiten erfolgen.}}
\end{center}
\end{figure}

Die bisherigen Darstellungen konzentrierten sich auf die Untersuchung des freien Induktionszerfalles nach einem Anregungspuls, durch den die Magnetisierung, welche urspr\"unglich keine transversale Komponente besa{\ss}, aus der $z$-Richtung ausgelenkt wurde. Sequenzen, bei denen jedes Mal die transversale Magnetisierung direkt vor der Hochfrequenzanregung zerst\"ort wird, werden als FLASH-Sequenzen (Fast-Low-Angle-Shot) bezeichnet \cite{Haase86}. Um ein Verschwinden der transversalen Komponente der Magnetisierung vor jedem weiteren Anregungspuls zu realisieren, k\"onnen Spoilergradienten die verbleibende transversale Magnetisierung dephasieren \cite{Frahm87,Wood87} oder es k\"onnen spezielle Zyklen der Pulsphase gew\"ahlt werden \cite{Crawley88}. Durch die Anregung mit dem Flipwinkel $\alpha$ entstehen die longitudinale Komponente $M_0 \cos \alpha$ und die transversale Komponente $M_0 \sin \alpha$ (siehe Abbildung \ref{Fig:FLASH}).
Die jetzt stattfindende Relaxation bringt die Gesamtmagnetisierung wieder in das Gleichgewicht. Dabei wird die transversale Komponente kleiner, die longitudinale Komponente nimmt zu und stellt sich parallel zum \"au{\ss}eren Magnetfeld ein. Der n\"achste Hochfrequenzpuls wird nach der Wiederholzeit $T_R$ eingestrahlt. Im Gleichgewicht wird durch die $T_1$-Relaxation genau so viel longitudinale Magnetisierung erzeugt, wie durch den Anregungspuls verloren geht. F\"ur den Betrag der transversalen Komponente, der letztlich eine messbare Spannung in der Empf\"angerspule induziert, gilt in diesem Fall die Ernst-Gleichung \cite{Ernst66}
\begin{align} \label{FLASH}
|M_x + \text{i} M_y| = M_0 \frac{1 - \text{e}^{-\frac{T_R}{T_1}}}{1 - \text{e}^{-\frac{T_R}{T_1}}\cos \alpha} \sin \alpha \,,
\end{align}
wobei diese transversale Komponente das Maximum annimmt, wenn die Anregung mit dem Ernst-Winkel $\alpha_E = \arccos \exp( - T_R / T_1)$ erfolgt.

Die freie Pr\"azession im station\"aren Zustand (SSFP) ist dadurch gekennzeichnet, dass direkt vor dem Hochfrequenzpuls auch transversale Magnetisierung vorliegt. F\"ur die transversale Komponente nach dem Hochfrequenzpuls ergibt sich der Ausdruck
\begin{align} \label{SSFP}
M_y(\omega) + \text{i} M_x(\omega) = (1-E_1) \sin\alpha \, \frac{1+E_2 \text{e}^{-\text{i} \omega T_R}}{p+q \cos \omega T_R} \,,
\end{align}
wobei $\alpha$ der Flipwinkel und $T_R$ die Repetitionszeit zwischen zwei Anregungspulsen ist \cite{Haacke99}. Die Parameter $p$ und $q$ sind gegeben durch \cite{Hinshaw76}
\begin{align}
p & = 1 - E_1 E_2^2 + (E_2^2 - E_1) \cos \alpha \;\;\;\text{und}\\[2ex]
q & = E_2 (1 - E_1) (1 + \cos \alpha) \,,
\end{align}
mit den Abk\"urzungen $E_1 = \exp(-T_R / T_1)$ und $E_2 = \exp(-T_R / T_2)$. Man sieht, dass in diesem Fall die transversale Magnetisierung von der Resonanzfrequenz $\omega$ abh\"angt. Da sowohl im Falle einer FLASH-Sequenz als auch einer SSFP-Sequenz die entstehende transversale Magnetisierung nur von Sequenzparametern ($\alpha$ und $T_R$) sowie von intrinsischen Parametern der Probe ($T_1$ und $T_2$) abh\"angt, wird die Gr\"o{\ss}e $M_x(\omega) + \text{i} M_y(\omega)$ oft auch als Responsefunktion der verwendeten Sequenz bezeichnet. 

\begin{figure}
\begin{center}
\includegraphics[width=12cm]{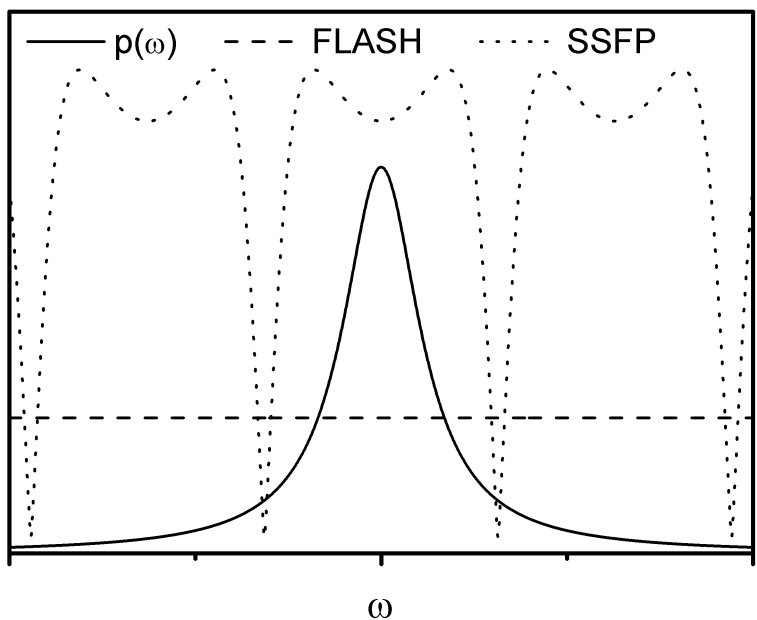}
\caption[Frequenzverteilung und Responsefunktion]{\label{Fig:Intro}{\footnotesize Frequenzverteilung und Responsefunktion. Dargestellt sind die Gr\"o{\ss}en, welche die Signalentstehung entsprechend Gleichung (\ref{Scheffler}) beeinflussen. Die Responsefunktion $M_x(\omega) +\mathrm{i}M_y(\omega)$ ist sequenzspezifisch und beschreibt das Anregungsverhalten f\"ur verschieden Frequenzen (gestrichelte Linie f\"ur eine FLASH-Sequenz nach Gleichung (\ref{FLASH}), gepunktete Linie f\"ur eine SSFP-Sequenz nach Gleichung (\ref{SSFP})). Mit der durchgezogenen Linie ist eine lorentzf\"ormige Frequenzverteilung entsprechend Gleichung (\ref{Lorentz}) dargestellt.}}
\end{center}
\end{figure}

Das Voxel, aus dem das Signal kommt, kann man gedanklich in viele kleine Subvoxel zerlegen. Jedes Subvoxel kann als homogen ansehen werden und diesem Subvoxel wird die lokale Resonanzfrequenz $\omega_i$ zugeordnet. Im Sinne eines Histogrammes kann nun ermittelt werden, wie viele Subvoxel die gleiche lokale Resonanzfrequenz $\omega_i$ besitzen. Somit erh\"alt man eine Wahrscheinlichkeitsverteilung $p(\omega)$, die der Frequenzverteilung in Gleichung (\ref{Mk}) entspricht. Jede lokale Resonanzfrequenz $\omega$ kommt mit einer bestimmten Wahrscheinlichkeit $p(\omega)$ vor. Des Weiteren erzeugt die gew\"ahlte Pulssequenz f\"ur jede lokale Resonanzfrequenz die Magnetisierung $M_x(\omega)+\text{i}M_y(\omega)$. Um das Signal vom gesamten Voxel zu erhalten, muss man jede Magnetisierung mit der entsprechenden Wahrscheinlichkeit multiplizieren und \"uber alle Subvoxel summieren. Im Riemannschen Sinne geht die Summe \"uber alle Subvoxel in ein Integral \"uber alle Frequenzen \"uber und es folgt \cite{Freeman71,Gyngell88,Scheffler03}:
\begin{equation} \label{Scheffler}
M(t) = \int_{-\infty}^{+\infty} p(\omega) \left[M_x(\omega)+\text{i}M_y(\omega) \right] \text{e}^{\text{i}\omega t} \; \text{d}\omega \,.
\end{equation}
Das Signal, das in einem NMR-Experiment entsteht, wird eindeutig durch die untersuchte Probe oder das zu untersuchende Gewebe (charakterisiert durch die Frequenzverteilung $p(\omega)$) und die benutzte Sequenz (charakterisiert durch die Responsefunktion $M_x(\omega) + \mathrm{i}M_y(\omega)$) festgelegt. Im Spezialfall einer FLASH-Sequenz \cite{Haase86} ist die Responsefunktion konstant ($M_x(\omega) + \text{i}M_y(\omega) = \text{const.}$) und deshalb ist das entstehende Signal die Fourier-Transformierte der Frequenzverteilung, wie aus Gleichung (\ref{Scheffler}) zu erkennen ist. In Abbildung \ref{Fig:Intro} sind die zur Signalentstehung beitragenden Gr\"o{\ss}en dargestellt. 

\section{K\"orper im Magnetfeld}

\subsection{Allgemeines Modell} \label{Subsec:Allgemeines Modell}
\begin{figure}
\begin{center}
\includegraphics[height=7cm]{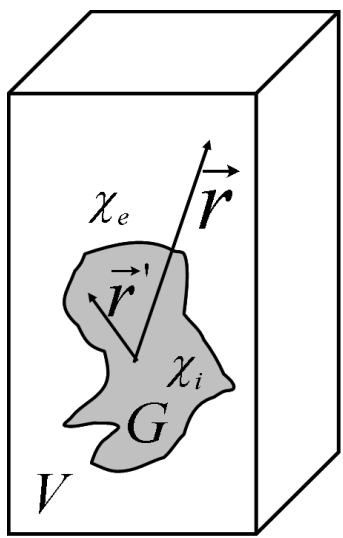}
\caption[Voxel mit magnetischem K\"orper]{{\footnotesize Voxel mit magnetischem K\"orper. Der magnetische K\"orper $G$ mit der Suszeptibilit\"at $\chi_i$ und das umgebende Dephasierunsvolumen $V$ mit der Suszeptibilit\"at $\chi_e$ befinden sich innerhalb eines Voxels. Die Koordinaten innerhalb des K\"orpers $G$ sind durch den Vektor ${\bf r'}$ gekennzeichnet und die Koordinaten des Relaxationsvolumens $V$ durch die Vektoren ${\bf r}$.}} \label{fig1}
\end{center}
\end{figure}

Zur Beschreibung der Diffusion und der damit verbundenen Relaxationsprozesse in den zu betrachtenden Geweben werden Modelle ben\"otigt, die einerseits die physiologischen Gegebenheiten richtig darstellen, andererseits aber auch eine mathematische Beschreibung erm\"oglichen. Um diesen Anforderungen gerecht zu werden, wird ein magnetischer K\"orper in einem Voxel betrachtet (siehe Abbildung \ref{fig1}). Diese beliebige Verteilung magnetischen Materials $G$ innerhalb eines Voxels verursacht einen Suszeptibilit\"atssprung $\Delta\chi = \chi_i - \chi_e$ zum umgebenden Medium mit dem Volumen $V$. Der Volumenanteil magnetischen Materials innerhalb des Voxels ist $\eta = G/(G+V)$. Die Dephasierung findet im verbleibenden Volumen des Voxels $V$ um den magnetischen St\"ork\"orper herum statt. Die Diffusion der Spins wird durch den Diffusionskoeffizienten $D$ beschrieben.
Das lokale inhomogene Magnetfeld, das durch den K\"orper $G$ erzeugt wird, berechnet sich nach den Gesetzen der Elektrodynamik \cite{Jackson04,Salomir03}. Im K\"orper $G$ sollen keine Str\"ome flie{\ss}en, d. h. die Stromdichteverteilung verschwindet (${\bf j} = {\bf 0}$). Die magnetischen Momente im K\"orper erzeugen die Magnetisierung ${\bf M}({\bf r}^{'})$. Ausgangspunkt der Berechnung des lokalen Magnetfeldes um den K\"orper $G$ sind die beiden Maxwell-Gleichungen f\"ur die magnetische Flussdichte ${\bf B}$ und die magnetische Feldst\"arke ${\bf H}$:
\begin{align}
\text{div} {\bf B} & = 0 \;\;\;\; \text{und} \\[2ex]
\text{rot} {\bf H} & = {\bf 0} \,.
\end{align}
Aufgrund der verschwindenden Rotation in der zweiten Maxwell-Gleichung ist es m\"oglich, ein skalares Potential $\varphi$ zu definieren, so dass die magnetische Feldst\"arke als
\begin{equation} \label{Pot}
{\bf H} = -\nabla \varphi
\end{equation}
geschrieben werden kann. Die magnetische Flussdichte ${\bf B}$ setzt sich aus der magnetischen Feldst\"arke ${\bf H}$ und der Magnetisierung ${\bf M}$ zusammen:
\begin{equation}
{\bf B} = \mu_0 ({\bf H} + {\bf M}) \,.
\end{equation}
Aus der ersten Maxwell-Gleichung ergibt sich nun
\begin{equation}
\text{div}({\bf H} + {\bf M}) = 0\,.
\end{equation}
Mit der Darstellung der magnetischen Feldst\"arke durch das skalare Potential ergibt sich ein Zusammenhang zwischen der vorgegebenen Magnetisierung ${\bf M}$ und dem gesuchten Potential $\varphi$:
\begin{equation}
\Delta \varphi = \text{div} {\bf M} \,.
\end{equation}
Diese Gleichung ist analog zur Poisson-Gleichung der Elektrostatik. Das gesuchte Potential l\"asst sich durch das Poisson-Integral ausdr\"ucken:
\begin{equation} \label{Poisson}
\varphi({\bf r}) = - \frac{1}{4\pi} \int_G \mathrm{d}^3{\bf r'} \frac{\text{div}{\bf M}({\bf r'})}{|{\bf r-r'}|} \,.
\end{equation}
Der Integrant kann noch entsprechend den Regeln der Vektoranalysis umgeformt werden:
\begin{align}
\frac{\text{div}{\bf M}({\bf r'})}{|{\bf r-r'}|} & = \text{div} \left( \frac{{\bf M}({\bf r'})}{|{\bf r-r'}|} \right) - {\bf M}({\bf r'}) \cdot \nabla_{r'} \frac{1}{|{\bf r-r'}|} \\[2ex]
& =  \text{div} \left( \frac{{\bf M}({\bf r'})}{|{\bf r-r'}|} \right) + {\bf M}({\bf r'}) \cdot \nabla_{r} \frac{1}{|{\bf r-r'}|} \,.
\end{align}
Der erste Summand f\"uhrt mit Hilfe des Gau{\ss}schen Satzes in Gleichung (\ref{Poisson}) zu einem Oberfl\"achenintegral, das wegen der Lokalisation von ${\bf M}$ verschwindet. Es bleibt
\begin{equation} \label{Poisson2}
\varphi({\bf r}) = - \frac{1}{4\pi} \nabla_r \int_G \mathrm{d}^3{\bf r'} \frac{{\bf M}({\bf r'})}{|{\bf r-r'}|} \,.
\end{equation}
Betrachtet werden K\"orper, die homogen in $z$-Richtung, d. h. parallel zum \"au{\ss}eren magnetischen Feld magnetisiert sind. Deshalb l\"asst sich die Magnetisierung in der Form ${\bf M}({\bf r'}) = M_0 {\bf e}_z$ schreiben, und folglich bleibt nur die Ableitung in $z$-Richtung erhalten:
\begin{equation} \label{Poisson3}
\varphi({\bf r}) = - \frac{M_0}{4\pi} \frac{\partial}{\partial z} \int_G \frac{\mathrm{d}^3{\bf r'}}{|{\bf r-r'}|} \,.
\end{equation}
Die magnetische Feldst\"arke ergibt sich nun nach Gleichung (\ref{Pot}). Die magnetische Flussdichte an einem Punkt im Voxel ergibt sich zu ${\bf B}({\bf r})=\mu_0 {\bf H}({\bf r}) = - \mu_0 \nabla \varphi ({\bf r})$. Es wird also durch den K\"orper eine lokale Flussdichte mit Komponenten in alle drei Raumrichtungen erzeugt. Die Komponente in $z$-Richtung \"uberlagert sich mit dem \"au{\ss}eren Magnetfeld zu $B_0 + B_z$. Die gesamte Flussdichte an einem Ort ergibt sich zu
\begin{align}
\nonumber
B & = \sqrt{ \left( B_0 + B_z \right)^2 + B_x^2 + B_y^2 } \\[3ex]
\nonumber
  & = \sqrt{ B_0^2 + 2 B_0 B_z + \underbrace{B_z^2 + B_x^2 + B_y^2}_{\approx 0}} \\[-3ex]
  & \\
\nonumber
  & \approx B_0 \underbrace{\sqrt{1 + 2 \frac{B_z}{B_0}}}_{\approx 1 + \frac{B_z}{B_0}} \\
\nonumber
  & \approx B_0 + B_z \,.
\end{align}
Aus dieser Ableitung wird ersichtlich, dass nur die $z$-Komponente des lokalen inhomogenen Feldes einen wesentlichen Beitrag zur Pr\"azession der umgebenden Spins leistet. Deshalb braucht nur die $z$-Komponente des Nabla-Operators in Gleichung (\ref{Pot}) ber\"ucksichtigt werden:
\begin{equation}
B_z({\bf r}) = \mu_0 H_z({\bf r}) = - \mu_0 \frac{\partial}{\partial z} \varphi ({\bf r}) \,.
\end{equation}
Mit dem Ausdruck f\"ur das Potential aus Gleichung (\ref{Poisson3}) ergibt sich 
\begin{equation}
\label{Eq1}
B_z({\bf r}) = \frac{\mu_0 M_0}{4 \pi} \frac{\partial^2}{\partial z^2}\int_G\frac{\mathrm{d}^3{\bf r'}}{|{\bf r-r'}|}\ ,
\end{equation}
wobei ${\bf r'}$ alle Punkte innerhalb des K\"orpers $G$ und ${\bf r}$ alle Punkte im umgebenden Volumen $V$ sind (siehe Abbildung \ref{fig1}). Es wird also nur das durch den K\"orper erzeugte inhomogene Magnetfeld und dessen Einfluss auf die Dephasierung der Spins im restlichen Volumen $V$ betrachtet. Die Effekte der benachbarten Voxel auf das Magnetfeld im signalgebenden Voxel werden vernachl\"assigt. Entsprechend der Larmor-Relation (\ref{Larmor}) erzeugt dieses ortsabh\"angige lokale Magnetfeld im Voxel die lokale Resonanzfrequenz $\omega({\bf r}) = \gamma B_z({\bf r})$, die in der Form 
\begin{equation}
\label{Eq2}
\omega({\bf r}) = \delta\omega\, f({\bf r})
\end{equation}
dargestellt werden kann. Die charakteristische Frequenz, welche die Suszeptibilit\"atseigenschaften des K\"orpers und die St\"arke des \"au{\ss}eren Magnetfeldes enth\"alt, kann als
\begin{equation}
\label{eEq3}
\delta \omega =\gamma \frac{\mu_0 M_0}{4 \pi}
\end{equation}
geschrieben werden, w\"ahrend die Geometriefunktion
\begin{equation}
\label{Eq3}
f({\bf r}) = \frac{\partial^2}{\partial z^2}\int_G\frac{\mathrm{d}^3{\bf r'}}{|{\bf r-r'}|}
\end{equation}
die Form der Magnetfeldinhomogenit\"at charakterisiert. Damit k\"onnen Suszeptibilit\"atseigenschaften und die r\"aumliche Anordnung des K\"orpers $G$ innerhalb des Voxels voneinander separiert werden. Dadurch ist es m\"oglich, den Einfluss des jeweiligen Anteiles unabh\"angig vom anderen Anteil zu untersuchen.

\subsection{Kugel} \label{Einleit-Kugel}
Ein kugelf\"ormiges Objekt in einem \"au{\ss}eren Magnetfeld dient als physikalisches Modell f\"ur einige biologische Gewebe bzw. Gegebenheiten. So betrachtet man beispielsweise die Lunge als Gewebe, das aus kleinen kugelf\"ormigen Alveolen zusammengesetzt ist. Jede Alveole ist eine luftgef\"ullte Kugel, die von Blut umgeben ist. Der Suszeptibilit\"atsunterschied zwischen Luft und Blut erzeugt das lokale inhomogene Magnetfeld. Des Weiteren k\"onnen durch das Kugelmodell Gewebe beschrieben werden, die magnetisch markierte Zellen enthalten. Die von den Zellen durch Phagozytose aufgenommenen Kontrastmittelteilchen bilden einen kugelf\"ormigen magnetischen Kern vom Radius $R_{\text{S}}$, der konzentrisch vom Dephasierungsvolumen mit dem Radius $R$ umgeben wird. Dieser Radius $R$ ist allerdings nicht mit dem Zellradius zu verwechseln, da das magnetische St\"orfeld der Kontrastmittelteilchen auch im Interzellularraum wirkt. Der Volumenanteil ist durch $\eta=R_{\text{S}}^3/R^3$ definiert. 

\begin{figure}
\begin{center}
\includegraphics[width=8cm]{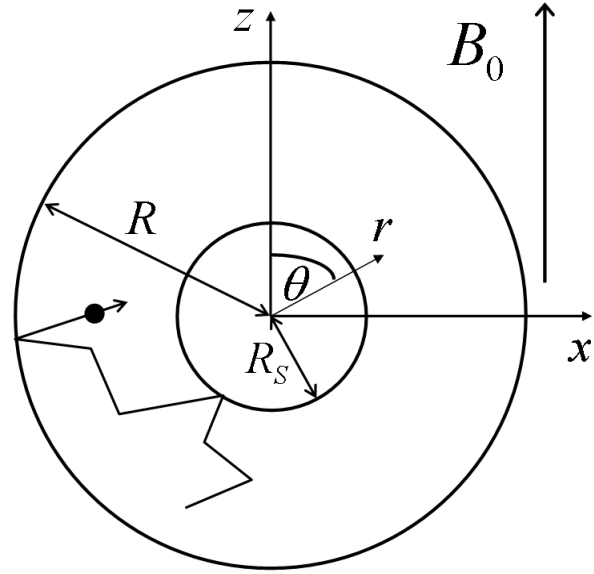}
\caption[Kugel im Magnetfeld]{{\footnotesize Kugel im Magnetfeld. Die Kugel mit dem Radius $R_{\text{S}}$ wird von einer konzentrischen Kugel mit dem Radius $R$ umgeben. Zwischen diesen beiden Kugeln diffundiert der Spin in einem Medium mit dem Diffusionskoeffizienten $D$.}} \label{Zellen}
\end{center}
\end{figure}

In vielen Situationen kann also als physikalisches Modell eine Kugel mit dem Radius $R_{\text{S}}$ angenommen werden, die von einem kugelf\"ormigen Relaxationsvolumen mit dem Radius $R$ umgeben wird (siehe Abbildung \ref{Zellen}). Dabei werden zwei N\"aherungen gemacht. Erstens erzeugen die anderen umliegenden Kugeln auch ein lokales Magnetfeld, das sich auch auf die Resonanzfrequenz des Spins auswirkt, der sich im Dephasierungsvolumen der betrachteten Kugel bewegt. Zweitens wird nur die Diffusion zwischen zwei konzentrischen Kugeln betrachtet, obwohl der umgebende Spin keine Grenzen hat und in seiner Bewegung nicht behindert wird. Diese Einschr\"ankung wird durch die Annahme reflektierender Randbedingungen an der \"au{\ss}eren Begrenzung des Dephasierungsvolumens gerechtfertigt. Ber\"uhrt also ein Spin die \"au{\ss}ere Begrenzung (siehe Abbildung \ref{Zellen}), wird er reflektiert und die Bewegung setzt sich im Dephasierungsvolumen fort.
Das inhomogene Magnetfeld um eine homogen magnetisierte Kugel ergibt sich nach Gleichung (\ref{Eq1}) als das Feld eines magnetischen Dipols, d. h. in Kugelkoordinaten $ (r,\theta,\phi) $\cite{Landau}:
\begin{equation} \label{B-Feld}
B_z(\mathbf{r}) = \mu_0 \, \frac{\Delta M}{3} \, R_{\text{S}}^3 \, \frac{3 \cos^2 \theta - 1}{r^3} \;,
\end{equation}
wobei $ \Delta M =M_{\text{S}}-M_{\text{T}} $ die Differenz der Magnetisierungen zwischen dem magnetischen Kern $ M_{\text{S}} $ und dem umgebenden Gewebe $ M_{\text{T}} $ ist. Der charakteristische \"aquatoriale Frequenzshift
\begin{align}
\nonumber
\delta\omega & = \gamma \, |B_z(r = R_{\text{S}},\theta = \pi/2)|
\\[-1ex] & \\[-1ex]
\nonumber & = \gamma \, \mu_0 \,\frac{\Delta M}{3}
\end{align}
beschreibt die St\"arke der Magnetfeldinhomogenit\"at, und die Formfunktion 
\begin{equation}
f({\bf r}) = R^2_{\text{S}}\frac{3\cos^2\theta-1}{r^3}
\end{equation}
charakterisiert die Abh\"angigkeit von den Koordinaten. Damit ergibt sich die lokale Resonanzfrequenz zu
\begin{equation} \label{om-Feld-Kugel}
\omega(\mathbf{r}) = \delta \omega \, R_{\text{S}}^3 \, \frac{3 \cos^2 \theta - 1}{r^3} \;.
\end{equation}

\subsection{Zylinder} \label{Einleit-Zylinder}
\begin{figure}
\begin{center}
\includegraphics[width=10cm]{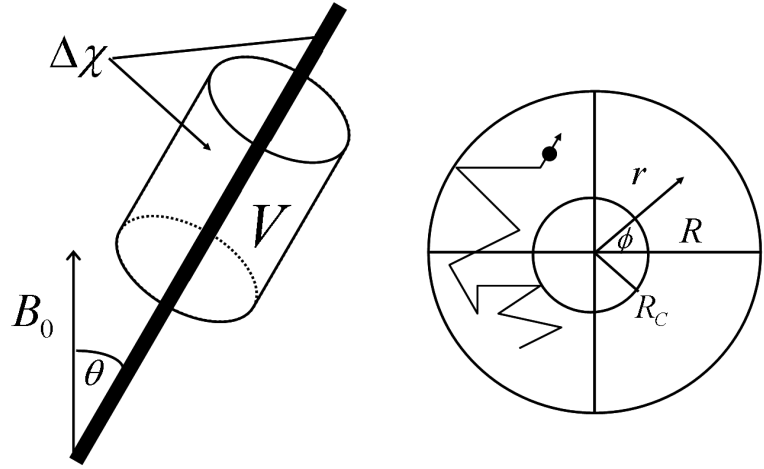}
\caption[Kroghsches Kapillarmodell]{{\footnotesize Kroghsches Kapillarmodell. Links: Kapillare mit Neigungswinkel $\theta$ zum  \"au{\ss}eren Magnetfeld $B_0$. Rechts: Querschnitt durch Voxel und Kapillare in Polarkoordinaten. Der diffundierende Spin ist dargestellt.}} \label{krogh}
\end{center}
\end{figure}

Als Modell, beispielsweise f\"ur eine Kapillare, wird ein d\"unner Zylinder mit dem Radius $R_{\text{C}}$ angenommen, der einen Neigungswinkel $\theta$ zum \"au{\ss}eren Magnetfeld $B_0$ hat (siehe Abbildung \ref{krogh}). Dabei wird das aus der Physiologie bekannten Kroghsche Zylindermodell zugrunde gelegt, das urspr\"unglich zur Beschreibung des Sauerstoffpartialdrucks in vaskularisiertem Gewebe entwickelt wurde \cite{Krogh19}. Dazu wird ein Gef\"a{\ss} als ein blutgef\"ullter Zylinder (mit Radius $R_{\text{C}}$) betrachtet, der konzentrisch von einem zylinderf\"ormigen Versorgungsgebiet (mit Radius $R$) umgeben wird (siehe linke Seite der Abbildung \ref{krogh}). Der Volumenanteil ist definiert als $\eta=R_{\text{C}}^2/R^2$. Im Raum zwischen den beiden konzentrischen Zylindern findet die Diffusion der Spins statt, deren Dephasierung untersucht wird. Deshalb wird dieses Gebiet als Dephasierungsvolumen bezeichnet. An der Oberfl\"ache dieses Dephasierungsvolumens werden reflektierende Randbedingungen angenommen, d. h., jede Trajektorie eines Spins, welche die Oberfl\"ache des Dephasierungsvolumens an der Stelle $R$ ber\"uhrt, wird durch eine symmetrische Trajektorie ersetzt, die wieder innerhalb des urspr\"unglich betrachteten Dephasierungsvolumens liegt (siehe rechte Seite der Abbildung \ref{krogh}).
An der Oberfl\"ache der Kapillare k\"onnen sowohl reflektierende als auch strahlende Randbedingungen betrachtet werden. Dabei kann im Falle strahlender Randbedingungen auch die Permeabilit\"at der Kapillare ber\"ucksichtigt werden. Der Einfluss der Randbedingungen wird in Kapitel \ref{Kap.Korr} ausf\"uhrlich untersucht.

Der Zylinder erzeugt zum umgebenden Gewebe die Suszeptibilit\"atsdifferenz $\Delta\chi$, was zu dem lokalen inhomogenen Magnetfeld
\begin{equation}
B_z({\bf r}) = \frac{\Delta\chi}{2}B_0\sin^2\theta\, R_{\text{C}}^2 \frac{\cos2\phi}{r^2}
\end{equation}
f\"uhrt, wobei $ (r,\phi) $ f\"ur Polarkoordinaten in einer Ebene senkrecht zur Zylinderachse stehen \cite{Reichenbach01}. Auf der Oberfl\"ache des Zylinders nimmt das lokale Magnetfeld f\"ur den Winkel $\phi = 0$ das Maximum $B_z(r=R_{\text{C}},\phi = 0)$ an. Die lokale Resonanzfrequenz ist durch die Larmor-Beziehung $\omega(\mathbf{r})=\gamma B(\mathbf{r})$ gegeben:
\begin{equation}
\omega({\bf r}) = \underbrace{ \gamma \frac{\Delta\chi}{2}B_0\sin^2\theta }_{\delta\omega_{\theta}}\, \underbrace{ R_{\text{C}}^2 \frac{\cos2\phi}{r^2} }_{f({\bf r})} \,.
\end{equation}
Der \"aquatoriale Frequenzshift
\begin{align} \label{Shift}
\nonumber 
\delta\omega_{\theta} & = \gamma \, B_z(r=R_{\text{C}},\phi = 0)
\\[-1ex] & \\[-1ex]
\nonumber & = \gamma \, \frac{\Delta\chi}{2} \, B_0 \, \, \sin^2 \theta
\end{align}
charakterisiert die St\"arke des Suszeptibilit\"atseffekts, und die Funktion 
\begin{equation}
\label{Eq18}
f({\bf r}) = R_{\text{C}}^2 \frac{\cos2\phi}{r^2}
\end{equation}
beschreibt die r\"aumliche Form des lokalen Magnetfelds. Damit kann man die lokale Resonanzfrequenz in der Form
\begin{equation} \label{om-Feld-Zyl}
\omega(\mathbf{r}) = \delta \omega_{\theta} \, R_{\text{C}}^2 \, \frac{\cos 2\phi}{r^2}
\end{equation}
schreiben.

\section{Diffusion} \label{Sec:Diffusion}

\subsection{Bloch-Torrey-Gleichung} \label{Subsec:BTG}
Das NMR-Signal entsteht durch die transversale Komponente der Magnetisierung innerhalb eines Voxels, die in der Empf\"angerspule eine Spannung induziert. Um die Zeitentwicklung dieses Signals zu beschreiben, wird wie in Abbildung \ref{fig1} veranschaulicht ein einzelnes Voxel betrachtet, in dem sich ein K\"orper $G$ befindet. Das durch den K\"orper erzeugte lokale Magnetfeld (\ref{Eq1}) induziert die dazugeh\"orige Larmor-Frequenz $\omega(\mathbf{r}) = \gamma \, B(\mathbf{r})$. In diesem inhomogenen Magnetfeld findet die Diffusion der Spins statt.

Um den Einfluss der Diffusion auf den Signalverlauf bzw. die Frequenzverteilung zu beschreiben, wurde von Torrey ein zus\"atzlicher Diffusionsterm in die Bloch-Gleichungen eingef\"uhrt \cite{Torrey56}. F\"ur die transversale Magnetisierung in polarer Form $ m(\mathbf{r},t) = M_x(\mathbf{r},t) - \text{i} \, M_y(\mathbf{r},t)$ ergibt sich somit die folgende Bewegungsgleichung:
\begin{align}
\label{BT}
\frac{\partial}{\partial t} m(\mathbf{r},t) \, = \, \underbrace{D \nabla^2 m(\mathbf{r},t)}_{\text{Diffusionsterm}} + \underbrace{\text{i} \, \omega (\mathbf{r}) m(\mathbf{r},t)}_{\text{Inhomogenit\"atsterm}}\, , 
\end{align}
wobei $\omega({\bf r})$ die lokale Larmor-Frequenz des inhomogenen Magnetfeldes ist und $D$ der Diffusionskoeffizienten des umgebenden Mediums ist. Mathematisch gesehen handelt es sich um eine partielle Differentialgleichung zweiter Ordnung vom Typ der Schr\"odingergleichung mit rein imagin\"arem Potential. Eine formale Zeitintegration der Bloch-Torrey-Gleichung liefert
\begin{equation}
\label{eEq6}
m({\bf r}, t) = m({\bf r}, 0) \exp\Big\{[D\Delta+\text{i} \omega({\bf r})]t \Big\}\, .
\end{equation}
Das resultierende Signal aus dem gesamten Voxel ist demnach
\begin{equation}
\label{Eq7}
M(t) = \frac{1}{V} \int_V \mathrm{d}^3{\bf r} \, m({\bf r}, 0) \exp \Big\{[D\Delta+\text{i} \omega({\bf r})]t \Big\} \,.
\end{equation}

\subsection{Strong-Collision-N\"aherung} \label{Subsec:SCA}
Um die Bloch-Torrey-Gleichung zu l\"osen, wird der Diffusionsprozess im lokalen inhomogenen Magnetfeld um den magnetisierten K\"orper im Sinne von \"Ubergangsdynamiken als stochastischer Prozess beschrieben. Aufgrund der Diffusion durch das lokale inhomogene Magnetfeld ist der betrachtete Spin verschiedenen Larmor-Frequenzen $\omega({\bf r})$ zu verschiedenen Zeiten ausgesetzt. Durch Einf\"uhrung einer Sprungdynamik zwischen den verschiedenen lokalen Larmor-Frequenzen kann man den Diffusionsprozess diskretisieren. Dies kann realisiert werden, indem der Diffusionsoperator durch die Ratenmatrix ${\bf R} = (r_{ij})$ ersetzt wird, wobei $r_{ij}$ die \"Ubergangsrate eines Spins von einem Ort mit der lokalen Larmor-Frequenz $\omega_i$ zu einem Ort mit der lokalen Larmor-Frequenz $\omega_j$ beschreibt. Unter der Annahme, dass diese \"Ubergangswahrscheinlichkeit nur von der aktuellen Position des Spins abh\"angt, kann der Diffusionsprozess durch einen Markov-Prozess ersetzt werden, der durch den Generator ${\bf R}$ beschrieben wird. In dieser N\"aherung wird die Zeitentwicklung der transversalen Magnetisierung des Spins in der Form $\partial_t m_i(t) = \sum_j r_{ij} m_j(t) + \mathrm{i}\omega_i m_i(t)$ geschrieben, wobei $m_i(t)$ die transversale Magnetisierung eines Spins an einem Ort mit der lokalen Larmor-Frequenz $\omega_i$ ist. Die durch den magnetisierten K\"orper verursachten lokalen Resonanzfrequenzen k\"onnen in Matrixform ${\bf\Omega} = (\omega_i\delta_{ij})$ geschrieben werden. Mit dem Vektor $|m(t)\rangle$, dessen $i$-tes Element die transversale Magnetisierung bei der Larmor-Frequenz $\omega_i$ beschreibt, ergibt sich eine Verallgemeinerung der Bloch-Torrey-Gleichung in Operatorform:
\begin{align}
\label{OBT}
\frac{\partial}{\partial t}\,|m(t)\rangle &= ({\bf R+\mathrm{i}\Omega})\,|m(t)\rangle \, .
\end{align}
In der Originalarbeit von Torrey \cite{Torrey56} wurden reine Diffusionsprozesse betrachtet, also ${\bf R} = \nabla D(x)\nabla$ und $\Omega = \omega(x)$, mit der lokalen Larmor-Frequenz $\omega(x)$ und dem lokalen Diffusionskoeffizienten $D(x)$. Demzufolge ist die urspr\"ungliche Bloch-Torrey-Gleichung (\ref{BT}) ein Spezialfall der allgemeinen Gleichung (\ref{OBT}). Die formale L\"osung von Gleichung (\ref{OBT}) ist durch $|m(t)\rangle = \exp[{\bf R + \mathrm{i}\Omega}]t\,|m(0)\rangle$ gegeben. Damit kann man den Zeitverlauf der \"uber das gesamte Dephasierungsvolumen gemittelten Magnetisierung berechnen:
\begin{align}
\label{OMagn}
M(t) = \langle m(0)|\,\exp[{\bf R + \mathrm{i}\Omega}]t\,|m(0)\rangle\,.
\end{align}

Um diesen allgemeinen Ausdruck zu l\"osen, wird die Strong-Collision-N\"aherung verwendet \cite{Bauer99,Dattagupta74}. Der Ausdruck \glqq Strong-Collision\grqq\ wurde von verschiedenen Autoren in unterschiedlichen Zusammenh\"angen benutzt. Hier wird das Strong-Collision-Modell im Sinne von Dattagupta und Blume \cite{Dattagupta76} angewandt, welches mit dem Modell \"ubereinstimmt, dass auch von Lynden-Bell \cite{Lynden-Bell71} benutzt wurde. Diese Approximation wurde von der statistischen Physik adaptiert und zur Charakterisierung ergodischer Markov-Prozesse genutzt. Dies sind Prozesse, bei denen Anfangs- und Endzustand stochastisch unabh\"angig sind und die \"Ubergangswahrscheinlichkeit zwischen Anfangs- und Endzustand proportional zur Gleichgewichtswahrscheinlichkeit des Endzustandes ist.

In der Strong-Collision-Approximation wird der Generator des stochastischen Prozesses in folgender Weise ersetzt:
\begin{equation} 
\label{SC}
{\bf R} \, \xrightarrow[\text{collision}]{\text{strong}} \lambda ({\bf \Pi - 1}) \,,
\end{equation}
wobei ${\bf \Pi} = |0\rangle\langle 0|$ ein Projektionsoperator auf einen Unterraum ist, der von dem Gleichgewichtseigenvektor $|0\rangle$ des Generators ${\bf R}$ aufgespannt wird. Der identische Operator wird mit ${\bf 1}$ bezeichnet. Der Magnetisierungsvektor $|m(0)\rangle$ zum Zeitpunkt $t=0$ ist dem Gleichgewichtseigenvektor proportional: $|m(0)\rangle\propto|0\rangle$. Der in der Approximation (\ref{SC}) eingef\"uhrte Fluktuationsparameter $\lambda$ ist derjenige Parameter, der die Zeitskala der \"Uberg\"ange des Markov-Prozesses beschreibt. Er verdeutlicht, wie schnell die Diffusion abl\"auft. Um diesen Parameter zu bestimmen, wird die Korrelationsfunktion $K(t)$ des urspr\"unglichen Diffusionsprozesses betrachtet, welcher der Bloch-Torrey-Gleichung (\ref{BT}) folgt. Diese Korrelationsfunktion ist folgenderma{\ss}en definiert:
\begin{equation}
\label{KF}
K(t) = \langle\omega(t)\omega(0)\rangle = \langle0|\,{\bf \Omega}\exp[{\bf R}t]{\bf \Omega}\,|0\rangle\, .
\end{equation}
Wie in fr\"uheren Arbeiten gezeigt wurde \cite{Bauer99,Ziener06a}, kann eine Korrelationszeit $\tau$ eines diffundierenden Spins eingef\"uhrt werden, die im Sinne einer Mean-Relaxation-Time-Approximation \cite{Nadler85} bestimmt werden kann \cite{Bauer99,Ziener06a}:
\begin{align} 
\label{taudev}
\tau :=& \int\limits_0^\infty\mathrm{d}t\, \frac{K(t)}{K(0)} \\
\label{tauint}
      =& \frac{1}{\langle \, \omega^2(\mathbf{r}) \, \rangle D V} \int_V \text{d}^3 \mathbf{r} \,\, \omega(\mathbf{r}) \left[ - \frac{1}{\nabla^2} \right] \omega(\mathbf{r}) \,.
\end{align}
Diese Korrelationszeit charakterisiert die Zeitskala der durch die molekulare Bewegung induzierten Feldfluktuationen. Um einen Ausdruck f\"ur die Korrelationszeit zu erhalten wird angenommen, dass die Diffusion im Dephasierungsvolumen zwischen der Oberfl\"ache des K\"orpers und der Oberfl\"ache des Voxels stattfindet. Wie in den Arbeiten \cite{Stables98,Ziener06a,Ziener06b} gezeigt wurde, ist die Integration problemlos ausf\"uhrbar und ergibt $\tau = L^2k(\eta)/D$, wobei $L$ eine charakteristische L\"ange des K\"orpers ist, wie z. B. der Radius einer Kapillare oder eines Zylinders. In der Arbeit \cite{Ziener06b} konnte dargestellt werden, dass die Form der Funktion $k(\eta)$ auch nur von der Gestalt der K\"orpers abh\"angt. Der Ausdruck $(1/\nabla^2)\omega({\bf r}) = c({\bf r})$ ist die L\"osung der inhomogenen Laplace-Gleichung $\nabla^2 c({\bf r}) = \omega({\bf r})$ mit den gleichen Randbedingungen wie die der urspr\"unglichen Bloch-Torrey-Gleichung (\ref{BT}). Dies ist m\"oglich, da beides Differentialgleichungen zweiter Ordnung sind. Die beiden Randbedingungen an der Oberfl\"ache des Voxels und der Oberfl\"ache des K\"orpers liefern die beiden Integrationskonstanten der Differentialgleichung zweiter Ordnung.
Auf diese Weise gelingt es, die Randbedingungen des urspr\"unglichen Problems in die Strong-Collision-N\"aherung einzuarbeiten. Ausf\"uhrlich wird dies in Kapitel \ref{Kap.Korr} sowie in den Gleichungen (6) und (7) in \cite{Ziener06a} diskutiert.

Um den Fluktuationsparameter $\lambda$ mit der Korrelationszeit $\tau$ zu verkn\"upfen, wird die Ratenmatrix ${\bf R}$ durch den Operator $\lambda({\bf \Pi -1})$ in der Definition der Korrelationsfunktion ersetzt: $K(t) = \langle0|\,{\bf \Omega}\exp[\lambda({\bf \Pi -1})t]{\bf \Omega}\,|0\rangle\,$. Das Einsetzen dieses Ausdruckes f\"ur die Korrelationsfunktion in die Definition der Korrelationszeit nach Gleichung (\ref{taudev}) ergibt
\begin{align}
\label{taulambda}
\tau = \int\limits_0^\infty\mathrm{d}t\, \frac{\langle0|\,{\bf \Omega}\exp[\lambda({\bf\Pi - 1})t]{\bf \Omega}\,|0\rangle}{\langle0|\,{\bf \Omega}^2\,|0\rangle} = \frac{1}{\lambda} \, .
\end{align}
Demzufolge kann man den Zeitverlauf der Magnetisierung $M(t)$ durch Anwendung der Strong-Collision-Approximation (\ref{SC}) mit $\lambda = \tau^{-1}$ auf die allgemeing\"ultige Gleichung (\ref{OMagn}) erhalten:
\begin{align}
\label{SCMagn}
M(t) = \langle m(0)|\,\exp[\{\tau^{-1}({\bf\Pi - 1}) + \mathrm{i}{\bf \Omega}\}t]\,|m(0)\rangle\,.
\end{align}
Zur weiteren Auswertung der Frequenzverteilung ist es zweckm\"a{\ss}ig, die Laplace-Transformierte des Magnetisierungs-Zeit-Verlaufes in der Form
\begin{equation}
\label{Laplacetransform}
\hat{M}(s) = \int_0^\infty \text{d}t \, \text{e}^{-st}M(t)
\end{equation}
zu betrachten. Berechnet man die Laplace-Transformierte von Gleichung (\ref{SCMagn}) mit Hilfe der Definitionsgleichung (\ref{Laplacetransform}), ergibt sich
\begin{align}
\label{LPSCMagn}
\hat{M}(s) = \langle m(0)|\,\frac{1}{\underbrace{(s+\tau^{-1}){\bf 1} - \mathrm{i}{\bf\Omega}\big.}_{{\displaystyle \bf A}}\underbrace{-\big.\tau^{-1}{\bf\Pi}\big.}_{{\displaystyle\bf B}}}\,|m(0)\rangle\,.
\end{align}
Anwendung der Operatoridentit\"at ${\bf (A+B)^{-1} = A^{-1}-A^{-1}\cdot B\cdot (A+B)^{-1} }$ mit den Abk\"urzungen ${\bf A} =  (s+\tau^{-1}){\bf 1} - \mathrm{i}{\bf\Omega}$ und ${\bf B} = -\tau^{-1}{\bf\Pi}$ auf Gleichung (\ref{LPSCMagn}) ergibt f\"ur die Laplace-Transformierte des Magnetisierungszerfalls
\begin{align}
\label{LPSCM2}
\hat{M}(s) &= \underbrace{\langle m(0)|\,\frac{1}{(s+\tau^{-1}){\bf 1} - \mathrm{i}{\bf\Omega}}\,|m(0)\rangle}_{\displaystyle\hat{M}_0(s+\tau^{-1})}\nonumber\\[-2ex]
& \\[-1ex] 
&+\underbrace{\langle m(0)|\,\frac{1}{(s+\tau^{-1}){\bf 1} - \mathrm{i}{\bf\Omega}}\,|m(0)\rangle}_{\displaystyle\hat{M}_0(s+\tau^{-1})}\cdot\tau^{-1}\cdot
\underbrace{\langle m(0)|\,\frac{1}{s{\bf 1} -{\bf R} - \mathrm{i}{\bf\Omega}}\,|m(0)\rangle}_{\displaystyle\hat{M}(s)}\, ,\nonumber
\end{align}
wobei der Index Null die Gr\"o{\ss}en im Static-Dephasing-Regime, in dem die Diffusionseffekte vernachl\"assigt werden (${\bf R = 0}$), kennzeichnet. Demzufolge ist die Magnetisierung im Static-Dephasing-Regime durch
 $M_0(t) = \langle m(0)|\,\exp[\mathrm{i}{\bf\Omega}]t\,| m(0)\rangle$ gegeben. Aufl\"osen von Gleichung (\ref{LPSCM2}) ergibt den Zusammenhang
\begin{equation} \label{Mdach}
\hat{M}(s) = \frac{\hat{M}_0(s + \tau^{-1})}{1 - \tau^{-1} \cdot \hat{M}_0(s + \tau^{-1})} \ ,
\end{equation}
der f\"ur die Erweiterung des Static-Dephasing-Grenzfalles auf alle Diffusionsregime wichtig ist. Dabei vermittelt die Korrelationszeit $\tau$ den Bezug zur Diffusion.

Die Aussage \"uber die stochastische Unabh\"angigkeit von Anfangs- und Endzustand ist eng mit der Annahme eines ergodischen Systems verbunden. F\"ur ergodische Markovprozesse gilt, dass w\"ahrend der Korrelationszeit alle m\"oglichen Resonanzfrequenzen von einem Spin mit der gleichen Wahrscheinlichkeit besucht werden. Nur unter dieser Annahme ist sichergestellt, dass die Trajektorie des Spins nahezu alle m\"oglichen lokalen Larmor-Frequenzen erfasst, was zur geforderten stochastischen Unabh\"angigkeit von Anfangs- und Endzustand f\"uhrt. Offensichtlich erfordert dies eine starke Diffusion, die durch einen gro{\ss}en Diffusionskoeffizienten charakterisiert ist. Deshalb funktioniert die Strong-Collision-N\"aherung auch im Motional-Narrowing-Regime. Andererseits wurde bei der Berechnung der Korrelationszeit der urspr\"ungliche komplette Diffusionsoperator ber\"ucksichtigt. Das f\"uhrt zur korrekten Abh\"angigkeit der Korrelationszeit vom Diffusionskoeffizienten $\tau\propto D^{-1}$ f\"ur alle Diffusionsregime. Dadurch wird auch das korrekte Verhalten f\"ur kleine Diffusionskoeffizienten eingearbeitet und deshalb k\"onnen die Ergebnisse der Strong-Collision-Approximation auf das Static-Dephasing-Regime erweitert werden. 

\subsection{Diffusionsregime} \label{Subsec:Diffusionsregime}
\begin{figure}
\begin{center}
\includegraphics[width=13cm]{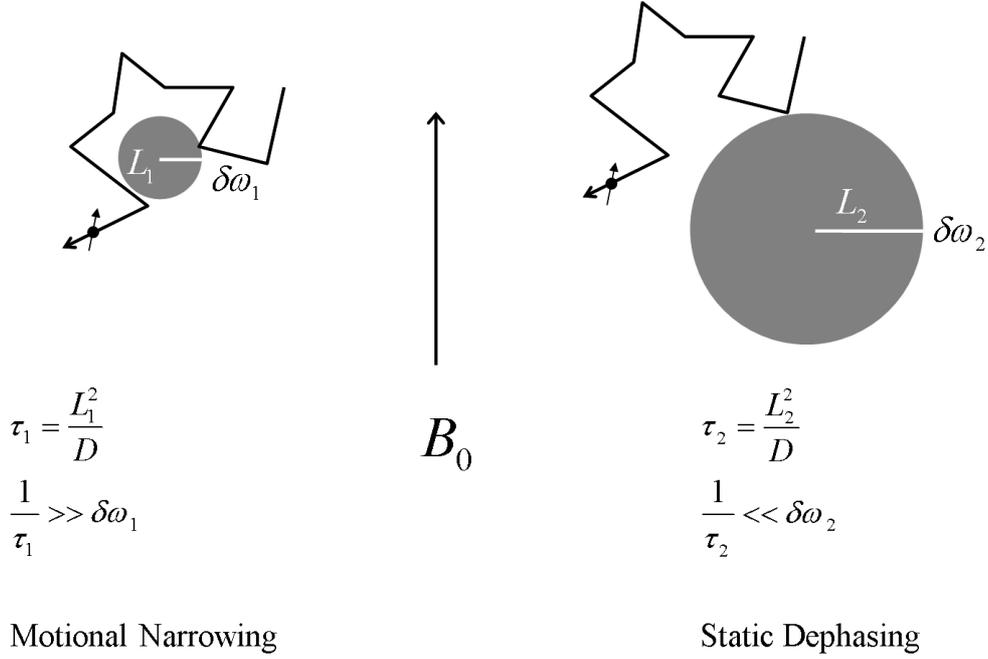}
\caption[Einteilung der Diffusionsregime]{{\footnotesize Einteilung der Diffusionsregime. Links: Motional-Narrowing-Regime: W\"ahrend der Diffusion um einen kleinen K\"orper mit der charakteristischen L\"ange $L_1$ besucht der Spin sehr viele unterschiedliche lokale Resonanzfrequenzen. Rechts: Static-Dephasing-Regime: Der Spin bewegt sich relativ zu dem gro{\ss}en K\"orper kaum und bleibt fast immer am gleichen Ort mit der gleichen Resonanzfrequenz.}} \label{Diffusionsregime}
\end{center}
\end{figure}

Basierend auf dem Bild diffundierender Spins in einem lokalen inhomogenen Magnetfeld um einen magnetisierten K\"orper wird der zugrunde liegende Relaxationsmechanismus durch zwei Frequenzskalen gekennzeichnet. Die dynamische Frequenzskala $ 1/\tau $ charakterisiert den Diffusionsprozess durch die Korrelation der sich bewegenden Spins. Die magnetische Frequenzskala wird von der St\"arke des lokalen Magnetfeldes bestimmt, die durch die \"aquatoriale Frequenz $\delta\omega$ charakterisiert ist. Ein Vergleich beider Frequenzen bestimmt das zugrunde liegende Diffusionsregime \cite{Yung03}. Wenn Diffusionseffekte vernachl\"assigt werden k\"onnen (d. h. $\delta\omega\gg1/\tau$), gilt das Static-Dephasing-Regime; im entgegengesetzten Fall $\delta\omega\ll1/\tau$ werden die N\"aherungen des Motional-Narrowing-Regimes angewandt. Je nach relativer St\"arke dieser beiden charakteristischen Frequenzen k\"onnen f\"unf Diffusionsregime eingef\"uhrt werden, die den Relaxationsprozess beschreiben:
\begin{enumerate}
\item \makebox[2cm][l]{$ 1/\tau \gg \delta\omega $} Motional-Narrowing-Regime,
\item \makebox[2cm][l]{$ 1/\tau \;\, > \; \delta\omega $} Fast-Diffusion-Regime,
\item \makebox[2cm][l]{$ 1/\tau \;\, = \; \delta\omega $} Intermediate Regime,
\item \makebox[2cm][l]{$ 1/\tau \;\, < \; \delta\omega $} Slow-Diffusion-Regime und
\item \makebox[2cm][l]{$ 1/\tau \ll \delta\omega $} Static-Dephasing-Regime.
\end{enumerate}
Die charakteristischen Parameter des zu untersuchenden Gewebes, insbesondere der Frequenzsprung $ \delta\omega $ und die dynamische Frequenz $ 1 / \tau $ entscheiden nun, welches Diffusionsregime das zugrunde liegende ist. In Abbildung \ref{Diffusionsregime} ist der qualitative Unterschied zwischen den beiden Grenzf\"allen dargestellt. Im Motional-Narrowing-Grenzfall besucht der Spin fast alle m\"oglichen lokalen Frequenzen, die durch den magnetischen K\"orper erzeugt werden. Im Static-Dephasing-Regime bleibt der Spin fast immer an einem Ort und pr\"azediert daher auch nur mit einer konstanten Larmor-Frequenz. 
Auf qualitativen Argumenten basierend konnten Yablonskiy und Haacke \cite{Yablonskiy94} das folgende Kriterium f\"ur die Anwendbarkeit des Static-Dephasing-Regime angeben:
\begin{equation}
\frac{(\bar{r}/2)^2}{D} \, \delta\omega \, \frac{\eta}{2d} \, \gg \, 1 \,,
\end{equation}
wobei $\bar{r}/2$ eine f\"ur den magnetisierten K\"orper charakteristische L\"ange und $d=1,2,3$ die zugrunde liegende Dimension ist. In Kapitel \ref{Kap.Korr} werden Korrelationszeiten, welche die Diffusion in inhomogenen Magnetfeldern charakterisieren, untersucht. Damit kann dann auch quantitativ eingeteilt werden, in welchem Diffusionsregime man sich gerade befindet.

\section{Signalformation in einem Voxel} \label{SigVox}

\subsection{Allgemeine Theorie zur Signalentstehung} \label{Allg-Theo}
Um die Zeitentwicklung der transversalen Magnetisierung zu beschreiben, wird ein einzelnes Voxel betrachtet in dem sich ein K\"orper $G$ befindet (siehe Abbildung \ref{fig1}). Der K\"orper wird von Gewebe mit konstantem Diffusionskoeffizienten $D$ umgeben. Das umgebende Gewebe, in dem Diffusion stattfinden, hat das Volumen $V$. Deshalb ist es m\"oglich, den Volumenanteil $\eta$ in der Form $\eta = G/(G+V)$ einzuf\"uhren. Der K\"orper mit der Magnetisierung $M_0$ erzeugt das inhomogene Magnetfeld $B_z({\bf r})$ und somit die r\"aumlich variierenden Larmor-Frequenz $\omega({\bf r})$ innerhalb des Voxels. Wie in Abschnitt \ref{Subsec:Allgemeines Modell} gezeigt wurde, ist die lokale Resonanzfrequenz durch
\begin{equation}
\label{local_frequency}
\omega({\bf r}) = \delta\omega \frac{\partial^2}{\partial z^2} \int_G \frac{\text{d}^3 \mathbf{r}^{'}}{|\mathbf{r}-\mathbf{r}^{'}|}
\end{equation}
gegeben, wobei $\delta\omega = \gamma \mu_0 M_0 /(4 \pi)$ den Frequenzshift charakterisiert, der durch den magnetisierten K\"orper erzeugt wird. Auf der anderen Seite wird die Signalentwicklung durch die Diffusion der Spins beeinflusst, die durch die Korrelationszeit $\tau\propto L^2/D$ veranschaulicht werden kann, wobei $L$ eine charakteristische L\"ange des K\"orpers $G$ und $D$ der Diffusionskoeffizient ist. Das Inverse der Korrelationszeit $\tau$ beschreibt eine Frequenz, die mit dem Frequenzshift der Feldinhomogenit\"at $\delta\omega$ verglichen werden muss \cite{Yung03}. Wenn die Diffusion vernachl\"assigt werden kann (d. h. $\delta\omega\gg1/\tau$), dann gilt das Static-Dephasing-Regime; im anderen Grenzfall $\delta\omega\ll1/\tau$ kann die N\"aherung des Motional-Narrowing-Regimes angewandt werden.

Wie in Abschnitt \ref{Subsec:BTG} beschrieben, kann die Zeitentwicklung der transversalen Magnetisierung in allen Diffusionsregimen durch die Bloch-Torrey-Gleichung \cite{Torrey56} beschrieben werden: $\partial_t \, m(\mathbf{r},t) \, = \, \left[ D \nabla^2 + \text{i} \, \omega (\mathbf{r}) \right] \, m(\mathbf{r},t)$, wobei $\omega({\bf r})$ die lokale Frequenz aus Gleichung (\ref{local_frequency}) darstellt. Hierbei ist $ m(\mathbf{r},t) = M_x(\mathbf{r},t) - \text{i} \, M_y(\mathbf{r},t) $ die Notation der Magnetisierung in Polarform, die am Punkt $\mathbf{r}$ erzeugt wird. Die \"uber das gesamte Voxel gemittelte Magnetisierung kann als Integral \"uber alle beitragenden Punkte ausgedr\"uckt und deshalb entsprechend Gleichung (\ref{Eq7}) geschrieben werden. Die formale Zeitintegration der Bloch-Torrey-Gleichung kann numerisch ausgewertet werden. Die Zeitabh\"angigkeit, wie sie in Gleichung (\ref{Eq7}) beschrieben wird, ist im Gradientenecho-Experimenten beobachtbar \cite{Haase86}. Ein \"aquivalenter Weg zur Beschreibung des Magnetisierungs-Zeit-Verlaufes f\"ur diese Art von Experimenten ist die Benutzung der Frequenzverteilung $p(\omega)$:
\begin{equation} \label{MC1}
M(t) \, = \, \int_{-\infty}^{+\infty} \text{d} \omega \,p(\omega) \text{e}^{\text{i} \,\omega t} \,.
\end{equation}
Um die Signal-Formation bei der Anwendung komplizierterer Sequenzen zu beschreiben (z. B. SSFP-Sequenzen \cite{Carr58,Lebel06}), muss Gleichung (\ref{MC1}) im Sinne von Gleichung (\ref{Scheffler}) verallgemeinert werden \cite{Haacke99}.

\subsection{Static-Dephasing-Regime}
\label{sdr-section}
Dieses Diffusionsregime ist anwendbar, wenn Diffusionseffekte im Vergleich zu den Suszeptibilit\"atseffekten vernachl\"assigbar sind, d. h. zur quantitativen Untersuchung kann in Gleichung (\ref{Eq7}) der Diffusionskoeffizient zu $D=0$ gesetzt werden und es resultiert folgender Ausdruck zur Beschreibung des Signals:
\begin{equation} \label{SignalSD} 
M_0(t) = \frac{1}{V} \int_V \text{d}^3 \mathbf{r} \,\, \rho ( \mathbf{r} ) \, \text{e}^{{\text{i} \, \omega(\mathbf{r}) \, t}} \,,
\end{equation} 
wobei $ V $ das Relaxationsvolumen, $\omega(\mathbf{r})$ die lokale Larmor-Frequenz und $ \rho ( \mathbf{r} ) $ die Spindichte \cite{Yablonskiy94,Haacke99,Cheng01,Bakker94} sind. Um zwischen dem Static-Dephasing-Regime und den anderen Diffusionsregimen unterscheiden zu k\"onnen, wird der Index Null zur Kennzeichnung des Static-Dephasing-Regimes verwendet. Wird die Dirac-Funktion in die obige Gleichung eingef\"uhrt, kann die Magnetisierung in folgender Form geschrieben werden:
\begin{align} \label{SignalDOS} 
M_0(t) & = \frac{1}{V} \int_V \text{d}^3 \mathbf{r} \,\, \rho ( \mathbf{r} )\int_{-\infty}^{+\infty} \text{d} \omega \,\,
\delta \left[ \omega - \omega ( \mathbf{r} )\right] \text{e}^{{\text{i} \, \omega \, t}}
\\[-1ex] & \nonumber \\[-1ex] 
\label{FT} & = \rho_0 \, \int_{-\infty}^{+\infty} \text{d} \omega \,\, p_0(\omega) \, \text{e}^{{\text{i} \, \omega \, t}} \,.
\end{align} 
In Analogie zu Methoden der Statistischen Physik \cite{Cheng01,Ziener05MAGMA,Landau5} kann die Zustandsdichte f\"ur die lokale Frequenz wie folgt definiert werden:
\begin{equation} \label{DOSInt} 
p_0(\omega) \, = \, \frac{1}{\rho_0 V} \int_V \text{d}^3 \mathbf{r} \,\, \rho ( \mathbf{r} ) \, \delta \left[ \omega - \omega ( \mathbf{r} )\right] \,,
\end{equation} 
wobei diese Zustandsdichte s\"amtliche Eigenschaften einer Wahrscheinlichkeitsdichte
\begin{equation} \label{Norm} 
\int_{-\infty}^{+\infty} \text{d} \omega \, p_0(\omega) \, = \, 1 \quad \text{und} \quad p_0(\omega) \geq 0
\end{equation}
besitzt. Wird die Fourier-Darstellung der Diracschen-$\delta$-Distribution genutzt, kann das Integral (\ref{DOSInt}) im Prinzip f\"ur jede beliebige lokale Larmor-Frequenz $\omega({\bf r})$ analytisch berechnet werden. F\"ur kompliziertere Geometrien kann die Frequenzverteilung numerisch bestimmt werden, indem das Dephasierungsvolumen in kleine Subvoxel aufgeteilt und f\"ur jedes Subvoxel die mittlere Larmor-Frequenz bestimmt wird. Die Frequenzverteilung ergibt sich dann als Histogramm der Anzahl der Subvoxel, dargestellt \"uber der zugeh\"origen Larmor-Frequenz. Jedoch k\"onnen durch diese Methode keine Diffusionseffekte ber\"ucksichtigt werden. Deshalb wird in den n\"achsten Kapiteln ein anderer Weg zur Beschreibung der Diffusionsabh\"angigkeit der Frequenzverteilung aufgezeigt.

\chapter{\label{Kap.Korr}Korrelationszeiten}
\pagestyle{headings}
\section{Diffusion in lokal inhomogenen Magnetfeldern}

Die in Abschnitt \ref{Subsec:Allgemeines Modell} betrachteten K\"orper im Magnetfeld erzeugen lokale inhomogene Magnetfelder. In diesen lokalen Magnetfeldern findet die Diffusion der signalgebenden Spins statt. Zur Beschreibung der Diffusionsbewegung nutzt man die Korrelationsfunktion, die den zeitlichen Verlauf der Korrelation eines sich bewegenden Spins beschreibt. Die Korrelationsfunktion wurde bereits in Abschnitt \ref{Subsec:SCA} benutzt, um die Diffusion um magnetische K\"orper im Magnetfeld zu beschreiben. Die Korrelationsfunktion wird von der Form der magnetisierten K\"orper und den Diffusionseigenschaften des umgebenden Mediums bestimmt. Die Korrelation nimmt mit zunehmender Zeit ab und kann \"ahnlich wie der Magnetisierungszerfall durch eine exponentielle Funktion approximiert werden, deren charakteristische Abklingkonstante die Korrelationszeit $\tau$ ist \cite{Stables98,Cowan97}. Auch die Einteilung der Diffusionsregime, die in Abschnitt \ref{Subsec:Diffusionsregime} vorgestellt wurde, basiert auf der Korrelationszeit $\tau$. Der Spin, der um den K\"orper im Magnetfeld diffundiert, besucht w\"ahrend seiner Bewegung viele verschiedene Orte und damit pr\"azediert er auch mit vielen verschiedenen Larmor-Frequenzen. Anschaulich beschreibt die Korrelationszeit nun diejenige Zeitspanne, in der sich die Larmor-Frequenz des Spins merklich \"andert. Betrachtet man die linke Seite der Abbildung \ref{Diffusionsregime}, so erkennt man, dass im Motional-Narrowing Grenzfall der Spin sich sehr schnell um den kleinen K\"orper bewegt und somit viele verschiedene Frequenzen besucht. Die Korrelationszeit ist in diesem Fall also sehr klein, da nur f\"ur einen kurzen Zeitraum eine nahezu konstante Resonanzfrequenz herrscht. Im Gegensatz dazu befindet sich im Static-Dephasing-Grenzfall der Spin fast immer am gleichen Ort, wie auf der rechten Seite der Abbildung \ref{Diffusionsregime} verdeutlicht ist. Dadurch pr\"azediert der Spin auch fast immer mit der gleichen Larmor-Frequenz und demzufolge ist die Korrelationszeit sehr lang.

Gew\"ohnlicherweise wird die transversale Relaxationszeit $T_2^*$ genutzt, um Gewebeparameter wie die Zelldichte oder den Oxygenierungszustand von Blut zu bestimmen. Die Angabe des Parameters $T_2^*$ beruht auf der Annahme eines exponentiellen Magnetisierungszerfalls der Form $\propto\text{exp}(-t/T_2^*)$. Wird jedoch der nicht-Gau{\ss}sche Charakter der Spindephasierung \cite{Bauer05} beachtet, f\"uhrt eine genauere Betrachtung des Relaxationsprozesses zu dem Ergebnis, dass der exakte Zeitverlauf der Magnetisierung w\"ahrend des freien Induktionszerfalls von der Relaxationszeit im Motional-Narrowing-Grenzfall $T_{2,0}$ und von der Korrelationszeit $\tau$ abh\"angig ist:
\begin{equation}
M(t) = \exp \left(- \frac{t}{T_{2,0}} \right) \; \exp \left[ \frac{\tau}{T_{2,0}} \left( 1 - \text{e}^{-t/\tau} \right) \right] \;,
\end{equation}
wobei die Relaxationsrate im Motional-Narrowing Regime $R_{2,0} = T_{2,0}^{-1} = \tau \langle \, \omega^2(\mathbf{r}) \, \rangle$ nur von der Form des lokalen Magnetfeldes abh\"angt. So wird z. B. der Magnetisierungszerfall, der vom inhomogenen Magnetfeld um Erythrozyten erzeugt wird, als nicht-gau{\ss}f\"ormig beschrieben \cite{Spees01}.

Nun stellt sich die Frage, wie die Korrelationszeit von der lokalen Frequenzverteilung und den charakteristischen Dimensionen des magnetisierten K\"orpers abh\"angt. Sph\"arische Objekte wurden zuerst in \cite{Abragam89} untersucht, wobei auch eine Korrelationsfunktion gefunden wurde. Die exakte Form der Korrelationsfunktion wurde ausf\"uhrlich sowohl von Sukstanskii und Yablonskiy \cite{Sukstanskii03} als auch von Jensen und Chandra \cite{Jensen00} untersucht. Sie konnten zeigen, dass die Zeitabh\"angigkeit der Korrelationsfunktion nicht exponentiell, sondern algebraisch ist. Sukstanskii und Yablonskiy \cite{Sukstanskii03} gaben einen Ausdruck f\"ur die Korrelationsfunktion an, wobei auch beliebig geformte K\"orper beschrieben werden k\"onnen.

Da in Bildgebungsexperimenten kurze Echozeiten relevant sind, ist ein exponentieller Zerfall der Korrelationsfunktion eine ausreichend genaue Beschreibung. Deshalb wird in der Magnetresonanzbildgebung \"ublicherweise ein exponentiellen Zerfall der Korrelationsfunktion der Form $\propto\text{exp}(-t/\tau)$ vorausgesetzt \cite{Stables98}. Die Korrelationszeit $\tau$ wird nun in Abh\"angigkeit von der lokalen Resonanzfrequenz $\omega(\mathbf{r})$ und den Randbedingungen auf der Oberfl\"ache des K\"orpers ermittelt. Ein einfacher Ausdruck f\"ur die Korrelationszeit, der leicht zu implementieren ist, wird im Folgenden angegeben, und f\"ur den Spezialfall von Kugeln und Zylindern k\"onnen sogar analytische Ausdr\"ucke ermittelt werden.

\section{Allgemeiner Ansatz}

\noindent
Betrachtet wird ein kompakter, homogen magnetisierter K\"orper in einem \"au{\ss}eren Magnetfeld (siehe Abbildung \ref{fig1}). Die Trajektorie eines um den K\"orper diffundierenden Spins wird verfolgt. Au{\ss}erhalb des magnetischen K\"orpers werden homogene Diffusionseigenschaften angenommen, d. h. \"au{\ss}ere Potentiale oder Einschr\"ankungen durch Membranen oder andere Strukturen werden vernachl\"assigt. Anstatt die Diffusion im gesamten Gewebe zu betrachten, wird nur das Relaxationsvolumen um das magnetische Objekt betrachtet (Abbildung \ref{fig1}). Die Form dieses Relaxationsvolumens ist abh\"angig von der Form des Objektes, d. h. im Falle einer homogen magnetisierten Kugel mit dem Radius $R_{\text{S}}$ ist das Relaxationsvolumen der Raum zwischen zwei konzentrischen Kugeln mit den Radien $R_{\text{S}}$ und $R$, wie in Abbildung \ref{Zellen} dargestellt. Der Volumenanteil ist dann gegeben durch $\eta=R_{\text{S}}^3/R^3$. Analog dazu ist im Falle von Zylindern das Relaxationsvolumen der Raum zwischen zwei konzentrischen Zylindern mit den Radien $R_{\text{C}}$ und $R$, und der Volumenanteil ist $\eta=R_{\text{C}}^2/R^2$ (siehe Abbildung \ref{krogh}). Die Diffusion wird damit eingeschr\"ankt auf den Raum zwischen zwei konzentrischen Objekten mit den Radien
 $R_{\text{C}}$ und $R$, d. h. periodische Randbedingungen werden an der \"au{\ss}eren Oberfl\"ache des Relaxationsvolumens angenommen. Es ist wichtig zu betonen, dass die \"au{\ss}ere Begrenzung mit dem Radius 
$R$ nur eine mathematische Grenze ist und nicht mit irgendwelchen biologischen Membranen ´verwechselt werden darf. Die Gr\"unde f\"ur diese Einschr\"ankungen und die daraus folgenden mathematischen Implikationen wurden bereits im Detail diskutiert \cite{Bauer92}.

Die Zweipunktkorrelationsfunktion der stochastischen Feldfluktuationen, denen ein Spin unterliegt, ist definiert als
\begin{equation} \label{KorrInt}
K(t) = \int_V \text{d}^3 \mathbf{r} \int_V \text{d}^3 \mathbf{r}_0 \,\, \omega ( \mathbf{r} ) \, p (\mathbf{r} , \mathbf{r}_0 , t) p ( \mathbf{r}_0 ) \, \omega ( \mathbf{r}_0 ) \,,
\end{equation} 
wobei $ p (\mathbf{r} , \mathbf{r}_0 , t) $ die Wahrscheinlichkeitsdichte daf\"ur ist, einen Spin an der Position $ \mathbf{r} $ nach der Zeit $t$ zu finden, wenn die urspr\"ungliche Position $ \mathbf{r}_0 $ zur Zeit $ t=0 $ angenommen wurde. Die Wahrscheinlichkeitsdichte der Gleichgewichtsverteilung $ p ( \mathbf{r}_0 ) $ ist in unserem Fall identisch mit der Spindichte, welche als homogen angenommen wird, d. h.
\begin{equation} \label{Spindi} 
p ( \mathbf{r}_0 ) \, = \, \frac{1}{V} \,,
\end{equation} 
wobei $ V $ das Dephasierungsvolumen ist (im Falle von Kugeln ist dieses das Volumen zwischen zwei konzentrischen Kugeln mit den Radien $R_{\text{S}}$ und $ R $). Die Wahrscheinlichkeit $ p (\mathbf{r} , \mathbf{r}_0 , t) $ erf\"ullt die Smoluchowski-Gleichung $\partial_t p (\mathbf{r} , \mathbf{r}_0 , t) = \nabla \cdot \mathbf{j} (\mathbf{r} , \mathbf{r}_0 , t)$, mit dem zugeh\"origen Wahrscheinlichkeitsstrom $\mathbf{j} (\mathbf{r} , \mathbf{r}_0 , t) = D \nabla p (\mathbf{r} , \mathbf{r}_0 , t) $. Unter der Annahme freier Diffusion innerhalb der Grenzen $ R_{\text{S}} \leq r \leq R $ ist die Wahrscheinlichkeit $ p (\mathbf{r} , \mathbf{r}_0 , t) $ die Greensche Funktion der Diffusionsgleichung
\begin{align}
\label{Diffgl}
\frac{\partial}{\partial t} \, p (\mathbf{r} , \mathbf{r}_0 , t) \, & = \, D \nabla^2 \, p (\mathbf{r} , \mathbf{r}_0 , t) \;\;\;\;\; \text{oder}\\[2ex]
\label{Wahr}
p (\mathbf{r} , \mathbf{r}_0 , t) \, & = \, \text{e}^{t \, D \nabla^2} \, \delta(\mathbf{r} - \mathbf{r}_0) \,\,,
\end{align} 
wobei $D$ der Diffusionskoeffizient ist.

W\"ahrend an der \"au{\ss}eren Oberfl\"ache des Relaxationsvolumens (im dreidimensionalen Fall die Kugel mit dem Radius $R$ oder im zweidimensionalen Fall der Zylinder mit dem Radius $R$) reflektierende Randbedingungen angenommen werden, wird an der inneren Oberfl\"ache zwischen reflektierenden und strahlenden Randbedingungen unterschieden. Im Falle einer impermeablen inneren Oberfl\"ache werden die Trajektorien der diffundierenden Spins durch eine symmetrische Trajektorie fortgesetzt, sobald der Spin die innere Oberfl\"ache ber\"uhrt, d. h. reflektierende Randbedingungen werden angenommen:
\begin{equation} \label{Randbed} 
\left. \frac{\partial p (\mathbf{r} , \mathbf{r}_0 , t)}{\partial r} \right|_{r=R_{\text{S}}} \, = 0 = \,\left. \frac{\partial p (\mathbf{r} , \mathbf{r}_0 , t)}{\partial r} \right|_{r=R} \,.
\end{equation}
Die reflektierenden Randbedingungen an der \"au{\ss}eren Oberfl\"ache $R$ umh\"ullen das System und verhindern das Entweichen der Spins ins Unendliche. Deshalb verschwindet der korrespondierende Strom an dieser Oberfl\"ache: $\mathbf{j}(R,t) = \mathbf{0}$. Wenn der Austausch von signalgebenden Protonen zwischen dem Relaxationsvolumen und dem magnetisierten Objekt ber\"ucksichtigt werden soll, m\"ussen an der inneren Oberfl\"ache strahlende Randbedingungen in der Form
\begin{equation} \label{radRandbed} 
\left. D \, \frac{\partial p (\mathbf{r} , \mathbf{r}_0 , t)}{\partial r} \right|_{r=R_{\text{S}}} \, = \, \left.k \, p (\mathbf{r} , \mathbf{r}_0 , t) \right|_{r=R_{\text{S}}}
\end{equation}
angenommen werden, wobei $k$ eine mikroskopische Reflexionsrate ist, welche die Permeabilit\"at der inneren Oberfl\"ache beschreibt \cite{Szabo80}.

Das Einsetzen der Wahrscheinlichkeitsdichten, die in den Gleichungen (\ref{Wahr}) und (\ref{Spindi}) gegeben sind, in die Definition der Korrelationsfunktion aus Gleichung (\ref{KorrInt}) liefert
\begin{equation} \label{Korr}
K(t) = \frac{1}{V} \, \int_V \text{d}^3 \mathbf{r} \, \omega ( \mathbf{r} ) \, \text{e}^{t \, D \nabla^2} \, \omega ( \mathbf{r} ) \,.
\end{equation}
Im Allgemeinen zeigt die Korrelationsfunktion $ K(t) $ keinen einfachen exponentiellen Abfall, wie oft angenommen wird \cite{Cowan97}. Dies erschwert eine einfache Bestimmung der Korrelationszeit. Eine geeignete Definition der Korrelationszeit besteht darin, sie gem\"a{\ss} \cite{Nadler85} als mittlere Relaxationszeit der Korrelationsfunktion zu definieren:
\begin{equation} \label{MRTA} 
\tau = \int_{0}^{\infty} \text{d}t \, \frac{K(t)}{K(0)} \,.
\end{equation}
Eine \"ubliche N\"aherung f\"ur die Korrelationsfunktion ist ein einfacher exponentieller Abfall der Form $K(t) \approx K(0) \,\cdot \text{e}^{- \, t/\tau} $. Im Falle uneingeschr\"ankter Diffusion konnte von Jensen und Chandra \cite{Jensen00} gezeigt werden, dass die Korrelationsfunktion durch eine algebraische Funktion besser approximiert werden kann. Einsetzen von Gleichung (\ref{Korr}) in Gleichung (\ref{MRTA}) ergibt
\begin{equation} \label{Tauint}
\tau = \frac{1}{\langle \, \omega^2(\mathbf{r}) \, \rangle D V} \int_V \text{d}^3 \mathbf{r} \,\, \omega(\mathbf{r}) \left[ - \frac{1}{\Delta} \right] \omega(\mathbf{r}) \,,
\end{equation}
wobei die Varianz des lokalen Magnetfeldes durch
\begin{equation} \label{Var}
\langle \, \omega^2(\mathbf{r}) \, \rangle = K(0) = \frac{1}{V} \, \int_V \text{d}^3 \mathbf{r} \, \omega^2 ( \mathbf{r} ) \propto \eta \, \delta\omega^2
\end{equation}
gegeben ist, mit $\eta$ als dem Volumenanteil des magnetischen Materials innerhalb des Voxels. Die Relation $ \langle \omega^2 ( \mathbf{r} ) \rangle \propto \, \eta \, \delta\omega^2 $ stimmt mit dem allgemeinen Ergebnis, wie es in Gleichung (1) in \cite{Sukstanskii03} angegeben wird, \"uberein. Wird der allgemeine Ausdruck in Gleichung (\ref{Tauint}) genutzt, kann ein Zusammenhang zwischen der Korrelationszeit $\tau$, dem Volumenanteil $\eta$, dem Radius des St\"ork\"orpers $R_{\text{S}}$ und dem Diffusionskoeffizienten $D$ gefunden werden. Dies f\"uhrt zu einer Relation, wie sie auch in Gleichung (15) der Ver\"offentlichung von Stables et al. \cite{Stables98} angegeben wird.

\section{\label{Sec:Spez}Spezielle Geometrien}
Der oben erhaltene allgemeine Ausdruck (\ref{Tauint}) zur Quantifizierung der Korrelationszeit wird auf zwei Formen von magnetischen Objekten angewandt, n\"amlich auf Kugeln und Zylinder. Um das Integral (\ref{Tauint}) zu l\"osen, muss zuerst eine Funktion $ f(\mathbf{r}) $ gefunden werden, welche die Beziehung
\begin{equation} \label{Lap}
\nabla^2 \, f(\mathbf{r}) \, = \, -\omega(\mathbf{r})
\end{equation}
erf\"ullt. Im Falle von Kugeln und Zylindern kann diese Funktion analytisch gefunden werden. F\"ur komplizierter geformte Objekte muss die Laplace-Gleichung numerisch gel\"ost werden. Des Weiteren muss die Funktion $f(\mathbf{r})$ die reflektierenden Randbedingungen nach Gleichung (\ref{Randbed})
\begin{equation} \label{Randbedf}
\left. \frac{\partial f (\mathbf{r})}{\partial r} \right|_{r=R} = 0
\end{equation}
an der \"au{\ss}eren Oberfl\"ache des Relaxationsvolumens erf\"ullen. Im Falle strahlender Randbedingungen an der inneren Oberfl\"ache gen\"ugt die Funktion $f(\mathbf{r})=f_{\text{rad}} (\mathbf{r})$ der Beziehung
\begin{equation} \label{radRandbedf}
\left. D \, \frac{\partial f_{\text{rad}} (\mathbf{r})}{\partial r} \right|_{r=R_{\text{S}}} \, = \,k \, f_{\text{rad}}(r=R_{\text{S}}) \,,
\end{equation}
wobei $k$ die Permeabilit\"at der Oberfl\"ache beschreibt. Wenn die Oberfl\"ache des magnetischen K\"orpers impermeabel ist ($k=0$), wird der Spin an der Oberfl\"ache reflektiert, d. h. im Falle reflektierender Randbedingungen an der inneren Oberfl\"ache gilt f\"ur $f(\mathbf{r})=f_{\text{ref}} (\mathbf{r})$ der Zusammenhang
\begin{equation} \label{refRandbedf}
\left. \frac{\partial f_{\text{ref}} (\mathbf{r})}{\partial r} \right|_{r=R_{\text{S}}} \, = 0 \,.
\end{equation}

\subsection{\label{Subsec:Spheres}Kugel}
Die lokale Resonanzfrequenz um eine homogen magnetisierte Kugel ist in Gleichung (\ref{om-Feld-Kugel}) gegeben. Die St\"arke der Feldinhomogenit\"at wird durch den charakteristischen \"aquatorialen Frequenzshift $\delta\omega=\gamma \mu_0 \Delta M / 3$ beschrieben. Der Erwartungswert des lokalen Magnetfeldes wird durch Einsetzen des lokalen Magnetfeldes in die Gleichung (\ref{Var}) erhalten:
\begin{equation} \label{varkug}
\langle \, \omega^2(\mathbf{r}) \, \rangle = \frac{4}{5} \, \eta \, \delta\omega^2 \,.
\end{equation}
Das Ergebnis $K(0)= \langle \, \omega^2(\mathbf{r}) \, \rangle = \frac{4}{5}\eta\delta\omega^2$ ist identisch mit dem von Jensen und Chandra \cite{Jensen00}, Gleichung (18), oder dem von Stables et al. \cite{Stables98}, Gleichung (8). Zur L\"osung der Laplace-Gleichung (\ref{Lap}) f\"ur ein Dipolfeld, wird der Ansatz
\begin{equation} \label{Ansatz}
f(\mathbf{r}) \, = \, g(r) \cdot (3\cos^2\theta -1)
\end{equation}
genutzt. Die Anwendung des Laplace-Operators $ \nabla^2 $ in Kugelkoordinaten auf die Funktion $ f(\mathbf{r}) $ ergibt f\"ur die radiale Funktion $ g(r) $ die Differentialgleichung $r^2 \partial^2_r g(r) + 2r \partial_r g(r) - 6 g(r) = - \delta\omega \, R_{\text{S}}^3 /r$. Diese Differentialgleichung vom Eulerschen Typ kann in eine inhomogene Differentialgleichung mit konstanten Koeffizienten transformiert werden. Zur L\"osung wird die Substitution $ r \, = \, \text{e}^t $ genutzt und damit ergibt sich f\"ur die radiale Funktion
\begin{equation}
g(r) \, = \, \frac{A}{r^3} \, + \, B \, r^2 \, + \, \frac{\delta\omega \, R_{\text{S}}^3}{6r} \,,
\end{equation}
wobei die Integrationskonstanten $A$ und $B$ durch die Randbedingungen an der inneren und \"au{\ss}eren Oberfl\"ache festgelegt sind. Wird der Ausdruck (\ref{Ansatz}) zur L\"osung der inhomogenen Laplace-Gleichung (\ref{Lap}) genutzt, reduziert sich der Ausdruck f\"ur die Korrelationszeit (\ref{Tauint}) auf die einfache Form
\begin{equation}
\tau \, = \, \frac{3}{D \delta\omega (1-\eta)} \,\int_{R_{\text{S}}}^R \text{d} r \, \frac{g(r)}{r} \,.
\end{equation}
Im Falle strahlender Randbedingungen an der inneren Grenzfl\"ache ergibt sich mit der Gleichung (\ref{radRandbedf}) f\"ur die Integrationskonstanten
\begin{equation}
A_{\text{rad}} \, = \, \frac{\delta\omega}{6} R_{\text{S}}^5 b \;\;\;\; \text{und} \;\;\;\; B_{\text{rad}} \, = \,\frac{\delta\omega}{4} \left( b \eta^{5/3}+ \frac{\eta}{3} \right)
\end{equation}
mit der Abk\"urzung
\begin{equation}
b = \frac{2 D (1 - \eta) + (2 + \eta)kR_{\text{S}}}{6 D \left(\eta^{5/3}-1 \right) - (3\eta^{5/3} + 2)kR_{\text{S}}} \,.
\end{equation}
Mit Gleichung (\ref{refRandbedf}) ergibt sich bei reflektierenden Randbedingungen an der inneren Grenzfl\"ache f\"ur die Integrationskonstanten:
\begin{equation}
A_{\text{ref}} \, = \, \frac{\delta\omega \, R_{\text{S}}^5}{18} \, \frac{\eta - 1}{1 - \eta^{5/3}} \;\;\;\; \text{und} \;\;\;\; B_{\text{ref}} \, = \, \frac{\delta\omega}{12} \, \left( 1 + \frac{\eta - 1}{1 - \eta^{5/3}} \right) \,.
\end{equation}
Die Korrelationszeit im Falle strahlender Randbedingungen wird also durch den Ausdruck
\begin{equation} \label{taukrad}
\tau_{\text{rad}} = \frac{R_{\text{S}}^2}{24\,D}\,\frac{12 - 9 \eta^{1/3} -3\eta + b (4+5\eta - 9\eta^{5/3}) }{1-\eta}
\end{equation}
beschrieben. Ist die innere Kugel impermeabel, h\"angt die Korrelationszeit nur vom Volumenanteil, vom Kugelradius und vom Diffusionskoeffizienten ab. Mit der Abk\"urzung
\begin{equation}
d_i = \sum_{j=0}^{i} \eta^{\frac{j}{3}}
\end{equation}
ergibt sich f\"ur die Korrelationszeit der Ausdruck:
\begin{equation} \label{tauk}
\tau_{\text{ref}} = \frac{R_{\text{S}}^2}{8\,D}\,\left[ \frac{3}{d_2} + \frac{5d_2}{9d_4} \right]\;.
\end{equation}
Gleichung (\ref{tauk}) erh\"alt man auch, wenn in der Gleichung f\"ur strahlende Randbedingungen der Koeffizient $k=0$ gesetzt wird. Eine Taylor-Entwicklung in $ \eta $ f\"uhrt zu einem einfachen Zusammenhang zwischen der Korrelationszeit und dem Volumenanteil $\eta$ im Falle eines permeablen Kerns, n\"amlich zu
\begin{equation} \label{taut2rad}
\tau_{\text{rad}} = \frac{R_{\text{S}}^2}{D} \, \left( \frac{4 D+k R_{\text{S}}}{3(3 D + k R_{\text{S}})} - \frac{3}{8}\sqrt[3]{\eta}\right) \,.
\end{equation}
Wird in dieser Gleichung den Koeffizienten $k=0$ gesetzt oder eine Taylor-Entwicklung von Gleichung (\ref{tauk}) durchgef\"uhrt, erh\"alt man
\begin{equation} \label{taut2}
\tau_{\text{ref}} = \frac{R_{\text{S}}^2}{D} \, \left( \frac{4}{9} - \frac{3}{8}\sqrt[3]{\eta} \right)
\end{equation}
f\"ur die Korrelationszeit im Falle eines impermeablen Kerns. 

\begin{figure}
\begin{center}
\includegraphics[width=10cm]{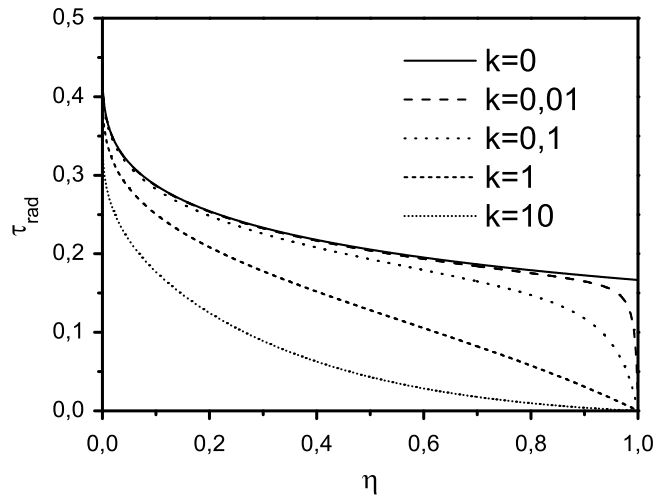}
\caption[Korrelationszeit bei der Diffusion um Kugeln]{\label{Fig:Korr-kugel}{\footnotesize Korrelationszeit bei der Diffusion um Kugeln. Dargestellt ist die Korrelationszeit bei der Diffusion um Kugeln nach Gleichung (\ref{taukrad}) f\"ur verschiedene Werte des Permeabilit\"atskoeffizienten $k$.}}
\end{center}
\end{figure}

Zur Veranschaulichung der Ergebnisse ist in Abbildung \ref{Fig:Korr-kugel} die Korrelationszeit in Abh\"angigkeit vom Volumenverh\"altnis $\eta$ f\"ur verschiedene Werte des Permeabilit\"atskoeffizienten $k$ dargestellt. Bei vorgegebenem Volumenverh\"altnis $\eta$ ist die Korrelationszeit am gr\"o{\ss}ten, wenn die Kugel impermeabel ist ($k=0$). In diesem Fall wird der Spin an der Oberfl\"ache der Kugel reflektiert und der Spin bewegt sich kurz nach der Reflexion in einem lokalen Magnetfeld, dessen Wert fast genau so gro{\ss} ist, wie kurz vor der Reflexion. Der Spin pr\"azediert also l\"anger mit der nahezu gleichen Resonanzfrequenz, und deshalb ist in diesem Fall die Korrelationszeit auch l\"anger. Ist die Kugel hingegen durchl\"assig, wird der Spin nur mit einer bestimmten Wahrscheinlichkeit reflektiert, ein gewisser Anteil dringt auch in die Kugel ein und tr\"agt dann nicht mehr zum Signal bei. Mit steigender Permeabilit\"at $k$ nimmt also die Korrelationszeit ab.

\subsection{\label{Subsec:Cylinders}Zylinder}
Wird ein homogen magnetisierter Zylinder untersucht, muss dessen Orientierung zum \"au{\ss}eren Magnetfeld ber\"ucksichtigt werden. Die lokale Resonanzfrequenz ist abh\"angig vom Neigungswinkel $\theta$ zwischen der Zylinderachse und der Richtung des \"au{\ss}eren Magnetfeldes $B_0$. Auf der linken Seite der Abbildung \ref{krogh} ist der Querschnitt durch den Zylinder und durch das Dephasierungsvolumen zu sehen, wobei zweidimensionale Polarkoordinaten $(r,\phi)$ genutzt werden. Die lokale Resonanzfrequenz um den Zylinder ist in Gleichung (\ref{om-Feld-Zyl}) gegeben. Die Varianz dieser lokalen Resonanzfrequenz folgt direkt aus Gleichung (\ref{Var}) als
\begin{equation} \label{varcyl}
\langle \, \omega^2(\mathbf{r}) \, \rangle = \frac{1}{2} \, \eta \, \delta\omega_{\theta}^2 \,,
\end{equation}
was mit dem Ergebnis von Stables at al. \cite{Stables98} \"ubereinstimmt. Wird der vom Winkel abh\"angige Anteil der lokalen Frequenz (\ref{om-Feld-Zyl}) betrachtet, kann in Analogie zum Ansatz (\ref{Ansatz}) im Falle von Zylindern der Ausdruck $f(\mathbf{r}) \, = \, g(r) \cdot \cos 2\phi$ verwendet werden, was zu der Differentialgleichung $r^2 \partial^2_r g(r) + r \partial_r g(r) - 4 g(r) = - \delta\omega_{\theta} \, R_{\text{C}}^2$ mit der L\"osung $g(r)=Ar^2+B/r^2-\delta \omega_{\theta}R_{\text{C}}^2/4$ f\"uhrt. Nach Ausf\"uhrung der gleichen Schritte wie im sph\"arischen Fall ergibt sich ein analoger Ausdruck f\"ur die Korrelationszeit, und zwar
\begin{equation} \label{taurad}
\tau_{\text{rad}} = \frac{R_\text{C}^2}{4 D} \cdot \left(\frac{1 + \eta}{\frac{2D}{kR_{\text{C}}}(\eta^2-1)-\eta^2-1} -\frac{\ln{\eta}}{1 - \eta} \right)
\end{equation}
f\"ur strahlende Randbedingungen und
\begin{equation} \label{tau}
\tau_{\text{ref}} = -\frac{R_\text{C}^2}{4 D} \cdot \frac{\ln{\eta}}{1 - \eta}
\end{equation}
f\"ur reflektierende Randbedingungen, was mit dem Ergebnis von Bauer et al. \cite{Bauer99} \"ubereinstimmt.

\section{\label{Sec:App}Anwendungen}

\subsection{Diffusionsregime}
In Abschnitt \ref{Subsec:Diffusionsregime} wurde gezeigt, dass die Korrelationszeit eine wesentliche Gr\"o{\ss}e ist, die zur Einteilung der Diffusionsregime notwendig ist. So gaben Yablonskiy und Haacke in ihrer Gleichung (55) der Arbeit \cite{Yablonskiy94} die empirische Bedingung
\begin{equation} \label{Haacke}
\frac{R_{\text{S}}^2}{D} \, \delta\omega \, \frac{\sqrt[3]{\eta}}{6} \, \gg \, 1
\end{equation}
an, die erf\"ullt sein muss, damit die Diffusion um die Kugel vernachl\"assigt werden kann. Anhand der Ungleichung (\ref{Haacke}) l\"asst sich somit feststellen, ob das Static-Dephasing-Regime das zu Grund liegende Diffusionsregime ist. In diese Betrachtungen geht jedoch die Permeabilit\"at des K\"orpers nicht mit ein. Mit den oben erhaltenen Resultaten kann nun die exakte Einteilung der Diffusionsregime auf mathematischer Basis erfolgen. Werden die Ergebnisse f\"ur die Korrelationszeit und deren Abh\"angigkeit vom Volumenanteil genutzt, k\"onnen \"ahnliche Ausdr\"ucke wie die von Yablonskiy und Haacke empirisch gefundenen ableitet werden. Im Falle von kugelf\"ormigen Objekten ergibt sich aus Gleichung (\ref{taut2}) f\"ur das Static-Dephasing-Regime ($\tau \, \delta\omega \gg 1$) das Kriterium
\begin{equation} \label{Ziener}
\frac{R_{\text{S}}^2}{D} \, \delta\omega \, \left( \frac{4}{9} - \frac{3}{8}\sqrt[3]{\eta} \right) \gg 1 \,.
\end{equation}
Analog dazu kann f\"ur zylindrische Objekte ein \"ahnliches Kriterium f\"ur die G\"ultigkeit des Static-Dephasing-Regimes angegeben werden.

Die Korrelationszeit kann nun zur Beantwortung konkreter Fragestellungen angewandt werden. Sie kann beispielsweise genutzt werden, um magnetisch markierte Zellen zu beschreiben. In der Arbeit \glqq Application of the static dephasing regime theory to superparamagnetic iron-oxide loaded cells\grqq\ untersuchten Bowen et al. \cite{Bowen02} das Relaxationsverhalten eisenbeladener Zellen. Typische Parameter waren in dieser Arbeit ein Volumenanteil von $\eta = 0,002$, ein Diffusionskoeffizient $D = 2,5 \, \mu\text{m}^2 \text{ms}^{-1}$ und ein Frequenzshift von $\delta\omega = 32944 \, \text{s}^{-1}$. Die zum Markieren benutzten Kontrastmittelteilchen wurden von Lymphozyten phagozytiert, die einen mittleren Radius von $R_{\text{S}} = 10 \, \mu \text{m}$ hatten. Aus Gleichung (\ref{taut2}) ergibt sich f\"ur die Korrelationszeit der Wert $\tau=15,9 \,\text{ms}$, was der dynamischen Frequenz $1/\tau=62,9 \, \text{Hz}$ entspricht. Diese dynamische Frequenz ist wesentlich kleiner als der charakteristische Frequenzshift, und deshalb kann das Static-Dephasing-Regime als das zugrunde liegende Diffusionsregime angesehen werden.

Die Korrelationszeit um Zylinder kann genutzt werden, um den Diffusionsprozess im Myokard zu quantifizieren \cite{Bauer99}. In diesem Fall ist der Volumenanteil \"aquivalent zum regionalen Blutvolumen $\eta=0,05$. Der typische Radius einer Kapillare im Myokard betr\"agt $R_{\text{C}}=2,5 \,\mu\text{m}$ \cite{Bassingwaighte74}. Mit einem typischen Wert f\"ur den Diffusionskoeffizienten von $D = 1 \, \mu\text{m}^2 \text{ms}^{-1}$ ergibt sich aus Gleichung (\ref{tau}) f\"ur die Korrelationszeit der Wert $\tau = 4,9 \, \text{ms}$, was der dynamischen Frequenz $1/\tau=203 \, Hz$ entspricht. Der charakteristische Frequenzshift auf der Oberfl\"ache einer Kapillare im Myokard betr\"agt $\delta\omega_{\theta} = 269 \, \text{Hz}$ \cite{Bauer99T2}. In diesem Fall haben die dynamische Frequenz $1/\tau$ und die charakteristische Frequenz $\delta\omega_{\theta}$ die gleiche Gr\"o{\ss}enordnung. Deshalb wird die Spindephasierung im Myokard dem intermedi\"aren Diffusionsregime zugeordnet. Hier sind weder die Ausdr\"ucke des Motional-Narrowing-Grenzfalls noch die des Static-Dephasing-Grenzfalls anwendbar.

\subsection{Relaxationsraten}
\label{Subsec:Relaxationsraten}
Die Spindephasierung, die durch die Magnetfeldinhomogenit\"aten verursacht wird, beeinflusst den Magnetisierungszerfall. Dieser kann f\"ur alle Diffusionsregime durch die Bloch-Torrey-Gleichung \cite{Torrey56} beschrieben werden. Die Signalentstehung im Static-Dephasing-Regime \cite{Yablonskiy94} wurde f\"ur den sph\"arischen Fall \cite{Cheng01} und den zylindrischen Fall \cite{Ziener05MAGMA} ausf\"uhrlich untersucht. Mittels der oben erhaltenen Ergebnisse f\"ur die Korrelationszeiten k\"onnen nun die Relaxationsraten im Motional-Narrowing-Regime bestimmt werden. In diesem Diffusionsregime wird der Signalverlust haupts\"achlich durch die Dephasierung aufgrund der Diffusion der Spins im inhomogenen Magnetfeld hervorgerufen. Im Gegensatz zum Static-Dephasing-Grenzfall kann der Signalverlust im Motional-Narrowing-Regime nicht durch Spinechos refokussiert werden.

\begin{figure}
\begin{center}
\includegraphics[width=12cm]{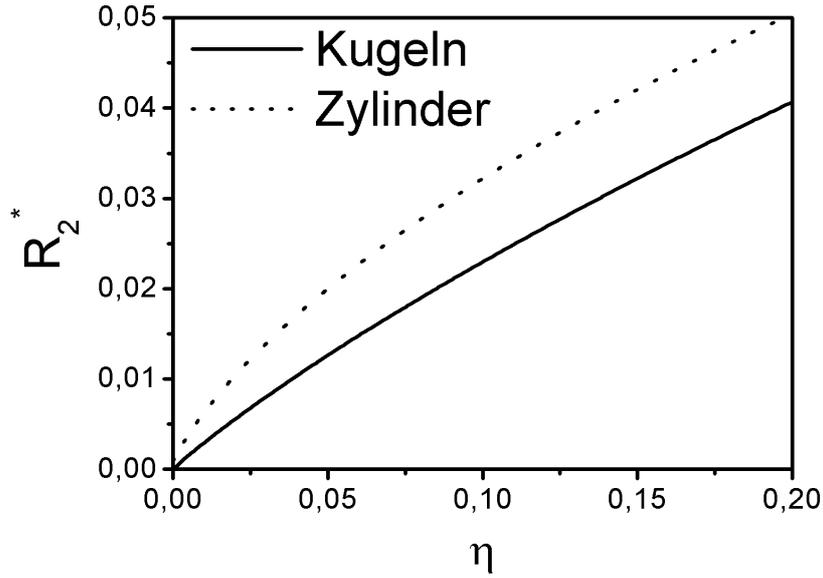}
\caption[Relaxationsrate f\"ur Kugeln und Zylinder]{\label{Fig:rel}{\footnotesize Relaxationsrate f\"ur Kugeln und Zylinder. Die Relaxationsrate in Abh\"angigkeit vom Volumenanteil $\eta$ ist nach Gleichung (\ref{R2stern}) f\"ur Kugeln und Zylinder dargestellt. In beiden F\"allen wurde f\"ur den Vorfaktor $R_\text{C}^2/D=1$ bzw $R_\text{S}^2/D=1$ angenommen.}}
\end{center}
\end{figure}

Im Motional-Narrowing-Regime sind die Frequenzfluktuationen, die durch den Diffusionsprozess verursacht und durch das Inverse der Korrelationszeit charakterisiert werden, gr\"o{\ss}er als die durch die Feldinhomogenit\"at erzeugte Frequenzverschiebung, d. h. $1/\tau\, \gg \, \delta\omega$. In diesem Fall ist die Relaxationsrate durch die einfache Beziehung
\begin{equation} \label{R2stern}
R_2^* \, = \, \tau \, \langle \omega^2 ( \mathbf{r} ) \rangle
\end{equation}
gegeben, wobei $\langle \omega^2 ( \mathbf{r} ) \rangle$ f\"ur Kugeln und Zylinder in Gleichung (\ref{varkug}) und Gleichung (\ref{varcyl}) gegeben sind. In Abbildung \ref{Fig:rel} werden die Relaxationsraten $R_2^*$ in Abh\"angigkeit vom Volumenanteil $\eta$ f\"ur Kugeln und Zylinder miteinander verglichen.
Im Falle kleiner Volumenanteile ergibt sich aus den beiden Gleichungen (\ref{R2stern}) und (\ref{taut2}) der einfache Ausdruck f\"ur die Relaxationsrate einer Suspension kleiner Kugeln
\begin{equation} \label{r2sternMN2}
R_2^* \, = \, \frac{16}{45} \, \eta \, \delta\omega^2 \, \frac{R_{\text{S}}^2}{D} \;,
\end{equation}
welcher mit den Ergebnissen von Moiny et al. \cite{Moiny92} und Brooks et al. \cite{Brooks01} \"ubereinstimmt. Mit der allgemeinen Beziehung (\ref{R2stern}) ergibt sich damit die Korrelationszeit
\begin{equation} \label{tau_mot_narr}
\tau = \frac{4}{9} \frac{R_{\text{S}}^2}{D} \,.
\end{equation}
Die auf diesem Wege erhaltene Korrelationszeit stimmt auch mit dem ersten Glied der Taylor-Entwicklung aus Gleichung (\ref{taut2}) \"uberein.

Jensen und Chandra \cite{Jensen00} bestimmten die Relaxationsrate im Falle eines durchl\"assigen Kerns zu
\begin{equation}
R_2^* \, = \, \frac{8}{25} \, \eta \, \delta\omega^2 \, \frac{R_{\text{S}}^2}{D} \,.
\end{equation}
Die Kombination dieses Ergebnisses mit dem allgemeinen Ausdruck (\ref{R2stern}) f\"uhrt zur Korrelationszeit, die den Diffusionsprozess um eine durchl\"assigen Kern beschreibt:
\begin{equation} \label{tauperm}
\tau = \frac{2}{5} \frac{R_{\text{S}}^2}{D} \,.
\end{equation}

Zur Ableitung der obigen Ergebnisse wurde die N\"aherung benutzt, dass benachbarte Feldinhomogenit\"aten keinen Einfluss auf das betrachtete Voxel haben. Um den G\"ultigkeitsbereich dieser N\"aherung zu quantifizieren, wird zuerst der Fall von Kapillaren betrachtet. Es wird eine regelm\"a{\ss}ige Anordnung der Kapillaren angenommen, d. h. im Querschnitt befindet sich die Kapillare im Zentrum eines Hexagons und hat sechs gleichnahe Nachbarn. Jedes Hexagon mit der Seitenl\"ange $a$ und der Fl\"ache $A=3\sqrt{3}a^2/2$ wird in diesem Modell durch einen Kreis mit dem Radius $R$ und dem gleichen Fl\"acheninhalt $A=\pi R^2$ ersetzt. Der Abstand zwischen zwei Kapillaren ist demzufolge $2a=2\sqrt{2\pi/(3\sqrt{3})}R$. Der Einfluss benachbarter Kapillaren kann vernachl\"assigt werden, wenn die charakteristische Frequenz an der Oberfl\"ache einer Kapillare $\delta \omega_{\theta}$ viel gr\"o{\ss}er ist als die Frequenz, die durch die sechs umgebenden Kapillaren im Abstand $2a$ erzeugt wird. Mit diesen geometrischen Anschauungen ergibt sich nach Gleichung (\ref{om-Feld-Zyl}) die Ungleichung $\delta \omega_{\theta} \gg 6\delta \omega_{\theta} R_{\text{C}}^2/(2a)^2$, was zu einer Absch\"atzung f\"ur den Volumenanteil f\"uhrt: $\eta \ll 4\pi/(9\sqrt{3})=0,81$.

Im dreidimensionalen Fall der regelm\"a{\ss}igen Anordnung ist jede Kugel im Zentrum eines Rhombendodekaeders (der Wigner-Seitz-Zell eines fcc-Bravais-Gitters) von zw\"olf gleichnahen Nachbarn umgeben. In diesem Modell wird jeder Rhombendodekaeder mit der Seitenl\"ange $s$ und dem Volumen $V=16 s^3/(3 \sqrt{3})$ durch eine Kugel mit dem Radius $R$ und dem gleichen Volumen $V=4\pi R^3/3$ ersetzt. Daraus folgt, dass der Abstand zur n\"achsten Kugel $2\sqrt{2/3}s$ betr\"agt. Der Einfluss des Magnetfeldes benachbarter Kugeln kann vernachl\"assigt werden, wenn die charakteristische Frequenz $\delta \omega$ auf der Oberfl\"ache einer Kugel viel gr\"o{\ss}er ist als die Frequenz, die durch die zw\"olf umgebenden Kugeln im Abstand $2\sqrt{2/3}s$ erzeugt wird. Mit diesen geometrischen Anschauungen ergibt sich nach Gleichung (\ref{om-Feld-Zyl}) die Ungleichung $\delta \omega \gg 12\delta \omega R_{\text{S}}^3/(2\sqrt{2/3}s)^3$, was zu einer Absch\"atzung f\"ur den Volumenanteil f\"uhrt: $\eta \ll \sqrt{2}\pi/9=0,49$.

Innerhalb des Diffusionsraumes werden also nur die Wechselwirkung eines Kernspins mit dem Magnetfeld einer Kugel oder eines Zylinders betrachtet, d. h. die Beitr\"age benachbarter K\"orper werden nicht ber\"ucksichtigt. In einem Gewebe, in dem der magnetisierte K\"orper einen kleinen Volumenanteil besitzt und dieser nur ein schwaches lokales Magnetfeld erzeugen, ist die Vernachl\"assigung der umgebenden St\"ork\"orper gerechtfertigt und die N\"aherungen in den Gleichungen (\ref{taut2rad}) und (\ref{taut2}) sind g\"ultig.
\begin{figure}
\begin{center}
\includegraphics[width=12cm]{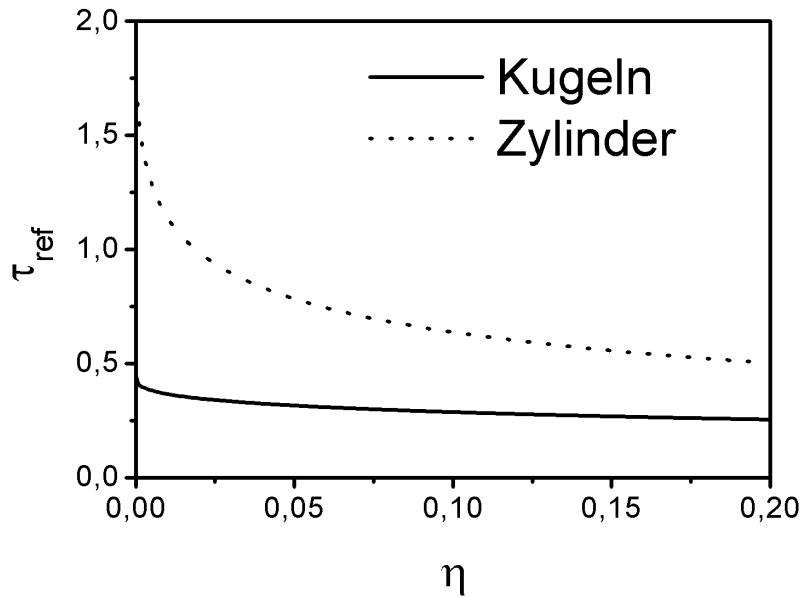}
\caption[Korrelationszeit f\"ur Kugeln und Zylinder]{\label{Fig:Korr}{\footnotesize Korrelationszeit f\"ur Kugeln und Zylinder. Nach Gleichung (\ref{tauk}) bzw. Gleichung (\ref{tau1}) wurde die Korrelationszeit in Abh\"angigkeit vom Volumenanteil dargestellt. In beiden F\"allen gilt f\"ur den Vorfaktor $R_\text{C}^2/D=1$ bzw. $R_\text{S}^2/D=1$.}}
\end{center}
\end{figure}

\chapter{\label{Kap:Frequenz}Frequenzverteilungen}
\pagestyle{headings}

\section{Formalismus zur Beschreibung der Frequenzverteilung}
Bereits in Abschnitt \ref{SigVox} wurde dargestellt, wie das Signal aus einem Voxel, das einen magnetischen K\"orper enth\"alt, untersucht werden kann. Der magnetische K\"orper erzeugt ein lokales inhomogenes Magnetfeld innerhalb des Voxels. Die Spins bewegen sich aufgrund der Diffusion in diesem Feld und beeinflussen dadurch das Dephasierungsverhalten. Diese Dephasierungseffekte k\"onnen genutzt werden, um Informationen \"uber das untersuchte Gewebe zu erhalten. So konnten aus diesen Informationen z. B. die Zelldichte \cite{Baklanov04} oder die Kapillardichte im Myokard \cite{Bauer99,Bauer99PRL} bestimmt werden. Um dies zu erreichen werden Modelle ben\"otigt, welche den zeitlichen Signalverlauf in Abh\"angigkeit von der Anordnung der magnetischen K\"orper beschreiben. Da diese magnetischen K\"orper ein inhomogenes Magnetfeld erzeugen, das die Relaxation der signalgebenden Spins beschleunigt, ist die Information \"uber die Anordnung der magnetischen K\"orper im Zeitverlauf des Magnetisierungszerfalls oder \"aquivalent dazu im zugeh\"origen Frequenzspektrum enthalten.

Um den Signal-Zeit-Verlauf exakt zu beschreiben, muss, wie in Abschnitt \ref{Subsec:Allgemeines Modell} beschrieben wurde, zuerst aus der Form des K\"orpers die lokale Resonanzfrequenz $\omega(\mathbf{r})$ bestimmt werden. Dann wird die Bloch-Torrey-Gleichung f\"ur diese lokale Resonanzfrequenz gel\"ost, wie es in Abschnitt \ref{Subsec:BTG} dargestellt wurde. Unter Ber\"ucksichtigung der Diffusionseffekte ist dies sehr m\"uhsam und nur in Spezialf\"allen m\"oglich, z. B. wenn $\omega(\mathbf{r})$ die Frequenz ist, welche durch einen linearen Gradienten erzeugt wird \cite{Torrey56,Stoller91}. Alternativ dazu ist es m\"oglich, die Bloch-Torrey-Gleichung numerisch zu l\"osen. Dies wurde beispielsweise genutzt, um die Effekte der Kantenverst\"arkung durch Diffusion zu beschreiben \cite{Puetz91,Puetz92,Barsky92}, und zwar unter Anwendung der Linienform-Theorie von Kubo \cite{Kubo62,Kubo63}. Eine weitere Methode zur Beschreibung der transversalen Relaxation stellten k\"urzlich Kiselev und Posse \cite{Kiselev99,Kiselev98} vor. Hierbei nutzten sie analytische Modelle f\"ur sehr lange und sehr kurze Korrelationszeiten. Das Modell f\"ur lange Korrelationszeiten erweitert das Static-Dephasing-Regime und das Modell f\"ur kurze Korrelationszeiten basiert auf einem St\"orungsansatz im lokalen Magnetfeld. Jedoch sind diese Ergebnisse auf Spezialf\"alle begrenzt, was deren Anwendbarkeit einschr\"ankt. Ein anderer Ansatz zur Beschreibung der Spindephasierung \"uber den gesamten Dynamikbereich ist die Gau{\ss}sche Dephasierung \cite{Sukstanskii02,Sukstanskii03,Sukstanskii04}, die jedoch nur anwendbar ist, solange die Dynamik auf eine eingeschr\"ankte Klasse stochastischer Prozesse beschr\"ankt bleibt \cite{Bauer05}. Die L\"osung der Bloch-Torrey-Gleichung kann auch durch die Strong-Collision-Approximation beschrieben werden \cite{Bauer02}. Um dem Nicht-Gau{\ss}schen Charakter der Diffusion gerecht zu werden, wird diese Approximation in der vorliegenden Arbeit genutzt.

Im folgenden Teil der Arbeit wird ein anderer Weg zur Beschreibung der Anordnung der Feldinhomogenit\"aten gegangen. Wie in Abschnitt \ref{Spin} dargestellt, beeinflussen verschiedene Parameter das Signal, das in dem Voxel entsteht, welches den magnetisierten K\"orper enth\"alt. Sowohl die genutzte Sequenz und ihre Parameter als auch die Verteilung der Feldinhomogenit\"aten im Gewebe beeinflussen das entstehende Signal (siehe auch Abbildung \ref{Fig:Intro}). Um Informationen \"uber die Signalentwicklung zu erhalten, wird die Frequenzverteilung, die durch die r\"aumliche Verteilung der Feldinhomogenit\"aten bestimmt wird, untersucht. 

F\"ur viel Probleme der Bildgebung stellt sich oft die Frage, welche Sequenz mit welchen Parametern zu nutzen ist. Um diese Frage zu beantworten ist es notwendig, das zu untersuchende Gewebe durch seine Frequenzverteilung zu charakterisieren. F\"ur eine vorgegebene Anordnung der magnetischen K\"orper innerhalb des Voxels hat die Frequenzverteilung eine typische Form. Die Kenntnis \"uber die Form der Frequenzverteilung erlaubt eine Wahl der Bildgebungssequenz mit den optimalen Parametern, um Informationen \"uber die Gewebeeigenschaften zu erhalten. 

Zuerst wird die Frequenzverteilung im Static-Dephasing-Regime untersucht, d. h. Diffusionseffekte um die magnetischen K\"orper werden vernachl\"assigt. F\"ur die beiden wichtigen Spezialf\"alle Kugel und Zylinder k\"onnen die Frequenzverteilungen explizit angegeben werden. In beiden F\"allen ist die typische Form der Frequenzverteilung abh\"angig von der Suszeptibilit\"atsdifferenz zwischen dem magnetisierten Objekt und dem umgebenden Gewebe, von der St\"arke des \"au{\ss}eren Magnetfeldes und auch vom Volumenanteil des magnetischen Materials innerhalb des Voxels.

Ein wichtiger Faktor, der die Frequenzverteilung und demzufolge auch die Signalentstehung im Voxel beeinflusst, ist die Diffusion der Wassermolek\"ule in den lokalen Magnetfeldinhomogenit\"aten \cite{Norris00}. W\"ahrend das Static-Dephasing-Regime gut verstanden ist \cite{Yablonskiy94,Haacke99}, sind Diffusionseffekte nur bei speziellen Geometrien untersucht worden. Im Gegensatz dazu wird hier eine rigorose Ableitung der Frequenzverteilung innerhalb des Voxels angegeben \cite{ZienerPRE}. Dabei wird sowohl die Form der K\"orper, welche die Suszeptibilit\"atsdifferenz erzeugen, als auch die Diffusion um diese K\"orper ber\"ucksichtigt. Um den Diffusionsprozess zu beschreiben, wurde eine Strong-Collision-Approximation genutzt, welche die Diffusionsdynamik durch einen einfacheren stochastischen Prozess ersetzt. Damit ist es m\"oglich einen Formalismus herzuleiten, der die Frequenzverteilung um beliebig geformte magnetische K\"orper beschreiben kann.

\subsection{Erweiterung auf alle andere Diffusionsregime}

Um einen Ausdruck f\"ur die Frequenzverteilung zu finden ist es zweckm\"a{\ss}ig, mit einer Verallgemeinerung von Gleichung (\ref{FT}) auf alle Diffusionsregime zu beginnen, wonach das Signal als Fourier-Transformierte der Frequenzverteilung geschrieben werden kann:
\begin{equation}
 \label{SignalD}
M(t) = \rho_0 \, \int_{-\infty}^{+\infty} \text{d} \omega \,\, p(\omega) \, \text{e}^{{\text{i} \, \omega \, t}}\,.
\end{equation}
Sowohl die Magnetisierung $M(t)$ als auch die Frequenzverteilung $p(\omega)$ sind nun vom Diffusionskoeffizienten $D$ abh\"angig. Im Grenzfall $D \to 0$ stimmen $M(t)$ und $M_0(t)$ bzw. $p(\omega)$ und $p_0(\omega)$ \"uberein. Die Frequenzverteilung wiederum ist die Fourier-Transformierte des Signals:
\begin{align} 
p(\omega) & = \frac{1}{2 \pi \rho_0} \int_{-\infty}^{+\infty} \text{d} t \, \text{e}^{- \text{i} \omega t} \, M(t)\\[1.5ex]
\label{dichte}
& = \frac{1}{2 \pi \rho_0} \left| \hat{M}(\text{i} \omega) + \hat{M}^*(\text{i} \omega) \right| \ ,
\end{align}
wobei die Laplace-Transformierte der Magnetisierung $\hat{M}(s)$ durch Gleichung (\ref{Laplacetransform}) definiert ist. Hierbei wurde der Zusammenhang zwischen Laplace-Transformation und Fourier-Transformation genutzt. F\"ur negative Zeiten wurde eine analytische Fortsetzung des Magnetisierungs-Zeit-Verlaufes in der Form $M(-t) = M(t)^*$ angenommen. Der Vorteil beim Benutzen der Laplace-Transformation besteht darin, dass nach Gleichung (\ref{Mdach}) eine Relation zwischen der Laplace-Transformierten des Magnetisierungs-Zeit-Verlaufs im Static-Dephasing-Regime $\hat{M}_0(s)$ und dem Magnetisierungs-Zeit-Verlauf in allen anderen Diffusionsregimen $\hat{M}(s)$ existiert. Um dieses Ergebnis zu erhalten, wurde die Strong-Collision-N\"aherung benutzt, die den Laplace-Operator in der Bloch-Torrey-Gleichung durch einen einfacheren stochastischen Prozess ersetzt \cite{Bauer99,Dattagupta74}. Obiges Ergebnis gilt, solange die Korrelationszeit $\tau$ kleiner ist als die transversale Relaxationszeit des Magnetisierungszerfalls. Wie in Kapitel \ref{Kap.Korr} gezeigt wurde, kann die Korrelationszeit nach Gleichung (\ref{Tauint})
\begin{equation}
\tau = \frac{1}{\langle \, \omega^2(\mathbf{r}) \, \rangle D V} \int_V \text{d}^3 \mathbf{r} \,\, \omega(\mathbf{r}) \left[ - \frac{1}{\Delta} \right] \omega(\mathbf{r})
\end{equation}
genutzt werden, um die Zeitskala der durch die molekulare Bewegung induzierten Feldfluktuationen zu beschreiben.

Die Laplace-Transformierte des Magnetisierungszerfalls im Static-Dephasing-Regime kann aus Gleichung (\ref{FT}) oder aus Gleichung (\ref{SignalSD}) ermittelt werden:
\begin{align}
\label{LT1}
\hat{M}_0 (s) & = \int_{0}^{\infty} \text{d} t \, \text{e}^{-st} \, M_0(t)\\[1ex]
\label{LT2}
& = \rho_0 \int_{-\infty}^{+\infty} \text{d} \omega \, \frac{p_0(\omega)}{s - \text{i} \omega}\\[1ex]
 \label{LT}
& = \frac{\rho_0}{V} \int_{V} \text{d}^3 \mathbf{r} \,\, \frac{1}{s - \text{i} \omega(\mathbf{r})} \,.
\end{align}
Einsetzen von Gleichung (\ref{LT}) in Gleichung (\ref{Mdach}) und Verwenden von Gleichung (\ref{dichte}) ergibt letztlich einen allgemeinen Ausdruck f\"ur die Frequenzverteilung:
\begin{equation}
\label{freqdist}
p(\omega)=\frac{\tau}{\pi} \left| \text{Re} \left\{ \left[ \frac{1}{V} \int_V \frac{\text{d}^3 \mathbf{r}}{1 + \text{i} \tau \left[ \omega- \omega(\mathbf{r}) \right]} \right]^{-1} - \rho_0 \right\}^{-1} \right| \,.
\end{equation}
Dieser Ausdruck ist die Verallgemeinerung von Gleichung (\ref{DOSInt}) auf alle Diffusionsregime. Die diffusionsabh\"angige Frequenzverteilung ist also nur von der Korrelationszeit $\tau$ und der lokalen Larmor-Frequenz $\omega(\mathbf{r})$ abh\"angig. Obwohl der Ausdruck f\"ur diese Frequenzverteilung eine einfache Struktur hat, sind analytische L\"osungen nur f\"ur wenige Spezialf\"alle m\"oglich.

Aus Gleichung (\ref{Mdach}) ergibt sich im Grenzfall kleiner Korrelationszeiten $\lim_{\tau \to 0} \hat{M}(s) = \rho_0/s$, was entsprechend dem Fourier-Theorem zu einem Delta-Peak f\"ur die Frequenzverteilung im Motional-Narrowing-Regime f\"uhrt: $\lim_{\tau \to 0} p(\omega) = \delta(\omega)$. Wie erwartet, streben alle Observablen im Grenzfall sich nicht bewegender Spins zu ihrem Grenzwert des Static-Dephasing-Regimes, d. h. $\lim_{\tau \to \infty} \hat{M}(s) = \hat{M}_0(s)$, was letztlich zu der Frequenzverteilung des Static-Dephasing-Regimes aus der Gleichung (\ref{DOSInt}) f\"uhrt: $\lim_{\tau \to \infty} p(\omega) = p_0(\omega)$.

\begin{figure}
\begin{center}
\includegraphics[width=10cm]{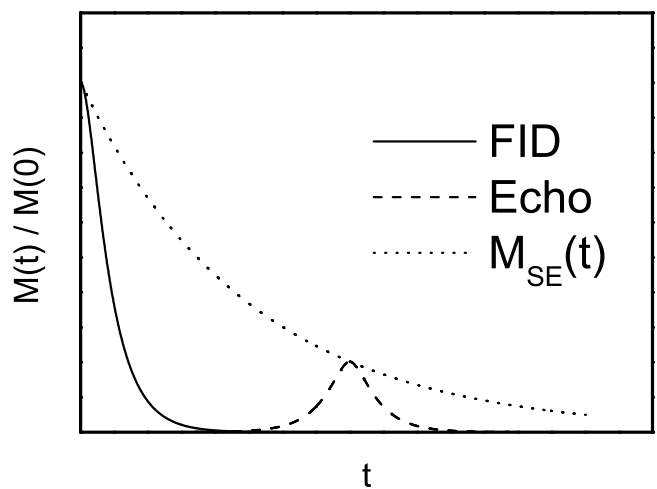}
\caption[Spin-Echo-Experiment]{\label{Fig:Echo}{\footnotesize Spin-Echo-Experiment. Einstrahlung eines $180^{\circ}$-Hochfrequenzpulses nach Abklingen des FID-Signals f\"uhrt zur Rephasierung und zum Entstehen des Spin-Echos (gestrichelte Linie). Die Amplituden des Spin-Echos liegen auf der Kurve $M_{\text{SE}}(t)$, die nach Gleichung (\ref{spin-echo}) aus dem freien Induktionsabfall $M(t)$ ermittelt werden kann.}}
\end{center}
\end{figure}

Bisher wurden nur Gradientenecho-Sequenzen oder Sequenzen, die im Sinne von Gleichung (\ref{Scheffler}) beschrieben werden k\"onnen, betrachtet. Um jedoch z. B. den Magnetisierungs-Zeit-Verlauf eines klassischen Hahnschen Spinecho-Experimentes (siehe Abbildung \ref{Fig:Echo}) beschreiben zu k\"onnen \cite{Hahn50}, kann man von den Ergebnissen der Gradientenecho-Sequenz ausgehen. Aus dem freien Induktionszerfall der Gradientenecho-Sequenz $M(t)$ nach Gleichung (\ref{SignalD}) kann man den Magnetisierungs-Zeit-Verlauf eines Spinechos folgenderma{\ss}en berechnen \cite{Bauer99T2}:
\begin{equation} \label{spin-echo}
M_\text{SE}(t) = \text{e}^{-t/\tau} + \frac{\text{e}^{-t/\tau}}{\tau} \int_0^t \text{d}\xi \, \text{e}^{\xi/\tau} \left| M \left( \frac{\xi}{2} \right) \right|^2 \,.
\end{equation}
Um beispielsweise eine SSFP-Sequenz zu beschreiben, kann der Ausdruck (\ref{Scheffler}) benutzt werden.

Die Bloch-Torrey-Gleichung hat die Form einer Schr\"odinger-Gleichung mit rein imagin\"arem Potential im Hamilton-Operator. Aufgrund dieses imagin\"aren Potentials ist der Hamilton-Operator nicht-hermitesch, was die Entwicklung der L\"osung in ein vollst\"andiges Orthonormalsystem verhindert. Damit liefern die oben beschriebenen Methoden ein praktisches Verfahren zum L\"osen der Bloch-Torrey-Gleichung in der Strong-Collision-N\"aherung. Weitere Analogien zur Quantenmechanik lassen sich aus Gleichung (\ref{DOSInt}) erkennen. Die lokale Resonanzfrequenz $\omega({\bf r})$ entspricht der Dispersionsrelation eines angeregten Ensembles (z. B. eines Elektronengases). In diesem Bild entspricht der Diffusionskoeffizient $D$ der Temperatur und Gleichung (\ref{freqdist}) ist die Verallgemeinerung von Gleichung (\ref{DOSInt}) auf endliche Temperaturen.

\section{Anwendung auf inhomogene Gewebe}
Die im vorigen Abschnitt abgeleitete allgemeine Theorie kann genutzt werden, um die Frequenzverteilung $p(\omega)$ f\"ur jedes Diffusionsregime zu berechnen, das durch einen beliebig geformten K\"orper erzeugt wird. Um die Allgemeing\"ultigkeit der Theorie zu demonstrieren, wird der Formalismus zuerst unabh\"angig von der Form des magnetischen K\"orpers entwickelt und im Nachhinein auf einige Spezialf\"alle angewandt, um auch analytische Ausdr\"ucke zu erhalten.

\subsection{Allgemeine Ausdr\"ucke}
Um allgemeine Ausdr\"ucke zu erhalten wird vorausgesetzt, dass das Voxel die gleiche Form wie der magnetische K\"orper besitzt, d. h. das Voxel entsteht aus der Form des K\"orpers durch eine zentrische Streckung. Dies bedeutet, dass im Falle eines zylinderf\"ormigen magnetisierten K\"orpers das Voxel auch zylinderf\"ormig ist (siehe Abbildung \ref{krogh}). Diese Vorgehensweise erfolgt analog zum aus der Physiologie bekannten Kroghschen Kapillarmodell \cite{Krogh19}, bei dem eine zylinderf\"ormige Kapillare von einem zylinderf\"ormigen Versorgungsgebiet umgeben ist. Es wird angenommen, dass der Volumenanteil sehr klein ist und die Bedingung $\eta\ll 1$ erf\"ullt. Deshalb ist die wahre Form des Voxels irrelevant und das urspr\"ungliche Voxel kann durch ein kubisches Voxel gleichen Volumens ersetzt werden \cite{Yablonskiy94,Ziener05MAGMA}. Durch diese Annahme ist es m\"oglich, den Magnetisierungs-Zeit-Verlauf im Static-Dephasimg-Regime entsprechend Gleichung (\ref{SignalSD}) in der Form
\begin{equation} \label{allg}
M_0(t) = \frac{\rho_0}{1-\eta}\left[ g(\eta \, \delta\omega \, t) - \eta \, g(\delta\omega \, t) \right]
\end{equation}
zu schreiben, wobei die Funktion $g$ nur von der Form des K\"orpers abh\"angig ist. Die Existenz und Eindeutigkeit dieser Funktion $g$ werden durch den Hauptsatz der Differential-und Integralrechnung gesichert. Obere und untere Grenze des Integrals in Gleichung (\ref{SignalSD}) sind die Oberfl\"ache des Voxels und die Oberfl\"ache des magnetisierten K\"orpers. Die Funktion $g$ kann sowohl numerisch als auch analytisch f\"ur jede beliebige Form des K\"orpers ermittelt werden. F\"ur den Spezialfall von kugelf\"ormigen und zylinderf\"ormigen Objekten kann die Funktion $g$ analytisch angegeben werden. In Analogie zur Gleichung (\ref{allg}), die den Signal-Zeit-Verlauf beschreibt, kann die Laplace-Transformierte des Signal-Zeit-Verlaufs in der Form
\begin{equation} \label{allgmd} 
\hat{M}_0(s) = \frac{\rho_0}{1-\eta} \, \frac{1}{s} \, \left[ G \left( \frac{s}{\eta \, \delta\omega} \right) - \eta \, G \left( \frac{s}{\delta\omega} \right) \right]
\end{equation}
geschrieben werden, wobei die Funktion $G$ auch einzig von der Form des Objektes abh\"angt.

Zur Charakterisierung der Diffusionseffekte muss die Korrelationszeit entsprechend Gleichung (\ref{Tauint}) ausgewertet werden. Wie in \cite{Stables98,Ziener06a,Ziener06b} gezeigt, ist die Integration direkt ausf\"uhrbar und liefert folgende Relation:
\begin{equation} \label{tau1}
\tau = \frac{L^2}{D}k(\eta) \,,
\end{equation}
wobei $L$ eine charakteristische L\"ange des K\"orpers ist (z. B. der Radius des Zylinders oder der Kugel). Wie in \cite{Ziener06b} dargestellt, ist die Funktion $k(\eta)$ auch nur von der Form des K\"orpers abh\"angig.

Um einen Ausdruck f\"ur die Frequenzverteilung zu finden, muss Gleichung (\ref{freqdist}) ausgewertet werden. Somit erh\"alt man letztlich
\begin{equation} \label{md} 
p(\omega) = \frac{\tau}{\pi \rho_0} \left| \text{Re}\frac{G \left( \frac{1 + \text{i}\tau \omega}{\eta \, \tau \, \delta\omega} \right) - \eta G \left( \frac{1 + \text{i} \tau \omega}{\tau \, \delta\omega} \right)}{\frac{1-\eta}{\rho_0}(1 + \text{i} \tau \omega) - G \left( \frac{1 + \text{i} \tau \omega}{\eta \, \tau \, \delta\omega} \right) + \eta G \left( \frac{1 + \text{i} \tau \omega}{\tau \, \delta\omega} \right)} \right| \,.
\end{equation}
Dieser Ausdruck stellt die Frequenzverteilung in Abh\"angigkeit von der Diffusion (charakterisiert durch die Korrelationszeit $\tau$), vom Suszeptibilit\"atseffekt (charakterisiert durch den in Gleichung (\ref{eEq3}) gegebenen Frequenzshift $ \delta \omega = \gamma \mu_0 M_0 /(4 \pi) $) und von der Form des K\"orpers (charakterisiert durch die Funktion $G$) dar. F\"ur zylinderf\"ormige und kugelf\"ormige K\"orper werden analytische Ausdr\"ucke f\"ur die Funktion $G$ in den n\"achsten Abschnitten angegeben.

\subsection{Kugel}
\begin{figure}
\begin{center}
\includegraphics[width=10cm]{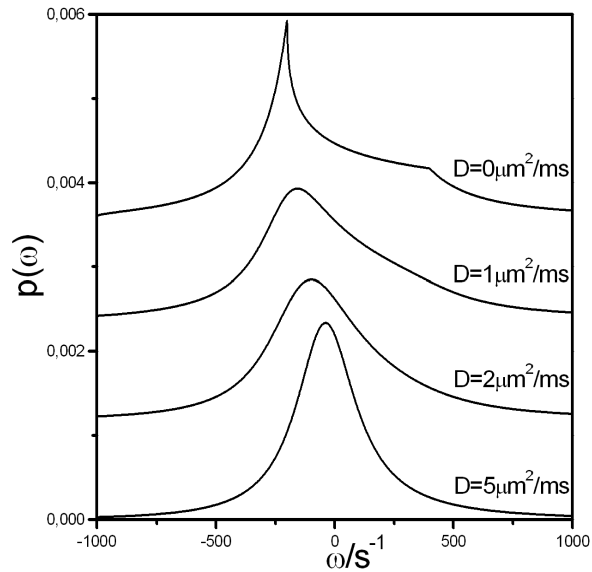}
\caption[Diffusionsabh\"angige Frequenzverteilung f\"ur Kugeln]{\label{Fig:dosdiffsph}{\footnotesize Diffusionsabh\"angige Frequenzverteilung f\"ur Kugeln. Die Kurven wurden nach Gleichung (\ref{md}) berechnet, wobei die kugelspezifische $G$-Funktion aus Gleichung (\ref{ggs}) genutzt wurde. Der Volumenanteil ist $\eta=0,2$, der \"aquatoriale Frequenzshift $\delta\omega=1000\text{s}^{-1}$, Radius $R_{\text{S}}=5\mu\text{m}$.}}
\end{center}
\end{figure}

Der erste analytisch l\"osbare Fall ist der einer homogen magnetisierten Kugel in einem \"au{\ss}eren Magnetfeld. Die Frequenzverteilung und die Signaleigenschaften im Static-Dephasing-Regime wurden von Cheng et al. \cite{Cheng01} analytisch untersucht. Diese Ergebnisse wurden von Seppenwoolde et al. \cite{Seppenwoolde05} experimentell best\"atigt. Um die Signalentstehung im Lungengewebe zu beschreiben, wurden die Alveolen als Kugeln betrachtet, die von einer wassergef\"ullten Kugelschale umgeben sind \cite{Case87,Durney89,Bertolina91,Cutillo96}. In diesen Arbeiten wurde die Frequenzverteilung entsprechend der in Abschnitt \ref{sdr-section} dargestellten Histogramm-Technik numerisch ermittelt. Bowen et al. \cite{Bowen02} haben das Static-Dephasing-Regime zur Beschreibung der Signalentstehung um magnetisch markierte Zellen angewandt. Die lokale Resonanzfrequenz um eine Kugel ist in Gleichung (\ref{om-Feld-Kugel}) gegeben. In Analogie zum zylindrischen Fall ist die Kugel im Zentrum eines kugelf\"ormigen Voxels mit dem Radius $R$ lokalisiert (siehe Abbildung \ref{Zellen}). Einsetzen der lokalen Resonanzfrequenz aus Gleichung (\ref{om-Feld-Kugel}) in Gleichung (\ref{DOSInt}) und Integration \"uber das Relaxationsvolumen liefert die Frequenzverteilung im Static-Dephasing-Regime, wie sie bereits in \cite{Cheng01} angegeben wurde:
\begin{align} \label{DOS-sph}
p_{0,\text{S}}(\omega) & = \left\{
\begin{array}{lll} 
\frac{\eta}{3 \sqrt{3} (1-\eta)} \: \left( \frac{\delta \omega}{\omega} \right)^2 \left( 2-\frac{\omega}{\delta \omega}\right) \sqrt{1+\frac{\omega}{\delta \omega}}
& \text{f\"ur} & \omega \leq -\eta \delta \omega \; \\
& \text{oder} & \omega \geq 2\eta \delta \omega \,,\\[2ex]
\frac{\eta}{3 \sqrt{3} (1-\eta)} \: \left( \frac{\delta \omega}{\omega} \right)^2 \left[ \left( 2-\frac{\omega}{\delta \omega}\right) \sqrt{1+\frac{\omega}{\delta \omega}} - \left(2-\frac{\omega}{\eta \, \delta \omega}\right) \sqrt{1+\frac{\omega}{\eta \, \delta \omega}} \right]
& \text{f\"ur} & \!\!\!\!\! - \eta \delta \omega \leq \omega \leq 2\eta \delta \omega \,, \\[6ex]
0 & & \text{sonst} \,,
\end{array}
\right.
\end{align}
wobei $ \eta = R_{\text{S}}^3 / R^3 $ der Volumenanteil ist ($0\leq\eta\leq 1$). Das Einsetzen dieser Frequenzverteilung (\ref{DOS-sph}) in die Fourier-Transformation (\ref{SignalDOS}) oder direktes Berechnen des Integrals (\ref{SignalSD}) liefert den Signal-Zeit-Verlauf, der im Sinne der allgemeinen Form nach Gleichung (\ref{allg}) geschrieben werden kann mit Hilfe der kugelspezifischen $g$-Funktion:
\begin{equation}
g_{\text{S}}(x) = \text{e}^{-\text{i} x} \,  _1F_1 \left\{ \begin{array}{c}\frac{1}{2}\\[1ex] \frac{3}{2}\end{array} \Bigg|3 \text{i} x  \right\} + \int_0^1 \text{d}z \, x (3z^2 - 1) \left[ \text{Si}(x(3z^2 - 1)) - \text{i} \, \text{Ci} (x |3z^2 - 1|) \right] \,.
\end{equation}
Die verallgemeinerte hypergeometrische Funktion oder auch Barnes erweiterte hypergeometrische Funktion ist durch
\begin{equation} \label{BEHF}
 _pF_q \left\{ \begin{array}{ccc} a_1, & .\,.\,. & ,a_p \\[1ex] b_1, & .\,.\,. & ,b_q \end{array} \Bigg| z \right\} = \sum_{k=0}^{\infty}\frac{(a_1)_{k} \cdot (a_2)_{k} \cdots (a_p)_{k}}{(b_1)_{k} \cdot (b_2)_{k} \cdots (b_q)_{k}} \, \frac{z^k}{k!}
\end{equation}
definiert \cite{Oberhettinger72}, und das Pochhammersymbol ist definiert als
\begin{equation} \label{PS}
(x)_k = \frac{\Gamma(x+k)}{\Gamma(x)} \,.
\end{equation}
Die Kummersche konfluente hypergeometrische Funktion $_1F_1$ ist ein Spezialfall von Barnes erweiterter hypergeometrischer Funktion (\ref{BEHF}) f\"ur $p=1=q$. Integralsinus und Integralcosinus sind folgenderma{\ss}en definiert \cite{Gradstein81}:
\begin{align} 
\text{Si}(x) & = \int_0^x \frac{\sin t}{t} \text{d}t \;\;\;\;\; \text{und} \\[1ex]
\text{Ci}(x) & = - \int_x^{\infty} \frac{\cos t}{t} \text{d}t \,.
\end{align}
Wird die Integration in einer der Gleichungen (\ref{LT1}) bis (\ref{LT}) ausgef\"uhrt, ergibt sich die Laplace-Transformierte des Magnetisierungszerfalles im Static-Dephasing-Regime, welche in der Form (\ref{allgmd}) mit der kugelspezifischen Funktion
\begin{equation} \label{ggs}
G_{\text{S}}(y) = \frac{1}{3} + \frac{2}{3} \left( 1 - \frac{2\text{i}}{y} \right) \sqrt{\frac{1-\text{i}y}{3}} \text{arccoth} \sqrt{\frac{1-\text{i}y}{3}}
\end{equation}
geschrieben werden kann. Die f\"ur die Berechnung der Frequenzverteilung notwendige Korrelationszeit $\tau$ ist in Gleichung (\ref{tauk}) angegeben, welche auch folgenderma{\ss}en geschrieben werden kann:
\begin{equation}
\tau = \frac{R_{\text{S}}^2}{2\,D\,(1-\eta)}\,\left[1-\eta^{1/3}+\frac{4(1-\eta)^2+9 \left(2\eta - \eta^{5/3} -\eta^{1/3} \right)}{36 \left(\eta^{5/3}-1\right)}\right] \,.
\end{equation}
Mit Kenntnis der Funktion $G_{\text{S}}$ und der Korrelationszeit $\tau$ kann durch Einsetzen in Gleichung (\ref{md}) ein analytischer Ausdruck f\"ur die Frequenzverteilung angegeben werden, der in allen Diffusionsregimen g\"ultig ist. Um die Abh\"angigkeit von der Diffusion zu illustrieren, ist in Abbildung \ref{Fig:dosdiffsph} die Frequenzverteilung f\"ur verschiedene Diffusionskoeffizienten $D$ dargestellt.
Erreicht der Diffusionskoeffizient den Wert $D=0$, dann stimmt die Frequenzverteilung mit der in Gleichung (\ref{DOS-sph}) angegebenen Frequenzverteilung im Static-Dephasing-Grenzfall \"uberein, welche auch in Abbildung 1 in \cite{Cheng01} zu sehen ist.

\begin{figure}
\begin{center}
\includegraphics[width=10cm]{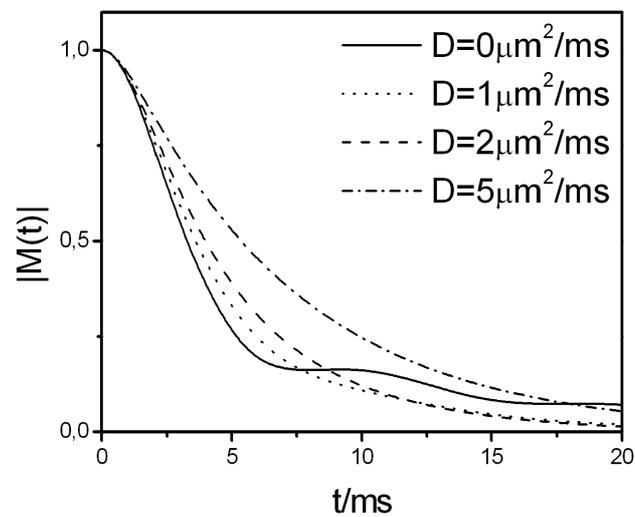}
\caption[Diffusionsabh\"angiger Signal-Zeit-Verlauf f\"ur Kugeln]{\label{Fig:sigdiffkug}{\footnotesize Diffusionsabh\"angiger Signal-Zeit-Verlauf f\"ur Kugeln. Der Volumenanteil ist $\eta=0,2$, der \"aquatoriale Frequenzshift $\delta\omega=1000\text{s}^{-1}$, Radius $R_{\text{S}}=5\mu\text{m}$.}}
\end{center}
\end{figure}

\begin{figure}
\begin{center}
\includegraphics[width=10cm]{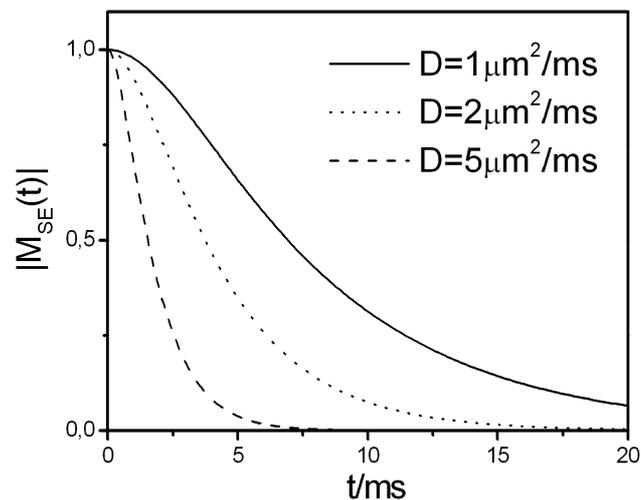}
\caption[Diffusionsabh\"angiger Signal-Zeit-Verlauf des Spin-Echos f\"ur Kugeln]{\label{Fig:sekug}{\footnotesize Diffusionsabh\"angiger Signal-Zeit-Verlauf des Spin-Echos f\"ur Kugeln. Die Kurven wurden nach Gleichung (\ref{spin-echo}) berechnet, wobei der Signal-Zeit-Verlauf des Gradienten-Echos $M(t)$ aus Abbildung \ref{Fig:sigdiffkug} eingesetzt wurde. Der Volumenanteil betr\"agt $\eta=0,2$, der \"aquatoriale Frequenzshift ist $\delta\omega = 1000\text{s}^{-1}$ und f\"ur den Radius wurde $R_{\text{S}}=5\mu\text{m}$ angenommen.}}
\end{center}
\end{figure}

Die Fourier-Transformation der Frequenzverteilung $p(\omega)$ entsprechend Gleichung (\ref{SignalD}) f\"uhrt zum Signal-Zeit-Verlauf wie er in Abbildung \ref{Fig:sigdiffkug} gezeigt wird.
Der Magnetisierungs-Zeit-Verlauf f\"ur ein Spin-Echo-Experiment kann aus dem Magnetisierungs-Zeit-Verlauf des Gradienten-Echos entsprechend Gleichung (\ref{spin-echo}) ermittelt werden. Die entsprechenden Verl\"aufe sind in Abbildung \ref{Fig:sekug} dargestellt.\\

\subsection{\label{Abschn.Zyl}Zylinder}

\subsubsection{Analytische L\"osung}
\begin{figure}
\begin{center}
\includegraphics[width=10cm]{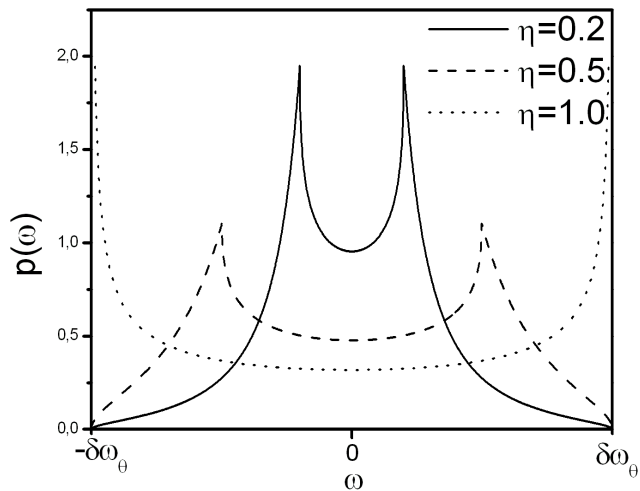}
\caption[Frequenzverteilung im Static-Dephasing-Regime]{\label{Fig:DOSsd}{\footnotesize Frequenzverteilung im Static-Dephasing-Regime. Dargestellt sind die Frequenzverteilungen f\"ur verschiedene Werte des Volumenanteils $\eta$. Die Peaks der Frequenzverteilung befinden sich bei den Frequenzen $\omega_{\text{Peak}}=\pm \eta\,\delta\omega_{\theta}$.}}
\end{center}
\end{figure}

Das zweite analytisch l\"osbare Beispiel ist das bereits in Abschnitt \ref{Einleit-Zylinder} betrachtete zylinderf\"ormige Objekt, das von einem zylindrischen Voxel umgeben ist (siehe Abbildung \ref{krogh}). Die Ausf\"uhrung der Integration in Gleichung (\ref{DOSInt}) unter Benutung der lokalen Resonanzfrequenz aus Gleichung (\ref{om-Feld-Zyl}) ergibt f\"ur die Frequenzverteilung im Static-Dephasing-Regime den Ausdruck
\begin{align} \label{DOS-cyl}
p_{0,\text{C}}(\omega) & = \left\{
\begin{array}{lll} 
\frac{\eta}{1-\eta} \: \frac{\delta \omega_{\theta}}{\pi \omega^2} \sqrt{1 - \left( \frac{\omega}{\delta \omega_{\theta}} \right)^2}
& \text{f\"ur} & \omega \leq - \eta \delta \omega_{\theta} \;\; \text{oder} \;\; \omega \geq \eta \delta \omega_{\theta} \,, \\[2ex]
\frac{\eta}{1-\eta} \: \frac{\delta \omega_{\theta}}{\pi \omega^2} \left[ \sqrt{1 - \left( \frac{\omega}{\delta \omega_{\theta}} \right)^2} - \sqrt{1 - \left( \frac{\omega}{\eta \delta \omega_{\theta}} \right)^2} \right]
& \text{f\"ur} & - \eta \delta \omega_{\theta} \leq \omega \leq \eta \delta \omega_{\theta} \;\;\;\;\;\;\text{und}\\[4ex]
0 & \text{sonst}, & 
\end{array}
\right.
\end{align}
wobei $ \eta = R_{\text{C}}^2 / R^2 $ der Volumenanteil ($ 0 \leq \eta \leq 1 $) des Zylinders innerhalb des Voxels ist. Im Grenzfall eines kleinen Volumenanteils konvergiert die Frequenzverteilung gegen die Delta-Funktion $\delta(\omega)$
\begin{equation}
\lim_{\eta \to 0} p(\omega)=\delta(\omega) \,,
\end{equation}
und im entgegengesetzten Grenzfall eines gro{\ss}en Volumenanteils, in dem das Relaxationsvolumen verschwindet, konvergiert die Frequenzverteilung gegen den folgenden Ausdruck:
\begin{equation} \label{p1}
\lim_{\eta \to 1} p(\omega)=\frac{1}{\pi \delta \omega_{\theta} \sqrt{1 - \left( \frac{\omega}{\delta \omega_{\theta}} \right)^2}} \,.
\end{equation}
In Abbildung \ref{Fig:DOSsd} ist die Frequenzverteilung im Static-Dephasing-Regime f\"ur verschiedene Werte des Volumenanteiles zu sehen. Aus der Gleichung (\ref{om-Feld-Zyl}), welche die lokale Resonanzfrequenz um einen Zylinder beschreibt, kann man erkennen, dass der minimale Wert der Frequenz bei $-\delta \omega_{\theta}$ und der maximale Wert von bei $+\delta \omega_{\theta}$ liegt. Diese beiden speziellen Werte werden an Punkten auf der Oberfl\"ache des Zylinders angenommen. Da Beitr\"age von Spins innerhalb des Zylinders vernachl\"assigt werden, sind Resonanzfrequenzen, die au{\ss}erhalb dieses Intervalls liegen, nicht m\"oglich. Dies bedeutet, dass die Frequenzverteilung im Static-Dephasing-Regime einen kompakten Tr\"ager hat. Eine weitere wichtige mathematische Eigenschaft besteht darin, dass die Frequenzverteilung eine rein reelle Funktion ist, was auch von einer Wahrscheinlichkeitsdichte erwartet wird. Des Weiteren ist die Frequenzverteilung eine gerade Funktion, d. h. $p(\omega)=p(-\omega)$, wohingegen die Frequenzverteilung um eine homogen magnetisierte Kugel \cite{Cheng01} keine Symmetrieeigenschaften besitzt.
\"Ahnliche Frequenzverteilungen wurden von Zimmerman und Foster (Abbildung 6 in \cite{Zimmerman56}) angegeben. Numerische Simulationen im Sinne des oben beschriebenen Histogramms best\"atigen die analytischen Ergebnisse f\"ur die Frequenzverteilung $p(\omega)$.

Da das Magnetfeld im Inneren des Zylinders konstant ist, erzeugt dieser Beitrag die konstante innere Frequenz $\omega_{\text{int}} = 2\delta\omega_{\pi/2}/3 - \delta \omega_{\theta}$, die zu einem Delta-Peak an der Position $\omega = \omega_{\text{int}}$ f\"uhrt (siehe Abbildung 5 in \cite{Scheffler01}).

Wird diese Frequenzverteilung des Static-Dephasing-Regimes (\ref{DOS-cyl}) in die Fourier-Transfor\-ma\-ti\-on (\ref{SignalDOS}) eingesetzt, ergibt sich ein Ausdruck f\"ur den Magnetisierungs-Zeit-Verlauf im Static-Dephasing-Regime, der in der Form von Gleichung (\ref{allg}) mit der zylinderspezifischen $g$-Funktion ausgedr\"uckt werden kann:
\begin{equation} \label{kgzyl}
g_{\text{C}}(x) = \, _1F_2 \left\{ \begin{array}{cc}-\frac{1}{2} & \\[1ex] \;\;\,\frac{1}{2}\,, & 1 \end{array} \Bigg|-\left(\frac{x}{2}\right)^2 \right\} \,.
\end{equation}
Um den Ausdruck f\"ur den Signal-Zeit-Verlauf im Static-Dephasing-Regime $M_0(t)$ zu erhalten, kann man auch die lokale Resonanzfrequenz $\omega_{\text{C}}(\mathbf{r})$ aus Gleichung (\ref{om-Feld-Zyl}) direkt in den allgemeinen Ausdruck (\ref{SignalSD}) einsetzen. Die Symmetrie $p(\omega)=p(-\omega)$, die Annahme rein reeller Werte und der kompakte Tr\"ager sind mathematische Eigenschaften der Frequenzverteilung, die dazu f\"uhren, dass die Fourier-Transformierte, also der Magnetisierungs-Zeit-Verlauf, eine rein reelle Funktion ist. Das Ergebnis f\"ur den Magnetisierungs-Zeit-Verlauf stimmt mit den von Yablonskiy und Haacke angegebenen Resultaten \"uberein (siehe Gleichung (36) in \cite{Yablonskiy94}).

Im Grenzfall gro{\ss}er Volumenanteile, in dem das Relaxationsvolumen gegen eine unendlich d\"unne R\"ohre konvergiert, ergibt sich aus Gleichung (\ref{allg}) der Ausdruck
\begin{equation} \label{eta1}
\frac{M(t)}{M(0)} = \, J_0(\delta \omega_{\theta} \, t) \,,
\end{equation}
wobei $J_0$ die Bessel-Funktion erster Art und nullter Ordnung ist. Dieses Ergebnis kann auch abgeleitet werden, wenn man die Frequenzverteilung im Grenzfall gro{\ss}er Volumenanteile aus Gleichung (\ref{p1}) in die Fourier-Transformation entsprechend Gleichung (\ref{FT}) einsetzt. Entsprechen dem allgemeinen Ausdruck f\"ur das NMR-Signal im Static-Dephasing-Regime (\ref{SignalSD}) f\"uhrt die Konvergenz des Relaxationsvolumens bis hin zu einer unendlich d\"unnen R\"ohre in der Folge zu einer verschwindenden Anfangsmagnetisierung $M(0)$, w\"ahrend jedoch das Verh\"altnis $M(t)/M(0)$ zur oben beschriebenen Bessel-Funktion konvergiert.

F\"ur kleine Zeiten f\"uhrt die Taylor-Entwicklung des Magnetisierungs-Zeit-Verlaufes aus Gleichung (\ref{allg}) zu dem Ausdruck
\begin{equation} \label{shorttime}
\frac{M(t)}{M(0)}=1 - \frac{\eta}{4} \delta \omega_{\theta}^2 t^2 \,,
\end{equation}
d. h. das Signal zerf\"allt quadratisch mit der Zeit. Dieses Ergebnis stimmt auch mit dem von Yablonskiy und Haacke in Gleichung (37) ihrer Ver\"offentlichung \cite{Yablonskiy94} angegebenen Kurzzeitverhalten des Magnetisierungszerfalles \"uberein.

\begin{figure}
\begin{center}
\includegraphics[width=10cm]{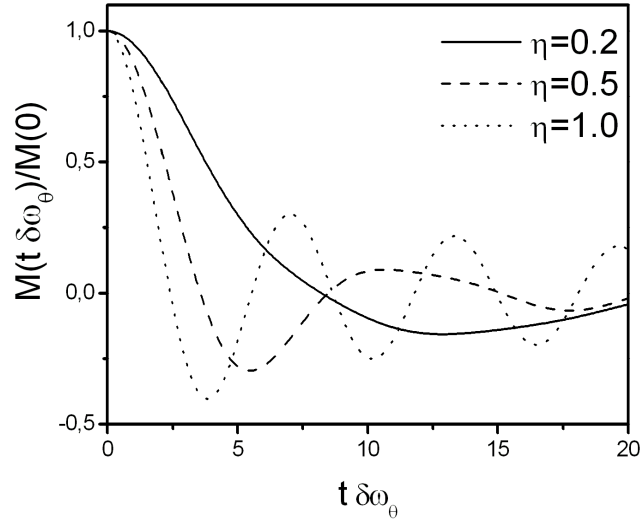}
\caption[Signal-Zeit-Verlauf im Static-Dephasing-Regime]{\label{Fig:SignalSD}{\footnotesize Signal-Zeit-Verlauf im Static-Dephasing-Regime. Der Magnetisierungszerfall nach Gleichung (\ref{allg}) ist f\"ur verschiedene Werte des Volumenanteils dargestellt. Wenn die exakte Form des Magnetisierungszerfalles im Sinne einer Mean-Relaxation-Time-Approximation durch einen exponentiellen Zerfall angen\"ahert wird, erh\"alt man f\"ur $\eta = 0,2; 0,5; 1,0$ nach Gleichung (\ref{T2stern}) die Relaxationszeiten $T_2^* \delta\omega_{\theta}=3,0; 1,5; 1,0$.}}
\end{center}
\end{figure}

Die Eigenschaften von Geweben, die Magnetfeldinhomogenit\"aten enthalten, werden oft durch ihre Relaxationszeit $T_2^*$ oder den Volumenanteil $\eta$ charakterisiert. Die Angabe einer Relaxationszeit $T_2^*$ setzt einen exponentiellen Signalzerfall der Form $M(t)=M(0) \, \text{exp}(-t/T_2^*)$ voraus. Eine Relaxationszeit $T_2^*$, die den exakten Verlauf des Magnetisierungszerfalles nach Gleichung (\ref{allg}) am besten durch einen exponentiellen Zerfall approximiert, kann mit Hilfe der Mean-Relaxation-Time-Approximation angegeben werden \cite{Nadler85}:
\begin{equation}
T_2^* = \int_0^{\infty} \text{d}t \frac{M(t)}{M(0)} \,.
\end{equation}
Nach Ausf\"uhrung der Integration ergibt sich ein einfacher Ausdruck f\"ur die transversale Relaxationszeit und ein Zusammenhang zur Frequenzverteilung kann gefunden werden:
\begin{align} \label{T2stern}
T_2^* & = \frac{1 + \eta}{2 \eta \delta \omega_{\theta}}
\\[-1ex] & \nonumber \\[-1ex] 
\label{Tp} & = \pi p(\omega = 0) \,.
\end{align}
In Abbildung \ref{Fig:SignalSD} wird der exakte Zeitverlauf des Magnetisierungszerfalles f\"ur verschiedene Werte des Volumenanteiles dargestellt und die dazugeh\"orige Relaxationszeit angegeben.
In Abbildung \ref{Fig:SignalSD} erkennt man f\"ur gro{\ss}e Zeiten einen oszillierenden Anteil, insbesondere f\"ur gro{\ss}e Volumenverh\"altnisse. Dieser oszillierende Anteil erkl\"art sich aus der charakteristischen Frequenzverteilung mit zwei Peaks (siehe Abbildung \ref{Fig:DOSsd}) an den Stellen $\pm \eta \delta\omega_{\theta}$. Im Sinne einer Schwebung \"uberlagern sich also zwei Schwingungen mit gleicher Frequenz und unterschiedlichem Vorzeichen. Es resultiert eine Schwebung mit der Frequenz $\eta \delta\omega_{\theta}$. Dies ist auch aus Gleichung (\ref{eta1}) ersichtlich, wenn man beachtet, dass die Besselfunktion f\"ur gro{\ss}e Argumente gegen die Sinusfunktion strebt.

\begin{figure}
\begin{center}
\includegraphics[width=10cm]{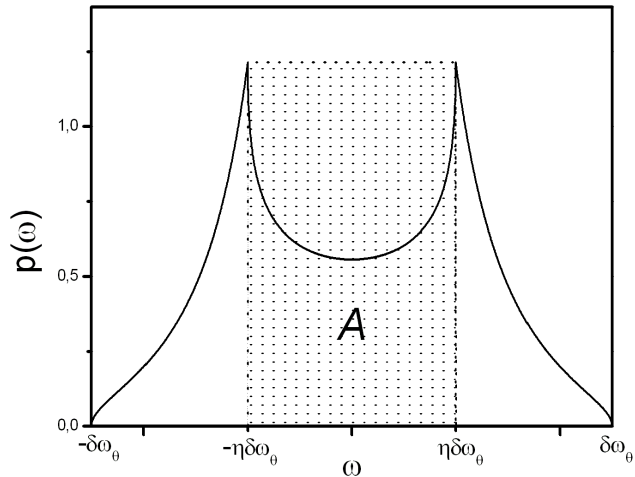}
\caption[Bestimmung des Volumenanteils]{\label{Fig:RBV}{\footnotesize Bestimmung des Volumenanteils. Die Fl\"ache des markierten Rechtecks $A$ ist in Gleichung (\ref{area}) gegeben.}}
\end{center}
\end{figure}

Im Gegensatz zu Kugeln haben Zylinder einen Neigungswinkel $\theta$ zum \"au{\ss}eren Magnetfeld. Sowohl die Frequenzverteilung als auch der Magnetisierungszerfall sind von dieser Orientierung des Zylinders abh\"angig. Ein von der Orientierung des Zylinders unabh\"angiger Wert ist die in Abbildung \ref{Fig:RBV} markierte Fl\"ache mit folgendem Fl\"acheninhalt:
\begin{equation} \label{area}
A=\frac{2}{\pi}\sqrt{\frac{1 + \eta}{1 - \eta}} \,.
\end{equation}
Wenn die Frequenzverteilung im Static-Dephasing-Grenzfall feststeht, dann ist es somit m\"oglich, den Volumenanteil $\eta$ unabh\"angig von der Orientierung $\theta$ zu berechnen. Ist erst einmal der Volumenanteil $\eta$ bekannt, kann der \"aquatoriale Frequenzshift $\delta \omega_{\theta}$ aus der Peakposition berechnet werden. Bei Kenntnis der St\"arke des \"au{\ss}eren Magnetfeldes ist es deshalb m\"oglich, die Suszeptibilit\"atsdifferenz zwischen Zylinder und umgebendem Medium zu ermitteln. So kann man z. B. im Fall von blutgef\"ullten Kapillaren auf den Oxygenierungsgrad des Blutes innerhalb des Gef\"a{\ss}es r\"uckschlie{\ss}en.
Zur Quantifizierung des Volumenanteils basierend auf der obigen Methode ist ein kleines Frequenzintervall notwendig, wof\"ur entsprechend dem Fourier-Theorem das Signalverhalten bei langen Zeiten ben\"otigt wird. Aus praktischen Gr\"unden wird der Magnetisierungszerfall jedoch f\"ur kurze Echozeiten aufgenommen, um ein ausreichend hohes SNR zu erhalten.

Um die Frequenzverteilung f\"ur alle Diffusionsregime zu erhalten, muss die Funktion $G$, die in Gleichung (\ref{md}) ben\"otigt wird, ermittelt werden. Deshalb wird zuerst die Laplace-Transformierte des Magnetisierungs-Zerfalles im Static-Dephasing-Regime $\hat{M}_0(s)$ ermittelt. Dazu kann eine der Gleichungen (\ref{LT1}) bis (\ref{LT}) genutzt werden. Die Auswertung aller drei Gleichungen f\"uhrt zu dem gleichen Ergebnis f\"ur die Laplace-Transformierte des Magnetisierungs-Zerfalles $\hat{M}_0(s)$, die in der allgemeinen Form (\ref{allgmd}) mit der zylinderspezifischen Funktion 
\begin{equation} \label{Gzyl}
G_{\text{C}}(y) = \sqrt{1 + \frac{1}{y^2}}
\end{equation}
geschrieben werden kann.

\begin{figure}
\begin{center}
\includegraphics[width=10cm]{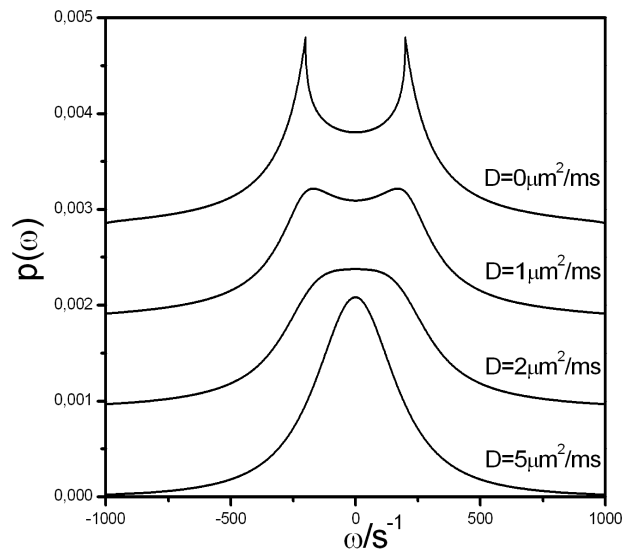}
\caption[Diffusionsabh\"angige Frequenzverteilung f\"ur Zylinder]{\label{Fig:dosdiff}{\footnotesize Diffusionsabh\"angige Frequenzverteilung f\"ur Zylinder. Die Kurven wurden nach Gleichung (\ref{md}) berechnet, wobei die zylinderspezifische $G$-Funktion aus Gleichung (\ref{Gzyl}) genutzt wurde. Der Volumenanteil ist $\eta=0,2$; der \"aquatoriale Frequenzshift $\delta\omega_{\theta}=1000\text{s}^{-1}$; Radius $R_{\text{C}}=5\mu\text{m}$.}}
\end{center}
\end{figure}

\begin{figure}
\begin{center}
\includegraphics[width=10cm]{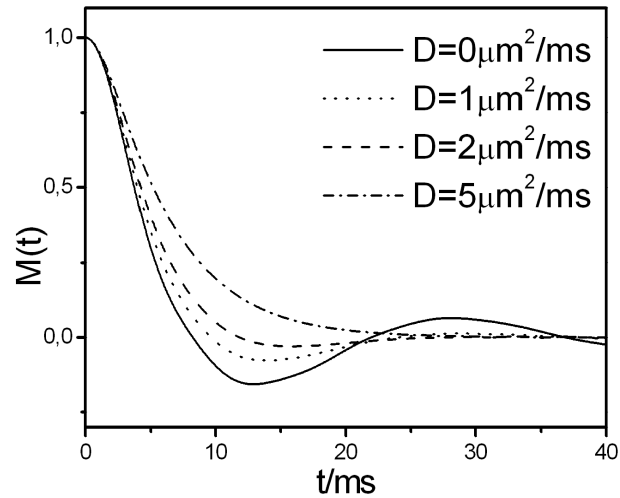}
\caption[Diffusionsabh\"angiger Magnetisierungs-Zeit-Verlauf f\"ur Zylinder]{\label{Fig:sigdiffzyl}{\footnotesize Diffusionsabh\"angiger Magnetisierungs-Zeit-Verlauf f\"ur Zylinder. Entsprechend der Fourier-Transformation nach Gleichung (\ref{SignalD}) erh\"alt man den Magnetisierungs-Zeit-Verlauf, wobei die Frequenzverteilung $p(\omega)$ aus Abbildung \ref{Fig:dosdiff} eingesetzt wurde. Der Volumenanteil betr\"agt $\eta=0,2$; der \"aquatoriale Frequenzshift ist $\delta\omega_{\theta}=1000\text{s}^{-1}$, und f\"ur den Radius wurde $R_{\text{C}}=5\mu\text{m}$ angenommen.}}
\end{center}
\end{figure}

\begin{figure}
\begin{center}
\includegraphics[width=10cm]{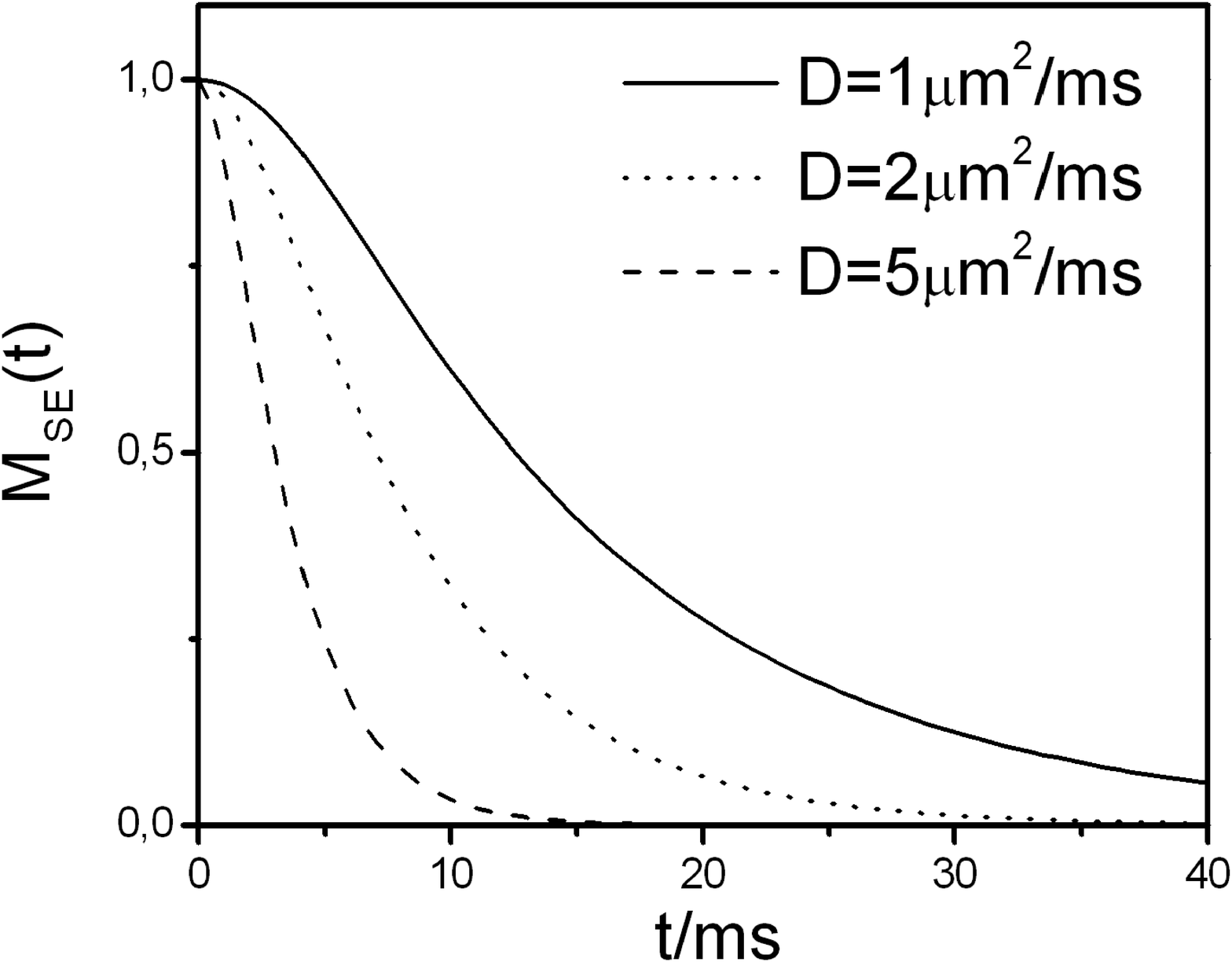}
\caption[Diffusionsabh\"angiger Signal-Zeit-Verlauf des Spin-Echos f\"ur Zylinder]{\label{Fig:sezyl}{\footnotesize Diffusionsabh\"angiger Signal-Zeit-Verlauf des Spin-Echos f\"ur Zylinder. Die Kurven wurden nach Gleichung (\ref{spin-echo}) berechnet, wobei der Signal-Zeit-Verlauf des Gradienten-Echos $M(t)$ aus Abbildung \ref{Fig:sigdiffzyl} eingesetzt wurde. Der Volumenanteil betr\"agt $\eta=0,2$; der \"aquatoriale Frequenzshift ist $\delta\omega_{\theta}=1000\text{s}^{-1}$, und f\"ur den Radius wurde $R_{\text{C}}=5\mu\text{m}$ angenommen.}}
\end{center}
\end{figure}

Die zweite Gr\"o{\ss}e, die zur Berechnung der Frequenzverteilung in Gleichung (\ref{md}) ben\"otigt wird, ist die Korrelationszeit $\tau$, welche die Diffusionseffekte beschreibt. Im Falle eines undurchl\"assigen Zylinders werden reflektierende Randbedingungen auf der Oberfl\"ache des inneren Zylinders angenommen. Die zugeh\"orige Korrelationszeit ist in Gleichung (\ref{tau}) gegeben. Werden die Ausdr\"ucke (\ref{Gzyl}) und (\ref{tau}) genutzt, kann aus Gleichung (\ref{md}) die Frequenzverteilung $p(\omega)$ erhalten werden. Um die Abh\"angigkeit von der Diffusion zu demonstrieren, ist in Abbildung \ref{Fig:dosdiff} die Frequenzverteilung f\"ur verschiedene Werte des Diffusionskoeffizienten $D$ dargestellt. Wenn die Diffusion vernachl\"assigbar ist, stimmt die Frequenzverteilung mit der des Static-Dephasing-Regimes aus Gleichung (\ref{DOS-cyl}) \"uberein, wie in Abbildung \ref{Fig:DOSsd} gezeigt. Eine \"ahnliche charakteristische Form der Frequenzverteilung bei zylinderf\"ormiger Geometrie wurde von Zimmerman und Foster numerisch beschrieben \cite{Zimmerman56}.
Eine Fourier-Transformation der Frequenzverteilung $p(\omega)$ entsprechend Gleichung (\ref{SignalD}) f\"uhrt zum Signal-Zeit-Verlauf wie er in Abbildung \ref{Fig:sigdiffzyl} dargestellt wird. Auch f\"ur den Signal-Zeit-Verlauf im Static-Dephasing-Regime stimmt das Ergebnis mit fr\"uheren Ergebnissen aus Gleichung (\ref{allg}) und der zylinderspezifischen Funktion (\ref{kgzyl}) \"uberein, wie aus Abbildung \ref{Fig:sigdiffzyl} hervorgeht.
Der Magnetisierungs-Zeit-Verlauf f\"ur ein Spin-Echo-Experiment kann aus dem Magnetisierungs-Zeit-Verlauf des Gradienten-Echos entsprechend Gleichung (\ref{spin-echo}) ermittelt werden. Die entsprechenden Verl\"aufe sind in Abbildung \ref{Fig:sezyl} dargestellt.\\

\subsubsection{Numerische L\"osung}
Um den Magnetisierungs-Zeit-Verlauf numerisch zu ermitteln wird die Bloch-Torrey-Gleichung (\ref{BT}) in der einfachen Form
\begin{equation} \label{Eq:Diff1}
\frac{\partial}{\partial t} \, f(\mathbf{r},t) = \Big[ \, \lambda \, \Delta +
\mu \, \I \, \omega(\mathbf{r}) \, \Big] f(\mathbf{r},t)
\end{equation}
geschrieben, wobei $\lambda$ dem Diffusionskoeffizienten entspricht und $\mu$ der St\"arke des Suszeptibilit\"atseffektes. Zum Zeitpunkt $t=0$ zeigt die Magnetisierung in jedem Punkt des Voxels in $z$-Richtung. Dies entspricht der Anfangsbedingung
\begin{equation}
f(\mathbf{r},t{=}0) = 1
\end{equation}
und den reflektierenden Randbedingungen $ \partial f(\mathbf{r},t) / \partial n = 0 $. Dabei seien $ \lambda \geqslant 0 $ und $ \mu $ reelle Konstanten sowie $ \omega(\mathbf{r}) $ eine reelle Funktion. Gesucht ist die Funktion
\begin{equation} \label{Eq:Integral2}
F(t) = \int_{0}^{2\pi} \text{d} \phi \int_{R_{\text{C}}}^{R} \text{d} r \,r \, f(\mathbf{r},t) \;.
\end{equation}
Durch Diskretisierung (z. B. mittels finiter Differenzen oder durch Entwicklung nach Basisfunktionen) geht (\ref{Eq:Diff1}) in die Vektor-Matrix-Gleichung
\begin{equation} \label{Eq:Diff2}
\frac{\partial}{\partial t} \, \mathbf{f}(t) = \Big[ \, \lambda \, \mathbf{D}
+ \mu \, \I \, \mathbf{W} \, \Big] \, \mathbf{f}(t)
\end{equation}
\"uber. Dabei sind $ \mathbf{D} $ und $ \mathbf{W} $ hermitesche Matrizen (im allgemeinen Sinne, siehe \cite{Glutsch04}), die sich durch die Diskretisierung von $ \Delta $ und $ \omega $ ergeben. Die L\"osung von (\ref{Eq:Diff2}) kann am einfachsten durch das folgende Schema erfolgen:
\begin{equation}
\mathbf{f}(t{+}\Delta t) = \E^{\frac{1}{2} \lambda \I \mathbf{W} \Delta t}
\Big[ 1 + \mathbf{D} \Delta t \Big] \E^{\frac{1}{2} \lambda \I \mathbf{W}
\Delta t} \, \mathbf{f}(t) \;.
\end{equation}
Dabei muss die Stabilit\"atsbedingung $ \Delta t < 2/(\lambda {\parallel}\mathbf{D}{\parallel}) $ erf\"ullt sein, wobei $ {\parallel}\mathbf{D}{\parallel} $ die Spektralnorm von $ \mathbf{D} $, d. h. der Betrag des gr\"o{\ss}ten Eigenwertes von $ \mathbf{D} $, ist. Um eine hohe Genauigkeit zu erzielen, muss au{\ss}erdem $ |\lambda| \, {\parallel} \mathbf{W} {\parallel} \ll 1 $ sein; die Aufteilung des Exponentials erh\"oht die Genauigkeit. Idealerweise ist $ \mathbf{D} $ schwachbesetzt und $ \mathbf{W} $ diagonal, dann ist die L\"osung nach (\ref{Eq:Diff2}) eine $ \mathrm{O}(N) $-Methode; aber auch vollbesetzte Matrizen mit bis zu $ N = 1000 $ sind noch tolerierbar.

Im zweidimensionalen Fall ist der Laplace-Operator in Polarkoordinaten durch
\begin{equation}
\Delta = \frac{1}{r} \frac{\partial}{\partial r} r
\frac{\partial}{\partial r} + \frac{1}{r^2} \frac{\partial^2}{\partial
\phi^2} = \Delta_r + \Delta_\phi \;,
\end{equation}
gegeben. Das Potential ergibt sich aus der lokalen Resonanzfrequenz in Gleichung (\ref{om-Feld-Zyl}) zu
\begin{equation} \label{Eq:Potential2}
\omega(r,\phi) = \frac{\cos(2\phi)}{r^2} \;.
\end{equation}
Die Schrittweiten f\"ur den Abstand $ h_r = (R-R_{\text{C}})/(m-1) $ sowie f\"ur den Winkel $ h_\phi = 2\pi/n $ ergeben die Gitterpunkte $ r_j = R_{\text{C}} + (j-1) \, h_r $; $ j = 1, \dots, m $ und $ \phi_k = k
h_\phi $; $ k = 1, \dots, n $. Um eine gleichm\"a{\ss}ige Diskretisierung zu gew\"ahrleisten, sollte $ R_{\text{C}} h_\phi \leqslant h_r \leqslant R h_\phi $ sein. Es folgt die Approximation des Integrals (\ref{Eq:Integral2}) durch
\begin{equation}
F(t) = \sum_{j=1}^m \sum_{k=1}^n \, g_j \, h_k \, f_{jk}(t)
\end{equation}
mit
\begin{equation}
g_j = \left\{ \begin{array}{ll} \frac{1}{2} \, h_r r_j & \mbox{f\"ur}
\;\;\; j = 1, n \\[2ex] h_r r_j & \mbox{sonst} \;. \end{array} \right.
\end{equation}
Die Diskretisierung des Operators $ \Delta_\phi $ ergibt
\begin{equation}
(\Delta_\phi f)_{jk} = \frac{1}{r_j^2} \,
\frac{f_{j,k-1}-2f_{j,k}+f_{j,k+1}}{h_\phi^2} \;,
\end{equation}
wobei die Punkte $ k = n $ und $ k = 0 $ \"ubereinstimmen.

Die Diskretisierung von $ \Delta_r $ ist aufw\"andiger, da die reflektierenden Randbedingungen eingearbeitet werden m\"ussen. Die entsprechende Matrix $ \mathbf{D}_r $ muss folgende Bedingungen erf\"ullen:
\begin{enumerate}
\item Wegen $ \Delta_r f(r) = 0 $ f\"ur $ f(r) = \mbox{const} $ muss gelten:
\begin{equation}
\sum_{j'} D_{r\,jj'} = 0 \;.
\end{equation}

\item Aus der Teilchenzahlerhaltung folgt:
\begin{equation}
\sum_j g_j \, D_{r\,jj'} = 0 \;.
\end{equation}

\item Aus der Hermitizit\"at von $ \Delta_r $ folgt:
\begin{equation}
g_j \, D_{r\,jj'} = g_{j'} \, D_{j'j} \;.
\end{equation}
Eine schw\"achere Forderung w\"are, dass Koeffizienten $ \tilde{g}_j > 0 $ existieren, welche die obige Gleichung erf\"ullen.

\item Das Differenzenschema muss konsistent sein, d. h. f\"ur $ h_r \rightarrow 0 $ m\"ussen die $ (\Delta_r f)_j $ gegen $ \Delta_r f(r_j) $ konvergieren. Daraus folgt insbesondere, dass f\"ur die Punkte, welche keine Randpunkte sind, sich der bekannte Ausdruck f\"ur $ \mathbf{D}_r $ \cite{Glutsch04} ergeben muss.
\end{enumerate}

Eine Tridiagonalmatrix, welche die obigen Bedingungen erf\"ullt, ist
\begin{equation}
\mathbf{D}_r = \frac{1}{h_r^2} \left( \begin{array}{llllllll}
-\frac{2r_{1\frac{1}{2}}}{r_1} & +\frac{2r_{1\frac{1}{2}}}{r_1}
& \\[1.5ex]
+\frac{r_{1\frac{1}{2}}}{r_2} & -\frac{2r_2}{r_2} &
+\frac{r_{2\frac{1}{2}}}{r_2} \\[1.5ex]
& +\frac{r_{2\frac{1}{2}}}{r_3} & -\frac{2r_3}{r_3} &
+\frac{r_{3\frac{1}{2}}}{r_3} \\[1.5ex]
& & \ddots & \ddots & \ddots\\[1.5ex]
& & & +\frac{r_{n-2\frac{1}{2}}}{r_{n-2}} &
-\frac{2r_{n-2}}{r_{n-2}} &
+\frac{r_{n-1\frac{1}{2}}}{r_{n-2}} \\[1.5ex]
& & & & +\frac{r_{n-1\frac{1}{2}}}{r_{n-1}} &
-\frac{2r_{n-1}}{r_{n-1}} &
+\frac{r_{n-\frac{1}{2}}}{r_{n-1}} \\[1.5ex]
& & & & & +\frac{2r_{n-\frac{1}{2}}}{r_{n}} &
-\frac{2r_{n-\frac{1}{2}}}{r_{n}}
\end{array} \right) \;.
\end{equation}
Die Diskretisierung des Potentials aus Gleichung (\ref{Eq:Potential2}) f\"uhrt auf eine Diagonalmatrix, und es ist
\begin{equation}
W_{jk,j'k'} = \delta_{jj'} \, \delta_{kk'} \, \omega(r_j,\phi_k) \;.
\end{equation}
F\"ur die Matrixnormen gilt nach \cite{Glutsch04}:
\begin{equation}
{\parallel} \mathbf{D}_r {\parallel} \leqslant \frac{4,84194}{h_r^2} \;;
\;\;\; {\parallel} \mathbf{D}_\phi {\parallel} = \frac{4}{R_{\text{C}}^2 \,
h_\phi^2} \;; \;\;\; {\parallel} \mathbf{W} {\parallel} = \max_{j,k} \,
\omega(r_j,\phi_k) \;.
\end{equation}

\begin{figure}
\begin{center}
\includegraphics[width=14cm]{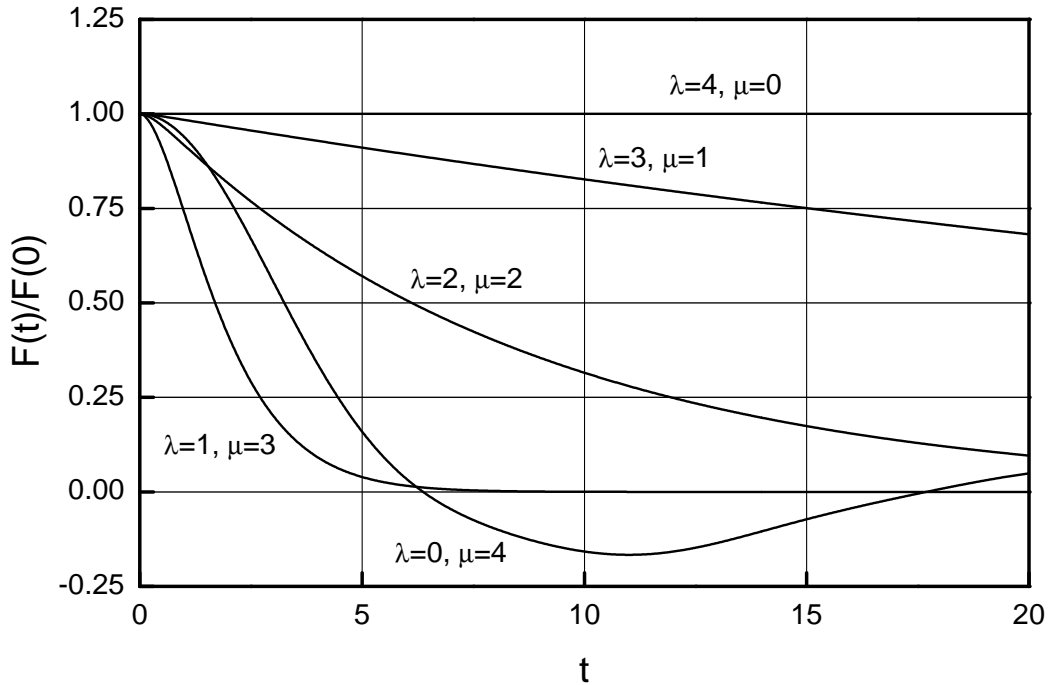}
\caption[Numerische L\"osung f\"ur Zylinder]{{\footnotesize Numerische L\"osung f\"ur Zylinder. Gleichung (\ref{Eq:Diff1}) f\"ur den zweidimensionalen Fall mit $ R_{\text{C}} = 1 $ und $ R = 2 $ wurde numerisch gel\"ost. Das Verh\"altnis $ F(t)/F(0) $ als Funktion von $ t $ ist dargestellt f\"ur verschiedene Parameter $ \lambda $ und $ \mu $.}}
\label{Fig:1}
\end{center}
\end{figure}

Die numerische L\"osung von Gleichung (\ref{Eq:Diff1}) f\"ur den zweidimensionalen Fall mit Hilfe der Methode (\ref{Eq:Diff2}) erfolgt mittels eines selbstgeschriebenen Fortran-Programms. Die L\"osung $ F(t)/F(0) $ ist in Abbildung \ref{Fig:1} f\"ur verschiedene Parameter $ \lambda $ und $ \mu $ dargestellt. Bei $ \mu = 0 $ bleibt $ F(t) = \mathrm{const} $. F\"ur bestimmte Parameter kann die Funktion $ F $ negativ werden, wie sich an dem Grenzfall $ \lambda = 0 $ zeigt.

F\"ur den dreidimensionalen Fall ist die Diskretisierung des Laplace-Operators komplizierter. Hier k\"onnte es vorteilhaft sein, $ f $ nach einem Orthonormalsystem zu entwickeln.

\subsubsection{Experimentelle L\"osung}
Um die analytischen Ergebnisse experimentell zu verifizieren, muss die exakte Form des freien Induktionsabfalls, der alleine durch das lokale inhomogene Magnetfeld verursacht wird, gemessen werden. Alle zus\"atzlichen Gradienten, wie Shimgradienten, beeinflussen die Form des freien Induktionszerfalls. Eine geeignete Methode, den Signal-Zeit-Verlauf in einem vorgegebenen Volumen zu messen, ist die voxelselektive PRESS-Sequenz (Point RESolved Spectroscopy) \cite{Bottomley84,Bottomley87}. 

\begin{figure}
\begin{center}
\includegraphics[width=9.8cm]{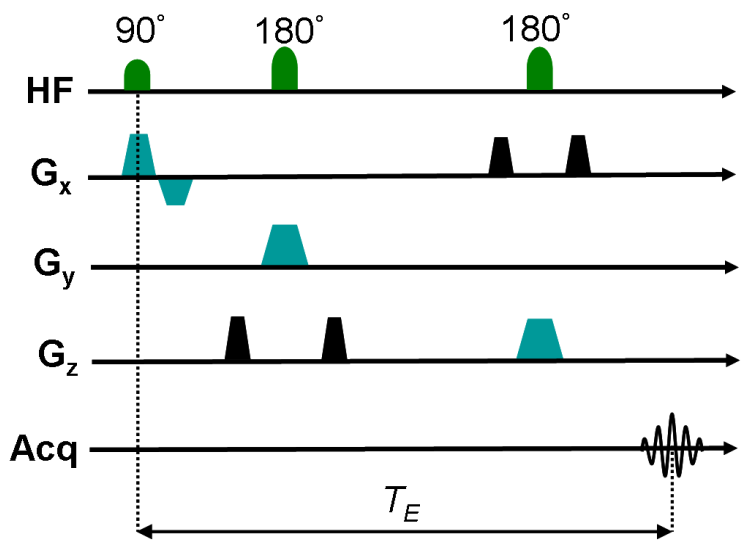}
\caption[Voxelselektive Pulssequenz]{\label{Fig:PRESS}{\footnotesize Voxelsselektive Pulssequenz. Dargestellt ist die Abfolge der eingestrahlten Hochfrequenzpulse (HF) und angelegten Magnetfeldgradienten (grau). Die $180^{\circ}$-Hochfrequenzpulse sind von Spoilergradienten (schwarz) umgeben.}}
\end{center}
\end{figure}

\begin{figure}
\begin{center}
\includegraphics[width=10cm]{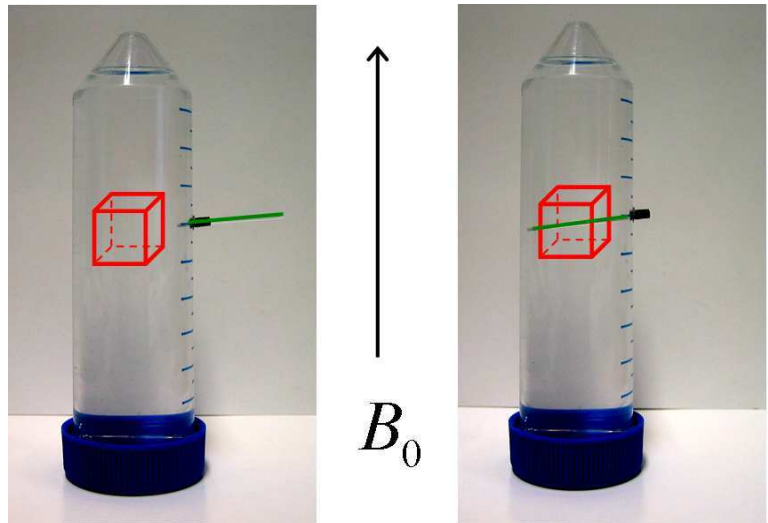}
\caption[Phantom zur Messung der Frequenzverteilung]{\label{Fig:Aufbau}{\footnotesize Phantom zur Messung der Frequenzverteilung. Links: wassergef\"ullte R\"ohre mit herausgezogener Glaskapillare und angedeutetem Lokalisierungsvoxel. Rechts: wassergef\"ullte R\"ohre mit hineingeschobener Glaskapillare.}}
\end{center}
\end{figure}

Die Lokalisierung des Voxels wird, wie in Abbildung \ref{Fig:PRESS} dargestellt, durch drei frequenzselektive Hochfrequenzpulse ($90^{\circ} - 180^{\circ} - 180^{\circ}$), die jeweils w\"ahrend eines angelegten Magnetfeldgradienten eingestrahlt werden, erreicht. Die Signalbeitr\"age von au{\ss}erhalb des selektierten Voxels werden durch Spoiler-Gradienten, welche die $180^\circ$-Pulse umgeben, dephasiert. Um die unerw\"unschten Einfl\"usse zus\"atzlicher Gradienten zu minimieren, wird zuerst auf das Voxel ohne Kapillare geshimmt. Dazu wurde ein Phantom konstruiert, das es erm\"oglicht, die Kapillare erst nach dem Shimmen in das selektierte Voxel einzuf\"uhren, ohne dabei das Phantom zu bewegen. Ein Phantom, welches diesen Anforderungen gen\"ugt, ist in Abbildung \ref{Fig:Aufbau} zu sehen. Es besteht aus einem wassergef\"ullten $50\mathrm{ml}$-Zentrifugenr\"ohrchen (Nunc GmbH $\&$ Co. KG, Thermo Fisher Scientific, Wiesbaden) mit einen Innendurchmesser von $28\, \mathrm{mm}$. An einer Seite ist eine luftgef\"ullte Glaskapillare mit einem Durchmesser von $2R_\mathrm{C} = 1\, \mathrm{mm}$ (Glas Nr. 140, Hilgenberg GmbH, Malsfeld) verschiebbar angebracht. Diese Glaskapillare mit vernachl\"assigbarer Wanddicke erzeugt einen Suszeptibilit\"atssprung von $\Delta \chi = 9\, \mathrm{ppm}$ zum umgebenden Wasser.

Die Experimente wurden an einem Bruker Biospec-System (Bruker BioSpin GmbH, Rheinstetten) mit einer Feldst\"arke von $B_0 = 7,05\, \mathrm{T}$ durchgef\"uhrt. Es wurde ein aktiv geschirmtes Gradientensystem ($397\, \mathrm{mT}/\mathrm{m}$ maximale Gradientenst\"arke) und ein $72\,\mathrm{mm}$-Quadratur-Birdcage-Resonator zum Senden und Empfangen benutzt. Die PRESS-Sequenz wurde in einem einzigen Scan mit einer Echozeit von $T_\mathrm{E} = 20\, \mathrm{ms}$ durchgef\"uhrt. Um das Voxel auszuw\"ahlen, wurden frequenzselektive Hermite-Pulse mit einer Bandbreite von $5,4\, \mathrm{kHz}$ genutzt. Der freie Induktionszerfall wurde f\"ur $4,1\,\mathrm{s}$ bei einer Bandbreite von $4\,\mathrm{kHz}$ aufgenommen, was einer spektralen Aufl\"osung von $0,24\,\mathrm{Hz}$ entsprach.

Zuerst wurde auf das Voxel ohne Glaskapillare geshimmt, wie es auf der linken Seite von Abbildung \ref{Fig:Aufbau} dargestellt ist. Somit konnte im gesamten Voxel ein nahezu homogenes Magnetfeld erreicht werden. Die Linienbreite des Wassersignals ohne Kapillare im untersuchten Voxel betrug $FWHM = 1,2 \, \text{Hz}$. Anschlie{\ss}end wurde die Glaskapillare in das Phantom eingef\"uhrt, wie auf der rechten Seite von Abbildung \ref{Fig:Aufbau} gezeigt. Jetzt erzeugt die Kapillare ein lokales inhomogenes Magnetfeld im Voxel. Der durch dieses lokale Magnetfeld beeinflusste Induktionszerfall kann nun mittels der oben beschriebenen voxelselektiven PRESS-Sequenz gemessen werden.

Entscheidend f\"ur die Qualit\"at des aufgenommenen Induktionszerfalles ist die Lage und Gr\"o{\ss}e des gew\"ahlten Voxels. Das Voxel darf nicht zu gro{\ss} gew\"ahlt werden, damit die Suszeptibilit\"atsspr\"unge vom Rand des Phantoms keinen Einfluss auf den Induktionszerfall nehmen. Andererseits darf das Voxel nicht zu klein gew\"ahlt werden, damit die Bedingung $\eta \ll 1$ nicht verletzt wird. Aus Symmetriegr\"unden sollte die Kapillare im Zentrum des Voxels liegen und senkrecht zum \"au{\ss}eren Magnetfeld orientiert werden ($\theta=90^{\circ}$), was die Suszeptibilit\"atseffekte maximiert.

\begin{figure}
\begin{center}
\includegraphics[width=10cm]{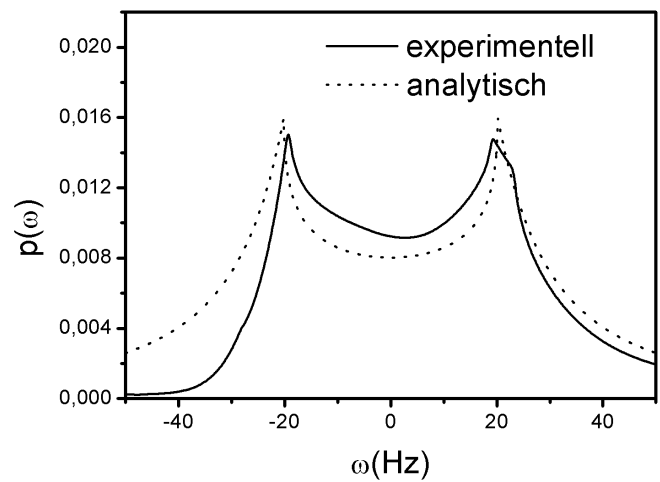}
\caption[Gemessene Frequenzverteilung]{\label{Fig:Spektrum}{\footnotesize Gemessene Frequenzverteilung. Vergleich der experimentell erhaltenen Frequenzverteilung (durchgezogene Linie) mit den analytischen Ergebnissen nach Gleichung (\ref{DOS-cyl}) f\"ur die Parameter $\eta=0,015$ und $\delta\omega_{\theta}=1341\, \mathrm{Hz}$ (gestrichelte Linie). Wie aus Gleichung (\ref{DOS-cyl}) zu erkennen ist, befinden sich die charakteristischen Peaks an den Positionen $\pm \eta \, \delta\omega_{\theta}$, d. h. in diesem Fall an den Stellen $\pm 20\, \mathrm{Hz}$. Dies stimmt auch gut mit den gemessenen Positionen bei $\pm 19,3\, \mathrm{Hz}$ \"uberein.}}
\end{center}
\end{figure}

Um den optimalen Kompromiss zwischen beiden Anforderungen zu finden, wurde ein w\"urfelf\"ormiges Voxel mit der Kantenl\"ange $7,2\, \mathrm{mm}$ gew\"ahlt. F\"ur diese gew\"ahlte Voxelgr\"o{\ss}e ergab sich ein Volumenanteil von $\eta=0,015$.
Entsprechend der Gleichung (\ref{Shift}) erzeugt die Glaskapillare theoretisch den Frequenzshift $\delta\omega_{\theta}=1341 \, \mathrm{Hz}$ auf der Oberfl\"ache der Kapillare. Damit ergibt sich, dass die beiden Peaks bei $\pm \eta \delta\omega_{\theta} = 20 \, \mathrm{Hz}$ liegen. Dies stimmt auch gut mit den experimentell ermittelten Werten von $\pm 19,3 \, \mathrm{Hz}$ \"uberein (siehe Abbildung \ref{Fig:Spektrum}).

\begin{figure}
\begin{center}
\includegraphics[width=10cm]{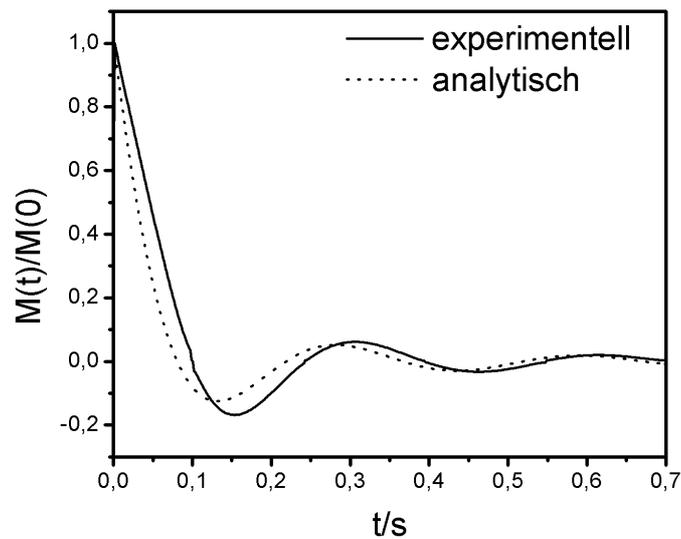}
\caption[Gemessener Signal-Zeit-Verlauf]{\label{Fig:FID}{\footnotesize Gemessener Signal-Zeit-Verlauf. Der experimentell erhaltene freie Induktionszerfall (durchgezogene Linie) wurde mit den analytischen Ergebnissen von Gleichung (\ref{allg}) und der zylinderspezifischen Funktion (\ref{kgzyl}) f\"ur die Parameter $\eta=0,015$ und $\delta\omega_{\theta}=1341\, \mathrm{Hz}$ verglichen (gestrichelte Linie).}}
\end{center}
\end{figure}

Mit den bekannten Werten f\"ur den Frequenzshift $\delta\omega_{\theta}$ und f\"ur den Volumenanteil $\eta$ kann das experimentell erhaltene Spektrum mit den Vorhersagen von Gleichung (\ref{DOS-cyl}) verglichen werden. In Abbildung \ref{Fig:Spektrum} ist eine sehr gute \"Ubereinstimmung zwischen der theoretisch berechneten und der experimentell gemessenen Frequenzverteilung \"uber den gesamten Frequenzbereich zu sehen.
Auch der experimentell ermittelte Signal-Zeit-Verlauf stimmt, wie Abbildung \ref{Fig:FID} zeigt, mit dem theoretisch nach Gleichung (\ref{allg}) berechneten gut \"uberein.
Die Frequenzverteilung um Kugeln wurde zuerst von Cheng et al. \cite{Cheng01} untersucht. Die experimentelle Best\"atigung wurde von Seppenwoolde et al. \cite{Seppenwoolde05} erbracht, wobei \"ahnliche Phantome und Methoden wie die hier vorgestellten verwendet wurden. Hiermit entspricht Abbildung 7 f\"ur eine kugelf\"ormige Geometrie in der Arbeit \cite{Seppenwoolde05} der Abbildung \ref{Fig:Spektrum} dieser Arbeit f\"ur eine zylindrische Geometrie.

\chapter{\label{Chap:Dephas}Nicht-Gau{\ss}f\"ormige Dephasierung}
\pagestyle{headings}

\section{Theorien zur Spindephasierung}
Die suszeptibilit\"atssensitive Bildgebung spielt bei der Untersuchung der Lokalisation magnetisch markierter Zellen \cite{Weissleder97} oder in der BOLD-Bildgebung \cite{Ogawa90} eine entscheidende Rolle. Lokale Variationen der Spindephasierung sind f\"ur den Kontrast bei der suszeptibilit\"atssensitiven Bildgebung verantwortlich.

F\"ur die Suszeptibilit\"atseffekte sind entweder ein intrinsisches Kontrastmittel (Desoxyh\"amoglobin \cite{Pauling36}) oder ein extrinsisches Kontrastmittel (USPIO) verantwortlich, die ein inhomogenes Magnetfeld erzeugen, welches die Spindephasierung beschleunigt. Dieser Effekt kann den Gewebestrukturen zugeordnet werden, die mit dem Kontrastmittel assoziiert sind, z. B. der vaskul\"aren Architektur beim BOLD-Effekt \cite{Yablonskiy94,Bauer99,Bauer99PRL,Bauer99T2,Kiselev98} oder der Zelldichte \cite{Weissleder97,Stroh05,Ziener05MRM}. Um diesen Effekt zu interpretieren, muss der Zusammenhang zwischen der Spinbeweglichkeit und den r\"aumlichen Dimensionen als auch der St\"arke der Feldinhomogenit\"aten beachten werden. Dadurch werden wie in Abschnitt \ref{Subsec:Diffusionsregime} beschrieben die sogenannten Diffusionsregime definiert. Die meisten Theorien zur Beschreibung der Spindephasierung beschr\"anken sich auf die Grenzf\"alle des Static-Dephasing-Regimes \cite{Yablonskiy94,Kiselev98} oder des Motional-Narrowing-Regimes \cite{Gillis87}. Jedoch liegt vielen Situationen ein intermedi\"ares Diffusionsregime zugrunde \cite{Bauer99,Bauer99PRL,Sukstanskii02,Sukstanskii03,Sukstanskii04,Bauer99T2}, in dem keine der beiden N\"aherungen g\"ultig ist.

Bisher gibt es nur zwei N\"aherungen, welche die Spindephasierung \"uber den gesamten Dynamikbereich beschreiben: die Strong-Collision-N\"aherung \cite{Bauer99,Bauer99PRL} mit ihrer Erweiterung \cite{Bauer02} und die Gau{\ss}sche N\"aherung. Die Strong-Collision-N\"aherung approximiert die Diffusionsdynamik der Spins durch eine einfachere stochastische Dynamik, wobei die urspr\"unglichen Parameter des Originalprozesses in selbstkonsistenter Weise erhalten bleiben. In der Gau{\ss}schen N\"aherung wird angenommen, dass die Wahrscheinlichkeitsverteilung der Phasenwinkel eine Gau{\ss}sche Funktion ist, was die Berechnung der Relaxation der Magnetisierung erleichtert. Obwohl die Voraussetzungen, die zur Anwendung der Gau{\ss}schen N\"aherung erf\"ullt sein m\"ussen, noch nicht ausreichend ermittelt wurden, konnte empirisch gezeigt werden, dass die Gau{\ss}sche N\"aherung in vielen F\"allen anwendbar ist \cite{Sukstanskii02,Sukstanskii03,Sukstanskii04} und auch durch die (erweiterte) Strong-Collision-N\"aherung \cite{Bauer02} approximierbar ist. Nun stellt sich die Frage, welche stochastische Dynamik der Phasenmodulation die Gau{\ss}sche N\"aherung rechtfertigt. Diese Fragestellung wird in den n\"achsten Abschnitten er\"ortert.

\section{\label{Sec:SteadyState}Gleichgewichtsverteilung der lokalen Frequenzen}
F\"ur die folgenden Untersuchungen wird ein Spin betrachtet, der im klassischen Sinne durch einen Hochfrequenzpuls in eine Ebene senkrecht zum \"au{\ss}eren Magnetfeld ausgelenkt wurde. Dieser Spin pr\"azediert jetzt in dieser senkrechten Ebene und wird deshalb transversal polarisierter Spin genannt. Der Phasenwinkel $ \varphi $ dieses pr\"azedierenden Spins, der in einer inhomogenen magnetischen Umgebung diffundiert, unterliegt stochastischen Modulationen, deren Zeitverlauf durch
\begin{equation} \label{phaseaccumulation}
\varphi(t) = \int_0^t \omega[x(\xi)] \text{d} \xi = \int_0^t \omega(\xi) \text{d} \xi
\end{equation}
gegeben ist, wobei $ x(\xi) $ die Trajektorie des Spins und $ \omega[x] $ die Pr\"azessionsfrequenz im lokalen Magnetfeld ist. Der Zeitverlauf des transversalen Anteils der Magnetisierung ist durch den Ensemblemittelwert des Phasenfaktors gegeben:
\begin{equation} \label{relaxation}
M(t)=\langle \, \text{e}^{\text{i} \varphi(t)} \, \rangle \;.
\end{equation}
Die Gau{\ss}sche Dephasierung impliziert, dass die Wahrscheinlichkeitsdichte, einen Phasenwinkel $ \varphi $ zur Zeit $ t $ zu finden, eine Gau{\ss}sche Funktion ist, d. h.
\begin{equation} \label{gaussian}
P(\varphi,t)=\frac{1}{\sqrt{2\pi\;\langle \, \varphi^2(t) \, \rangle}}\;\exp\left({-\frac{\varphi^2}{2 \; \langle \, \varphi^2(t) \, \rangle}}\right)\;,
\end{equation}
wobei $ \langle \, \varphi^2(t) \, \rangle $ die Ensemblevarianz des Phasenwinkels ist. Unter diesen Bedingungen liefert Gleichung (\ref{relaxation}) den bekannten Ausdruck \cite{Cowan97}
\begin{equation} \label{gaussianrelaxation}
M(t) = \exp \left(- \frac{\langle \, \varphi^2(t) \, \rangle}{2}\right) \;.
\end{equation}
Wie \"ublich, werden Gleichgewichtsbedingungen f\"ur die Wechselwirkung des Spinensembles mit seiner Umgebung angenommen, d. h. die Dephasierung ist unabh\"angig von einer Zeitverschiebung. Um die Varianz des Phasenwinkels $\langle \, \varphi^2(t) \, \rangle$ zu erhalten, wird Gleichung (\ref{phaseaccumulation}) auf beiden Seiten quadriert:
\begin{equation} 
\varphi^2(t) = \left[ \int_0^t \omega(\xi) \text{d} \xi \right] \left[ \int_0^t \omega(\xi) \text{d} \xi \right] \,.
\end{equation}
Um die Klammern zu entfernen, wird die Integrationsvariable in der ersten Klammer $\eta$ und in der zweiten Klammer $\xi$ genannt. Damit ergibt sich 
\begin{equation} 
\varphi^2(t) = \int_0^t \text{d} \eta \int_0^t \text{d} \xi \omega(\eta) \omega(\xi) \,.
\end{equation}
Die Bildung der Varianz kann nun ausgef\"uhrt werden und ergibt
\begin{align}
\nonumber
\langle \, \varphi^2(t) \, \rangle & = \int_0^{t} \text{d} \eta \int_0^{t} \text{d}\xi \; \langle \, \omega(\eta)\omega(\xi) \, \rangle
\\[-1ex] & \\[-1ex]
\nonumber
& = \int_0^{t} \text{d} \eta \;(t-\eta)\;c_2(\eta)\;.
\end{align}
Hierbei ist $ c_2 $ die Zweipunktkorrelationsfunktion, die unter Gleichgewichtsbedingungen durch
\begin{equation}
c_2(t) = \langle \, \omega(t) \omega(0) \, \rangle
\end{equation}
definiert ist.

Die Verteilung des Phasenwinkels legt die Wahrscheinlichkeitsverteilung der lokalen Frequenzen fest. Gleichung (\ref{phaseaccumulation}) liefert f\"ur kleine Zeitintervalle $ t \to \text{d}t $:
\begin{equation}
\varphi(\text{d}t) = \omega[x(0)] \; \text{d}t \;.
\end{equation}
Demzufolge ist die Verteilung des Phasenwinkels proportional zur Gleichgewichtsverteilung der lokalen Frequenzen $ p_0(\omega) $
\begin{equation}
P(\varphi,\text{d}t) \sim p_0(\omega)\;.
\end{equation}
Diese Relation impliziert, dass im Falle Gau{\ss}scher Dephasierung (siehe Gleichung (\ref{gaussian})) die lokale Frequenzverteilung $ p_0(\omega) $ eine Gau{\ss}sche Funktion ist. 

Diese Gau{\ss}sche N\"aherung zur Beschreibung der Spindephasierung wurde erstmals von Anderson und Weiss eingef\"uhrt \cite{Callaghan,Anderson53}.

\section{\label{Sec:Non}Nicht-Gau{\ss}f\"ormige Phasenakkumulation}
Von Anderson und Weiss wurde angenommen, dass die Verteilung der lokalen Frequenzen $ p_0(\omega) $ gau{\ss}f\"ormig ist. Die entscheidende weitere Annahme basiert auf der stochastischen Akkumulation des Phasenwinkels (\ref{phaseaccumulation}). Der Phasenwinkel kann als Summe lokaler Frequenzen multipliziert mit dem Zeitintervall $ \text{d}t $ betrachtet werden. Anderson und Weiss behaupteten, dass die Summe gau{\ss}verteilter lokaler Frequenzen auch wieder gau{\ss}verteilt ist. Das Gleiche gilt f\"ur den Phasenwinkel, was den einfachen Zusammenhang zum Magnetisierungs-Zeit-Verlauf nach Gleichung (\ref{gaussianrelaxation}) impliziert.

Jedoch ist die Aussage \glqq eine Summe gau{\ss}verteilter Variablen ist gau{\ss}verteilt\grqq\ im Allgemeinen nicht richtig. Diese Aussage gilt, wenn die Variablen stochastisch unabh\"angig sind. Jedoch ist diese stochastische Unabh\"angigkeit f\"ur die Dynamik der lokalen Frequenzen offensichtlich nicht erf\"ullt, es sei denn, die Fluktuationen sind extrem schnell. Um dies zu zeigen, wird ein Beispiel der Strong-Collision-N\"aherung mit gau{\ss}verteilten lokalen Frequenzen betrachtet, das eine Nicht-Gau{\ss}sche Phasenakkumulation zeigt.

Anstatt sich auf das gesamte Integral der Phasenakkumulation in Gleichung (\ref{phaseaccumulation}) zu konzentrieren, werden nur zwei lokale Frequenzen $ \omega_1 $ und $ \omega_2 $ betrachtet, die durch das Zeitintervall $ \Delta t $ getrennt sind. Der Phasenwinkel kann als
\begin{equation} \label{Linearkombi}
\varphi = \omega_1 \Delta t + \omega_2 \Delta t
\end{equation}
geschrieben werden. In der Dynamik eines Strong-Collision-Prozesses ist die \"Ubergangsrate zwischen verschiedenen lokalen Frequenzen proportional zur Gleichgewichtswahrscheinlichkeit der Frequenz. F\"ur die Wahrscheinlichkeitsdichte im Gleichgewicht wird eine Gau{\ss}sche Form angenommen:
\begin{equation} \label{Gauss-equi}
p_0(\omega) = \frac{1}{\sqrt{2 \pi \sigma_0^2}} \exp \left( -\frac{\omega^2}{2 \sigma_0^2} \right) \,.
\end{equation}
Die Greensche Funktion, d. h. die bedingte Wahrscheinlichkeit, einen Spin unter dem Einfluss des lokalen Feldes mit der Frequenz $ \omega_2 $ nach der Zeit $ \Delta t $ zu finden, wenn er urspr\"unglich die Frequenz $ \omega_1 $ besa{\ss}, hat in der Strong-Collision-N\"aherung die Form \cite{Bauer99,Bauer99PRL}
\begin{equation} \label{strongcollision1}
G(\omega_2,\omega_1,\Delta t) = (1-\text{e}^{-\Delta t/\tau}) \; p_0(\omega_2) + \text{e}^{-\Delta t/\tau} \delta(\omega_2-\omega_1)\;,
\end{equation}
wobei $ \tau $ f\"ur die Korrelationszeit der Zweipunktkorrelationsfunktion steht und $ \delta $ die Dirac-Distribution darstellt. Nach dieser Gleichung nimmt der Anfangszustand $ \omega_1 $ mit der Korrelationszeit $ \tau $ ab und der Gleichgewichtszustand w\"achst entgegengesetzt dazu.

\begin{figure}
\begin{center}
\includegraphics[width=10cm]{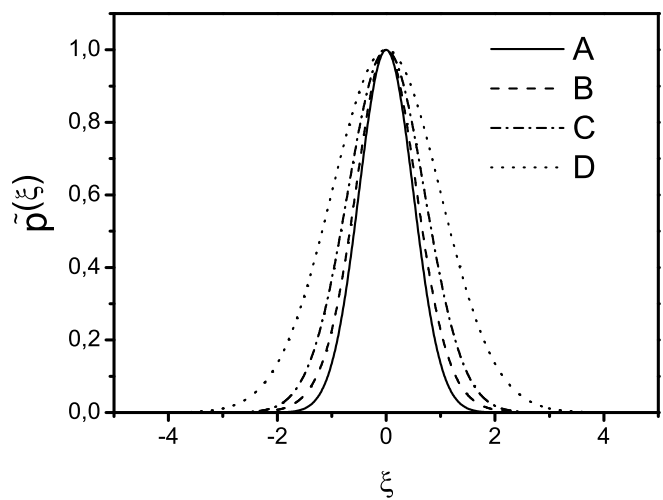}
\caption[Fourier-Transformierte der Phasenwinkelwahrscheinlichkeit]{\label{Fig:Nicht-Gauss}{\footnotesize Fourier-Transformierte der Phasenwinkelwahrscheinlichkeit. Dargestellt ist die Fourier-Transformierte der Wahrscheinlichkeit des Phasenwinkels nach Gleichung (\ref{FTSC2}) f\"ur verschiedene Werte $\Delta t / \tau$ (A: $\Delta t / \tau = 0$, B: $\Delta t / \tau = 0,5$, C: $\Delta t / \tau \to \infty$). F\"ur die Varianz wurde $\sigma_0^2 = 1$ angenommen. Nur die Fourier-Transformierte der Gleichgewichtsdichte (dargestellt in D) $ \tilde{p}_0(\xi) = \exp(- \xi^2 \sigma_0^2 /2) $ ist eine Gau{\ss}sche Funktion, w\"ahrend alle anderen Kurven Linearkombinationen von Gau{\ss}funktionen zeigen.}}
\end{center}
\end{figure}

Die Wahrscheinlichkeit, den Phasenwinkel $ \varphi = \chi \Delta t $ zu finden, ist
\begin{equation} \label{prob}
p(\chi) = \int_{-\infty}^{+\infty} \text{d}\omega_1 \int_{-\infty}^{+\infty} \text{d} \omega_2 \; \delta(\chi - \omega_2 - \omega_1) G(\omega_2,\omega_1,\Delta t) \; p_0(\omega_1) \;.
\end{equation}
Wird die Greensche Funktion (\ref{strongcollision1}) eingesetzt, ergibt sich f\"ur die Fourier-Transformierte der Wahrscheinlichkeit des Phasenwinkels
\begin{align} \label{FTSC}
\tilde{p}(\xi) & = \int_{-\infty}^{+\infty} \text{d}\chi \; \text{e}^{\text{i} \xi \chi} \; p(\chi) \\[2ex]
\label{FTSC2}
& = (1-\text{e}^{-\Delta t/\tau}) \; \tilde{p}_0^2(\xi) + \text{e}^{-\Delta t/\tau} \tilde{p}_0(2\xi)\;,
\end{align}
wobei $ \tilde{p}_0(\xi) = \exp(- \xi^2 \sigma_0^2 /2) $ die Fourier-Transformierte der Gleichgewichtsdichte (\ref{Gauss-equi}) darstellt. Da sich aus der Fourier-Transformation einer Gau{\ss}funktion wieder eine Gau{\ss}funktion ergibt, zeigt Gleichung (\ref{FTSC}), dass die Fourier-Transformierte des Phasenwinkels aus der Summe der Gau{\ss}schen Funktionen $ \tilde{p}_0^2(\xi) $ und $ \tilde{p}_0(2\xi) $ entsteht. Diese Summe stellt jedoch im Allgemeinen keine Gau{\ss}sche Funktion dar. Deshalb ist die Originalfunktion, also die Verteilung des Phasenwinkels im Allgemeinen auch keine Gau{\ss}sche Funktion. In Abbildung \ref{Fig:Nicht-Gauss} ist die Fourier-Transformierte des Phasenwinkels f\"ur verschiedene Werte $\Delta t / \tau$ dargestellt. Weder f\"ur sehr kurze noch f\"ur sehr lange Zeitintervalle $\Delta t$ ist die Fourier-Transformierte des Phasenwinkels eine rein Gau{\ss}sche Funktion, sondern immer eine Linearkombination aus $\tilde{p}_0^2(\xi)$ und $\tilde{p}_0(2\xi)$.

Unter der Annahme einer gau{\ss}f\"ormigen Gleichgewichtsdichte der Form (\ref{Gauss-equi}) stellt sich die Frage, welche Greensche Funktion zu einer gau{\ss}f\"ormigen Verteilung $p(\chi)$ f\"uhrt. Wenn beide Verteilungen gau{\ss}f\"ormig sind, dann sind sie ein Fourier-Paar und deshalb ist das innere Integral in Gleichung (\ref{prob}) der Kern der Fourier-Transformation, d. h. $G(\omega_1,\omega_2,\Delta t) = \sqrt{\sigma_0^2/(2\pi)} \exp{[-\text{i}(\omega_1 + \omega_2)\omega_1]}$.

\section{\label{Sec:Transition}\"Ubergangsdynamik zwischen lokalen Frequenzen}
\begin{figure}
\begin{center}
\includegraphics[width=10cm]{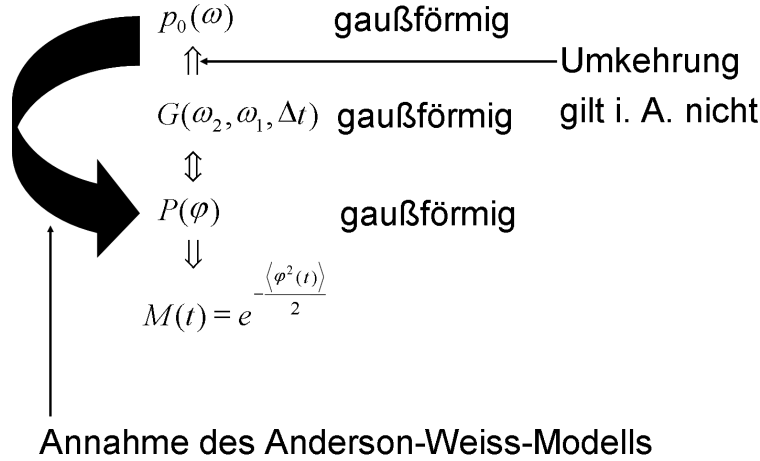}
\caption[Widerspr\"uche im Anderson-Weiss-Modell]{\label{Fig:Gauss}{\footnotesize Widerspr\"uche im Anderson-Weiss-Modell. Aus einer gau{\ss}f\"ormigen Resonanzfrequenzverteilung im Gleichgewicht $p_0(\omega)$ folgerten Anderson und Weiss, dass auch der Phasenwinkel gau{\ss}verteilt ist. Dies ist aber nicht gerechtfertigt, da aus einer gau{\ss}f\"ormigen Resonanzfrequenzverteilung im Gleichgewicht $p_0(\omega)$ nicht folgt, dass die Dichte der bedingten Wahrscheinlichkeit $ G(\omega_2,\omega_1,\Delta t) $ auch gau{\ss}f\"ormig ist.}}
\end{center}
\end{figure}

Im letzten Abschnitt wurde gezeigt, dass eine gau{\ss}f\"ormige Gleichgewichtsverteilung der lokalen Frequenzen nicht automatisch eine gau{\ss}f\"ormige Verteilung des Phasenwinkels impliziert. Dieser Sachverhalt ist noch einmal in Abbildung \ref{Fig:Gauss} illustriert.

In diesem Abschnitt wird der entgegengesetzte Fall behandelt. Die stochastische Abh\"angigkeit der aufeinanderfolgenden lokalen Frequenzen, welche die Annahme der Gau{\ss}schen Dephasierung impliziert, wird abgeleitet, d. h. die \"Ubergangsdynamik der lokalen Frequenzen wird definiert.

Wie auch im vorangehenden Abschnitt werden nur zwei lokale Frequenzen $ \omega_1 $ und $ \omega_2 $ betrachtet, die durch das Zeitintervall $ \Delta t $ getrennt sind, anstatt das gesamte Integral der Phasenakkumulation in Gleichung (\ref{phaseaccumulation}) zu untersuchen. Der Phasenwinkel ergibt sich dann nach Gleichung (\ref{Linearkombi}) und f\"ur diesen Winkel wird eine Gau{\ss}verteilung vorausgesetzt. Zur Verallgemeinerung wird angenommen, dass nicht nur die Summe, sondern jede Linearkombination
\begin{equation}
\chi = a \omega_1 + b \omega_2
\end{equation}
der lokalen Frequenzen $ \omega_1 $ und $ \omega_2 $ gau{\ss}verteilt ist. Die Wahrscheinlichkeitsdichte f\"ur diese Variable ist
\begin{equation}
p(\chi) = \int_{-\infty}^{+\infty} \text{d}\omega_1 \int_{-\infty}^{+\infty} \text{d}\omega_2 \;
p(\omega_2,\omega_1,\Delta t) \; \delta(\chi - a \omega_1 - b \omega_2) \,,
\end{equation}
wobei $ p(\omega_2,\omega_1,\Delta t) $ die Wahrscheinlichkeitsdichte ist, die lokalen Frequenzen $ \omega_1 $ und $ \omega_2 $ im Zeitabstand $ \Delta t $ zu finden; mit $ \delta $ ist die Dirac-Distribution bezeichnet. Die Fourier-Transformierten von $ p(\chi) $ und $ p(\omega_2,\omega_1,\Delta t) $ gen\"ugen der Beziehung
\begin{equation}
\tilde{p}(\xi) = \tilde{p}(a\xi,b\xi,\Delta t) \;.
\end{equation}
Die Fourier-Transformierte einer Gau{\ss}funktion ist wieder eine solche, d. h. die Funktionen $ \tilde{p}(\xi) $ und $ \tilde{p}(a\xi, b\xi, \Delta t) $ sind Gau{\ss}funktionen. Da die Parameter $ a $ und $ b $ beliebig gew\"ahlt werden k\"onnen, ist $ \tilde{p}(\eta_2,\eta_1,\Delta t) $ eine Gau{\ss}funktion in den Variablen $ \eta_2 $ und $ \eta_1 $. Die inverse Fourier-Transformation impliziert, dass $ p(\omega_2,\omega_1, \Delta t) $ auch eine Gau{\ss}funktion ist.

Die Gau{\ss}sche Dephasierung impliziert also, dass die Wahrscheinlichkeitsdichte, die Frequenz $ \omega_1 $ zur Zeit $ t=0 $ und die Frequenz $ \omega_2 $ zum sp\"ateren Zeitpunkt $ t=\Delta t $ zu finden, eine Gau{\ss}funktion ist. 

Als n\"achstes wird die Greensche Funktion untersucht, d. h. die Dichte der bedingten Wahrscheinkichkeit
 $ G(\omega_2,\omega_1,\Delta t) $, dass sich ein Spin unter dem Einfluss eines lokalen Feldes mit der Frequenz $ \omega_2 $ nach dem Zeitintervall $ \Delta t $ befindet, wenn er urspr\"unglich unter Einfluss des lokalen Feldes mit der Frequenz $ \omega_1 $ stand. Solange die Gleichgewichtsbedingungen erf\"ullt sind, gilt die Relation
\begin{equation}
p(\omega_2,\omega_1,\Delta t) = G(\omega_2,\omega_1,\Delta t) \; p_0(\omega_1)\;,
\end{equation}
und da $ p_0 $ eine Gau{\ss}funktion ist, impliziert diese Relation, dass auch die bedingte Wahrscheinlichkeit eine Gau{\ss}funktion ist. Diese Greensche Funktion ist der Propagator der stochastischen Dynamik der Frequenzen und hat die Eigenschaft
\begin{equation} \label{delta}
G(\omega_2,\omega_1,0) = \delta(\omega_2-\omega_1)\;.
\end{equation}

Nun stellt sich die Frage, welche Form diese Greensche Funktion annimmt. Um diese Frage zu beantworten, werden einige fundamentale Eigenschaften der Greenschen Funktion genutzt. Eine Eigenschaft ist, dass die Wahrscheinlichkeit, einen Spin nach dem Zeitintervall $ \Delta t $ unter dem Einfluss irgendeiner lokalen Frequenz zu finden, auf Eins normiert ist, d. h.
\begin{equation} \label{normalization}
\int_{-\infty}^{+\infty} \text{d}\omega_2 \; G(\omega_2,\omega_1,\Delta t) = 1 \;.
\end{equation}
Wird die gau{\ss}f\"ormige Greensche Funktion
\begin{equation}
G(\omega_2,\omega_1,\Delta t)= N \; \exp \left[ -\frac{1}{2}\left(\frac{\omega_2}{\sigma(\Delta t)}\right)^2 - \frac{1}{2} \left(\frac{\omega_1}{\varsigma(\Delta t)}\right)^2 + \kappa (\Delta t) \; \omega_2 \; \omega_1 \right]
\end{equation}
mit den zeitabh\"angigen Funktionen $ \sigma $, $ \varsigma $, $\kappa $ und dem Normalisierungsfaktor $ N $ geschrieben, folgt aus Gleichung (\ref{normalization}) der Zusammenhang
\begin{equation}
\sigma\;\varsigma = \frac{1}{\kappa}\;.
\end{equation}
Mit der Beziehung $ \gamma = \kappa \sigma^2 $ nimmt der Propagator demzufolge die Form
\begin{equation} \label{Greensfunction}
G(\omega_2,\omega_1,\Delta t) = \frac{1}{\sqrt{2\pi\sigma^2(\Delta t)}} \; \exp \left[ -\frac{1}{2}\;\left(\frac{\omega_2 - \omega_1 \; \gamma(\Delta t)}{\sigma(\Delta t)}\right)^2 \right]
\end{equation}
an, die eine verschobene gau{\ss}f\"ormige Funktion in $ \omega_2 $ ist. Entsprechend Gleichung (\ref{delta}) und aufgrund der Tatsache, dass die Greensche Funktion f\"ur $ \Delta t \to \infty $ in die Gleichgewichtswahrscheinlichkeit $ G(\omega_2,\omega_1,\Delta t) \to p_0(\omega_2) $ \"ubergeht, ergibt sich
\begin{equation} \label{boundarygamma}
\gamma(0) = 1, \quad \sigma(0) = 0, \quad \gamma(\infty) = 0, \quad \text{und} \quad \sigma(\infty) = \sigma_0,
\end{equation}
wobei $ \sigma_0 $ die Varianz der Gleichgewichtsverteilung $ p_0(\omega) $ ist.

\begin{figure}
\begin{center}
\includegraphics[width=\textwidth]{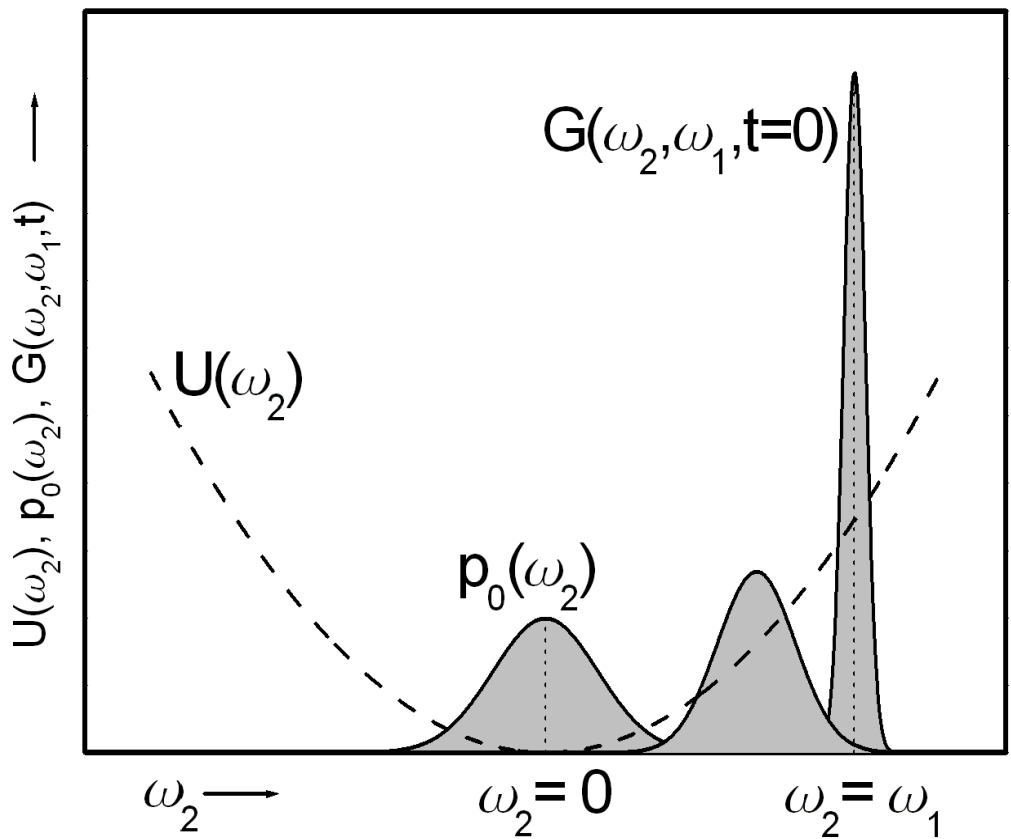}
\caption[Zeitentwicklung der Greenschen Funktion]{\label{Fig:potential}{\footnotesize Zeitentwicklung der Greenschen Funktion. Die Greensche Funktion nach Gleichung (\ref{Greensfunction2}) entwickelt sich im harmonischen Potential (\ref{harmpot}) (gestrichelte Linie) vom Anfangspunkt $\omega_2=\omega_1$, in dem die Greensche Funktion durch $G(\omega_2,\omega_1,t=0)=\delta(\omega_2-\omega_1)$ gegeben ist, zum Gleichgewichtszustand, in dem die Greensche Funktion durch die gau{\ss}f\"ormige Gleichgewichtsverteilung (\ref{Gauss-equi}) $G(\omega_2,\omega_1,t\to\infty)=p_0(\omega_2)$ gegeben ist.}}
\end{center}
\end{figure}

Eine weitere fundamentale Eigenschaft ergibt sich, wenn angenommen wird, dass die stochastische Dynamik der lokalen Feldfluktuationen ein Markovprozess ist, d. h., dass der stochastische \"Ubergang zwischen zwei Frequenzen unabh\"angig vom vorhergehenden \"Ubergang ist. Damit k\"onnen f\"ur die Zeitintervalle $ \Delta t_1 $ und $ \Delta t_2 $ folgende Relation gefunden werden:
\begin{equation} \label{solutionGreen}
G(\omega_3,\omega_1,\Delta t_1 + \Delta t_2) = \int_{-\infty}^{+\infty} \text{d} \omega_2 \; G(\omega_3,\omega_2,\Delta t_2) \; G(\omega_2,\omega_1,\Delta t_1) \;.
\end{equation}
Das Einsetzen der Greenschen Funktion (\ref{Greensfunction}) ergibt:
\begin{equation} \label{gamma}
\gamma(\Delta t_1 + \Delta t_2) = \gamma (\Delta t_2) \; \gamma(\Delta t_1)
\end{equation}
und
\begin{equation} \label{sigma}
\sigma^2(\Delta t_1 + \Delta t_2) = \gamma^2(\Delta t_2) \; \sigma^2(\Delta t_1) + \sigma^2(\Delta t_2) \;.
\end{equation}
Die Gleichungen (\ref{gamma}) und (\ref{sigma}) werden unter Beachtung der Randbedingungen (\ref{boundarygamma}) gel\"ost. Zuerst ergibt sich
\begin{equation}
\gamma(t) = \text{e}^{-rt}\;,
\end{equation}
wobei $ r $ f\"ur die Zerfallsrate steht. Einsetzen in Gleichung (\ref{sigma}) ergibt
\begin{equation}
\sigma^2(t) = \sigma^2_0 \; \left(1-\text{e}^{-2rt} \right) \;.
\end{equation}
Deshalb gilt f\"ur beliebige Zeiten $ t $:
\begin{equation} \label{Greensfunction2}
G(\omega_2,\omega_1,t) = \frac{1}{\sqrt{2 \pi \sigma_0^2 (1 - \text{e}^{-2rt})}} \;
\exp \left[ -\frac{(\omega_2 - \omega_1 \text{e}^{-rt})^2}{2 \sigma_0^2 (1 - \text{e}^{-2rt})} \right] \;.
\end{equation}
Dieser Ausdruck zeigt, dass die Greensche Funktion eine Gau{\ss}funktion ist, die zum Zeitpunkt $t=0$ mit verschwindender Varianz um die Anfangsfrequenz $ \omega_1 $ zentriert ist. Mit zunehmender Zeit erreicht diese Gau{\ss}funktion den Gleichgewichtszustand, d. h. die Varianz stimmt mit der des Gleichgewichtszustandes \"uberein und das Zentrum verlagert sich von $ \omega_1 $ zum Nullpunkt (Abbildung \ref{Fig:potential}).
Die Dynamik dieses Prozesses wird durch die Rate $ r $ bestimmt. Dieser Typ der Greenschen Funktion ist als L\"osung der Diffusionsgleichung 
\begin{equation}
\partial_t G(\omega_2,\omega_1,t) = D \; \partial_{\omega_2} \left[ \partial_{\omega_2} + \sigma_0^{-2} \omega_2 \right] \; G(\omega_2,\omega_1,t)
\end{equation}
mit dem Diffusionskoeffizienten $ D $ bekannt. Dies ist eine Fokker-Planck-Gleichung, welche die Diffusion in einem harmonischen Potential
\begin{equation} \label{harmpot}
U(\omega) = \frac{1}{2} \; \frac{\omega^2}{\sigma_0^2}
\end{equation}
beschreibt (siehe Abbildung \ref{Fig:potential}). Die dazugeh\"orige treibende Kraft ist
\begin{equation}
F(\omega) = -\partial_{\omega}U = -\sigma_0^{-2} \omega \;.
\end{equation}
Der Zusammenhang zwischen der Rate $ r $ und dem Diffusionskoeffizienten
\begin{equation}
r = \frac{D}{\sigma_0^2} \;,
\end{equation}
zeigt, dass ein gr\"o{\ss}erer Diffusionskoeffizient und ein steileres Potential (kleines $ \sigma_0^2 $) zu einer schnelleren Einstellung des Gleichgewichts f\"uhren. Die Diffusion findet im Frequenzraum statt und die vom harmonischen Potential abgeleiteten Kr\"afte sind Entropiekr\"afte, die durch die Gleichgewichtswahrscheinlichkeitsdichte und die \"Ubergangsdynamik zwischen den Frequenzzust\"anden festgelegt sind.

Die Zweipunktkorrelationsfunktion der lokalen Frequenzen dieses stochastischen Prozesses ist durch
\begin{align}
\nonumber c_2(t) & = \int_{-\infty}^{+\infty} \text{d} \omega_2 \int_{-\infty}^{+\infty} \text{d} \omega_1 \; \omega_2 \; G(\omega_2,\omega_1,t) \; p_0(\omega_1) \; \omega_1
\\[-1ex] & \\[-1ex]
\nonumber & = \langle \, \omega^2 \, \rangle \, \text{e}^{-rt} \;,
\end{align}
gegeben (siehe Gleichung (\ref{solutionGreen})), wobei $ \langle \, \omega^2 \, \rangle $ die Varianz der lokalen Frequenz ist. Diese Gleichung zeigt, dass die Zweipunktkorrelationsfunktion exponentiell mit der Korrelationszeit
\begin{equation}
\tau = \frac{1}{r}
\end{equation}
abf\"allt. Daher ergibt sich der Magnetisierungs-Zeit-Verlauf nach Gleichung (\ref{gaussianrelaxation}) zu
\begin{equation} \label{decay}
M(t) = \exp \left(- \frac{t}{T_{2,0}} \right) \; \exp \left[ \frac{\tau}{T_{2,0}} \left( 1 - \text{e}^{-t/\tau} \right) \right] \;,
\end{equation}
wobei die Relaxationszeit im Motional-Narrowing-Regime ($ \tau \to 0 $)
\begin{equation}
\frac{1}{T_{2,0}} = \tau \langle \, \omega^2 \, \rangle
\end{equation}
eingesetzt wurde. Dies zeigt, dass f\"ur gro{\ss}e Zeiten der Signalzerfall rein exponentiell ist und durch die Relaxationszeit des Motional-Narrowing-Regimes $ T_{2,0} $ charakterisiert ist, wohingegen f\"ur kleine Zeiten $ \tau \gg t $ der Signalzerfall gau{\ss}f\"ormig $\sim \exp[-t^2/(2 \tau T_{2,0})]$ ist. Das stimmt auch mit Gleichung (2) der Arbeit von Sukstanskii und Yablonskiy \cite{Sukstanskii04} \"uberein.

\section{\label{Sec:Dis}Zusammenfassung der Implikationen}
Entscheidend ist, dass die Gleichung
\begin{equation}
M(t) = \exp \left[ - \int_{0}^{t} (t - \xi) c_2(\xi) \text{d} \xi \right] \,,
\end{equation}
die den Magnetisierungszerfall beschreibt, nur dann anwendbar ist, wenn die Greensche Funktion $ G(\omega_2,\omega_1,\Delta t) $, welche die Wahrscheinlichkeit des \"Uberganges zwischen den beiden Frequenzen $ \omega_1 \to \omega_2 $ charakterisiert, gau{\ss}f\"ormig ist. Diese Gau{\ss}sche N\"aherung schr\"ankt die Vielfalt der m\"oglichen Zerfallskurven betr\"achtlich ein.

Unter der Annahme einer Gau{\ss}schen Verteilung $ P ( \varphi ) $ des Phasenwinkels $ \varphi $ konnte in Abschnitt \ref{Sec:Transition} gezeigt werden, dass dies unter Gleichgewichtsbedingungen zu einer gau{\ss}f\"ormigen bedingten Wahrscheinlichkeitsdichte $ G(\omega_2,\omega_1,\Delta t) $ f\"uhrt:
\begin{equation} \label{R1}
P(\varphi) \;\; \text{gau{\ss}f\"ormig} \;\; \Longrightarrow \;\; G(\omega_2,\omega_1,\Delta t) \;\; \text{gau{\ss}f\"ormig} \,.
\end{equation}

Umgekehrt kann eine gau{\ss}f\"ormige Dichte $ G(\omega_n,\omega_{n-1},\Delta t) $ daf\"ur angenommen werden, dass ein Spin unter dem Einfluss des lokalen Feldes mit der Frequenz $ \omega_n $ nach der Zeit $ \Delta t $ steht, wenn er urspr\"unglich unter dem Einfluss des lokalen Feldes mit der Frequenz $ \omega_{n-1} $ stand. Mit dieser Annahme kann die Verteilung $ P ( \varphi ) $ des Phasenwinkels $ \varphi $ in folgender Weise berechnet werden: w\"ahrend eines kleinen Zeitintervalls $ \Delta t $ ergibt sich der Phasenwinkel zu $ \varphi = \sum_{k=0}^n \omega(k \, \Delta t) $. Demzufolge ist die Wahrscheinlichkeitsdichte eines Phasenwinkels durch die Beziehung
\begin{equation} \label{convolution}
P(\varphi) = \int\limits_{-\infty}^{+\infty} \text{d}\omega_n \cdot \cdot \cdot \int\limits_{-\infty}^{+\infty} \text{d}\omega_0 \; G(\omega_n, \omega_{n-1}, \Delta t) \cdot \cdot \cdot G(\omega_1, \omega_0, \Delta t) \; p_0(\omega_0) \; \delta \left( \varphi - \Delta t \sum_{k=0}^n \omega_k \right)
\end{equation}
gegeben. Dieser Ausdruck ist eine verkettete Faltung von Gau{\ss}funktionen, woraus folgt, dass $ P ( \varphi ) $ selbst eine Gau{\ss}funktion ist. Daher folgt die Implikation
\begin{equation} \label{R2}
G(\omega_2,\omega_1,\Delta t) \;\; \text{gau{\ss}f\"ormig} \;\; \Longrightarrow \;\; P(\varphi) \;\; \text{gau{\ss}f\"ormig} \,.
\end{equation}

Mit den Implikationen (\ref{R1}) und (\ref{R2}) konnte die \"Aquivalenz zwischen einer Gau{\ss}verteilung des Phasenwinkels $ P ( \varphi ) $ und einer gau{\ss}f\"ormigen Greenschen Funktion $ G(\omega_2,\omega_1,\Delta t) $ demonstriert werden.

Die einzige Annahme, die im Modell von Anderson und Weiss gemacht wurde, war eine gau{\ss}f\"ormige Gleichgewichtsverteilung der Frequenzen $ p_0( \omega ) $. Aus Gleichung (\ref{convolution}) kann man sehen, dass diese Annahme nicht ausreichend ist. Das Anderson-Weiss-Modell ist f\"ur einen speziellen Fall von Fluktuationen anwendbar, und zwar dann und nur dann, wenn die Greensche Funktion $ G(\omega_2,\omega_1,\Delta t) $, welche die Wahrscheinlichkeit des \"Uberganges zwischen zwei Frequenzen beschreibt, gau{\ss}f\"ormig ist.

Wird der Gleichgewichtszustand durch die Greenschen Funktion $ p_0( \omega ) = G( \omega , 0 , t \to \infty) $ beschrieben, bedeutet dies, dass eine gau{\ss}f\"ormige Greensche Funktion eine gau{\ss}f\"ormige Gleichgewichtsverteilung der Frequenzen $ p_0( \omega ) $ impliziert. Damit kann die Implikation (\ref{R2}) erweitert werden:
\begin{equation} \label{R3}
G(\omega_2,\omega_1,\Delta t) \;\; \text{gau{\ss}f\"ormig} \;\; \Longrightarrow \;\; p_0(\omega) \;\; \text{gau{\ss}f\"ormig} \,.
\end{equation}
Die Umkehrung dieser Implikation ist jedoch im Allgemeinen falsch. Es gibt stochastische Prozesse wie den der
 Strong-Collision und seine Erweiterung (ESC), die eine gau{\ss}f\"ormige Gleichgewichtsverteilung $ p_0(\omega) $ besitzen, deren Dynamik jedoch durch eine nicht-gau{\ss}f\"ormige Greensche Funktion $ G(\omega_2,\omega_1,\Delta t) $ beschrieben wird \cite{Bauer02}. Zusammengefasst ergeben sich nun die folgenden Implikationen:
\begin{equation} \label{R4}
\nonumber
p_0( \omega ) \; \text{gau{\ss}f\"ormig} \; \Longleftarrow \; G(\omega_2,\omega_1,\Delta t) \; \text{gau{\ss}f\"ormig} \; \Longleftrightarrow \; P(\varphi) \; \text{gau{\ss}f\"ormig} \; \Longrightarrow \; M(t) = \text{e}^{- \frac{1}{2}\langle \varphi^2(t) \rangle} \,,
\end{equation}
die sowohl den Zusammenhang zwischen der Greenschen Funktion und der Gleichgewichtsverteilung der Frequenzen als auch den Zusammenhang zwischen der Verteilung der Phasenwinkel und dem Magnetisierungszerfall zeigen.

Es wurde also gezeigt, dass die Annahme der Gau{\ss}schen Dephasierung des Phasenwinkels die stochastische Dynamik zwischen den lokalen Frequenzen auf eine sehr begrenzte Klasse von \"Ubergangsdynamiken beschr\"ankt, die durch Gleichung (\ref{Greensfunction2}) beschrieben werden. Die \"Ubergangsdynamiken sind die eines Teilchens, dass in dem harmonischen Potential diffundiert. Der Magnetisierungszerfall folgt aus Gleichung (\ref{decay}). Obwohl Gleichung (\ref{gaussianrelaxation}) eine Vielzahl m\"oglicher Zerfallskurven suggeriert, sind nur die in der Form von Gleichung (\ref{decay}) mit der Gau{\ss}schen N\"aherung vereinbar.

\chapter{\label{Kap:Relaxation}Relaxationszeiten magnetisch markierter Zellen}
\pagestyle{headings}

\section{Bedeutung magnetisch markierter Zellen}
Viele Krankheiten beruhen auf dem irreversiblen Absterben von gewebespezifischen Zellen. So kann z. B. nach einem Myokardinfarkt isch\"amisch gesch\"adigtes Gewebe nicht mehr regenerieren, es kann eine Herzmuskelinsuffizienz entstehen. Zur Regeneration des abgestorbenen Gewebes k\"onnen noch nicht differenzierte Stammzellen in das betroffene Gewebe eingebracht werden, um sich in gewebespezifische Zellen weiter zu differenzieren. Um diese Zellen lokalisieren zu k\"onnen, werden sie vorher mit kleinen magnetischen Kontrastmittelteilchen (USPIOs) markiert, um eine f\"ur die Kernspinresonanzbildgebung nutzbare Suszeptibilit\"atsdifferenz zum umgebenden Gewebe zu erzeugen. Diese Kontrastmittelteilchen k\"onnen nun als kleine magnetisierte Objekte betrachtet werden, die aufgrund der Suszeptibilit\"atsdifferenz ein lokales inhomogenes Magnetfeld erzeugen. W\"ahrend bisher die Signalentstehung in einem Voxel untersucht wurde, wird nun die transversale Relaxationszeit $T_2^*$ untersucht, die durch die Anwesenheit von magnetischen Kontrastmittelteichen beeinflusst wird.

\section{Relaxationsverhalten}
Als physikalisches Modell f\"ur magnetisch markierte Zellen wird angenommen, dass eine unbekannte Anzahl von Nanopartikeln mit bekanntem Radius von der Zelle phagozytiert wurden und sich zu einem Kern innerhalb der Zelle zusammenlagern. Um dieses Problem mathematisch zu erfassen, wurde vorerst angenommen, dass s\"amtliche magnetische Kerne den gleichen Radius haben und der Volumenanteil magnetischen Materials am gesamten untersuchten Gewebe proportional zur Konzentration der Nanopartikel sei. Diese geometrische Anordnung mitsamt der um die magnetischen Kerne stattfindenden Diffusion bestimmt das Signalverhalten in einem Bildgebungsexperiment.

Um die transversale Relaxationszeit $T_2^*$ zu bestimmen wird angenommen, dass die Zeitentwicklung der Magnetisierung in der Form
\begin{equation}
M(t) \, = \, \text{exp}(- \, t/T_2^* \, + \, \text{i} \Omega t)
\end{equation}
darstellbar ist, wobei $\Omega t$ ein unbekannter Phasenfaktor ist. Wird dieser Zeitverlauf in die Definition der Laplace-Transformation (\ref{LT}) eingesetzt, ergibt sich
\begin{equation}
\hat{M}(s) \, = \, 1/(R_2^* - \text{i} \, \Omega + s) \,.
\end{equation}
Daraus folgt, dass die Relaxationsrate entsprechend der Gleichung
\begin{equation}
R_2^* \, = \, \frac{1}{T_2^*} \, = \, \text{Re} \left[ \frac{1}{\hat{M}(0)} \right]
\end{equation}
bestimmt werden kann. Um den Wert $\hat{M}(0)$ zu erhalten, wird der bekannte Zusammenhang
\begin{align} \label{Md}
\hat{M} (0) = \, \frac{\hat{M}_0(\tau^{-1})}{1 - \tau^{-1} \cdot \hat{M}_0(\tau^{-1})}
\end{align}
(siehe Gleichung (\ref{Mdach}) mit $s=0$) genutzt, wobei $\hat{M}_{0}(s)$ die Laplace-Transformierte des Zeitverlaufs der Magnetisierung im Static-Dephasing-Regime ist. Diese Laplace-Transformierte wurde schon in Gleichung (\ref{allgmd}) in Kapitel \ref{Kap:Frequenz} untersucht, wobei in diesem Fall die f\"ur sph\"arische K\"orper spezifische $G$-Funktion aus Gleichung (\ref{ggs}) relevant ist. Wird dieses Ergebnis in Gleichung (\ref{Md}) eingesetzt, ergibt sich letztlich f\"ur die Relaxationsrate der Ausdruck
\begin{equation} \label{MRT}
R_2^* \, = \, \frac{1}{T_2^*} \, =\frac{1}{\tau} \,
\text{Re} \left[ \frac{1 - \eta}{G_{\text{S}}(\frac{1}{\eta \, \tau \, \delta\omega})-\eta \, G_{\text{S}}(\frac{1}{\tau \, \delta\omega})} - 1 \right] \,.
\end{equation}
Diese Methode zur Bestimmung der Relaxationszeit $ T_2^* $ ist der schon fr\"uher verwendeten Mean-Relaxation-Time-Approximation \cite{Nadler85} analog, in welcher die Relaxationszeit das erste Langzeitmoment des Magnetisierungszerfalls ist: $T_2^* = \mu_{-1}[M(t)]$. Um die Relaxationszeit zu bestimmen, wurde die Laplace-Transformierte f\"ur den Grenzfall kleiner $s$ genutzt. Dies reflektiert jedoch das Signalverhalten f\"ur lange Zeiten, obwohl die Messwerte f\"ur $T_2^*$ \"ublicherweise aus Messungen mit kurzen Echozeiten ermittelt werden. Deshalb w\"are die Definition $R_2^* = - \text{ln}[M(t)/t]$ sicherlich sinnvoller. Eine genaue Beschreibung des Magnetisierungszerfalls f\"ur kurze Echozeiten erfolgt durch die Extended-Strong-Collision-Approximation \cite{Bauer02}.

Die Korrelationszeit um kugelf\"ormige Objekte wurde schon in Abschnitt \ref{Subsec:Spheres} bestimmt. F\"ur die weiteren Untersuchungen wird der Zusammenhang (\ref{tau_mot_narr}) genutzt, um die Korrelationszeit zu berechnen.

Die Relaxationsrate magnetisch markierter Zellen im Motional-Narrowing-Regime wurde bereits in Abschnitt \ref{Subsec:Relaxationsraten} untersucht. Die Relaxationsrate $R_2^* = 1 / T_2^*$ ist in diesem Fall vom Radius der Teilchen $R_{\text{S}}$ und vom Diffusionskoeffizienten des umgebenden Mediums abh\"angig (siehe Abschnitt \ref{Subsec:Relaxationsraten}):
\begin{equation} \label{R2stern_mot_narr}
R_2^* \, = \, \frac{16}{45} \, \eta \, \delta\omega^2 \, \frac{R_{\text{S}}^2}{D} \,.
\end{equation}

\begin{figure}
\begin{center}
\includegraphics[width=14cm]{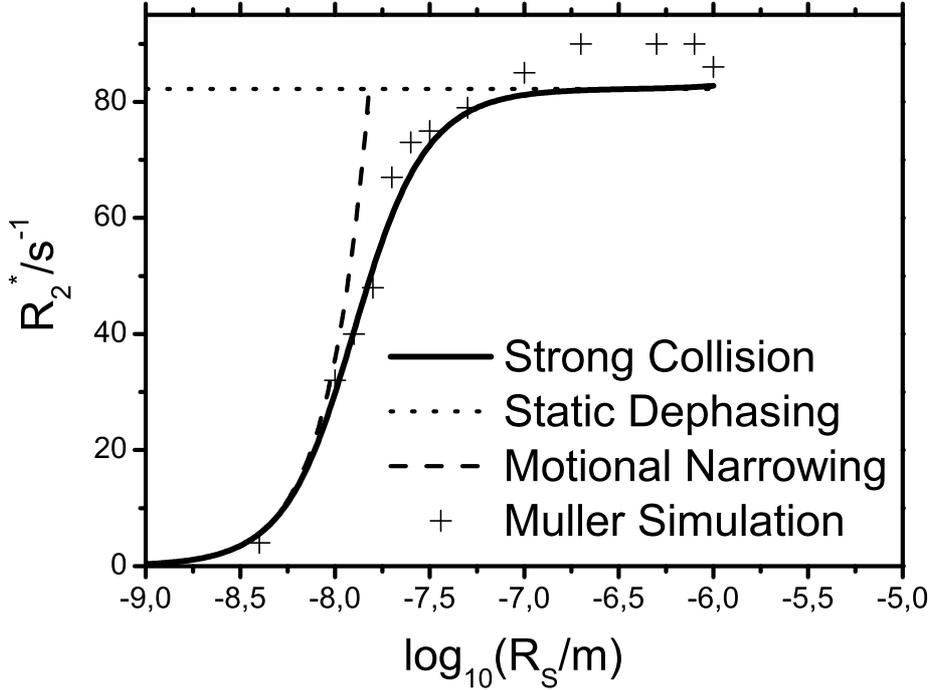}
\caption[Relaxationsraten f\"ur verschiedene Kugelradien]{\label{Fig:r2sternrad}{\footnotesize Relaxationsraten f\"ur verschiedene Kugelradien. Die Relaxationsrate $ R_2^* $ ist in Abh\"angigkeit vom Kugelradius $ R_{\text{S}} $ dargestellt. Die durchgezogene Linie wurde aus Gleichung (\ref{MRT}) mit den Parametern $ \eta \, = \, 2 \cdot 10^{-6} $, $ \delta\omega \, = \, 34 \cdot 10^6 \,\text{Hz} $ und $ D = 2,3 \,\mu \text{m}^2 \,\text{ms}^{-1} $ erhalten. Die Kreuze sind die Werte f\"ur $ 1/T_2^* $ in Abh\"angigkeit vom Kugelradius nach den Simulationen von Muller \cite{Muller91}. Wenn die Bedingung $ \text{log}_{10}(R_{\text{S}}/\text{m}) \gg -6,746 $ nach Ungleichung (\ref{Haacke}) erf\"ullt ist, stimmen die Ergebnisse des Static-Dephasing-Regimes (Gleichung (\ref{r2sternSD})) mit den Ergebnissen des Strong-Collision-Modells \"uberein. F\"ur kleine Radien stimmen die Werte gut mit den Vorhersagen des Motional-Narrowing-Grenzfalls nach Gleichung (\ref{R2stern_mot_narr}) \"uberein.}}
\end{center}
\end{figure}

Die Relaxationszeiten im Static-Dephasing-Regime wurden bereits ausf\"uhrlich von Yablonskiy und Haacke untersucht \cite{Yablonskiy94}. Sie geben f\"ur die Relaxationsrate im Static-Dephasing-Regime die Beziehung
\begin{equation} \label{r2sternSD}
R_2^* \, = \, \frac{2 \pi}{3 \sqrt{3}} \, \eta \, \delta\omega
\end{equation}
an, w\"ahrend Jensen und Chandra \cite{Jensen99} das Ergebnis
\begin{equation} \label{r2sternJ}
R_2^* \, = \, 3 \, k \, \eta \, \delta\omega
\end{equation}
erhalten, wobei f\"ur kompakte Objekte $k \approx 0,4031 \approx 2\pi/(9\sqrt{3})$ gilt. Sowohl die Gleichung (\ref{r2sternSD}) als auch die Gleichung (\ref{r2sternJ}) sind vom Kugelradius unabh\"angig und stimmen im Grenzfall gro{\ss}er Radien (d. h. $ \tau \, \delta\omega \gg 1 $ ) mit den Ergebnissen, die aus der Strong-Collision-N\"aherung erhalten werden, gut \"uberein (siehe Abbildung \ref{Fig:r2sternrad} f\"ur gro{\ss}e Radien).

Wie in Abschnitt \ref{Subsec:Diffusionsregime} beschrieben, wird das zugrunde liegende Diffusionsregime durch die Werte von $ \tau $ und $ \delta \omega $ festgelegt. Da die Oberfl\"achenfrequenz $\delta\omega = \gamma \mu_0 \Delta M /3 $ vom Kugelradius unabh\"angig ist, wird das Diffusionsregime alleine durch die Korrelationszeit $ \tau \, \propto \, R_{\text{S}}^2/D $ festgelegt. Deshalb ist f\"ur kleine Kugeln der Motional-Narrowing-Grenzfall ($ 1/\tau \gg \delta\omega $) anzuwenden und f\"ur gro{\ss}e Kugeln der Static-Dephasing-Grenzfall ($ 1/\tau \ll \delta\omega $). Abbildung \ref{Fig:r2sternrad} zeigt, dass die Resultate des Strong-Collision-Modells nach Gleichung (\ref{MRT}) gut mit den numerischen Simulationen von Muller et al. \"ubereinstimmen \cite{Muller91}. Auch die Werte f\"ur die Grenzf\"alle Static-Dephasing-Regime und Motional-Narrowing-Regime stimmen mit den jeweiligen N\"aherungsformeln \"uberein.
Die Abh\"angigkeit der Relaxationsrate vom Kugelradius l\"asst sich sehr gut an Abbildung \ref{Fig:r2sternrad} veranschaulichen. Im Motional-Narrowing-Regime (kleine Radien) ist die Relaxationsrate vom Radius abh\"angig (siehe Gleichung (\ref{R2stern_mot_narr})). Wenn die Bedingungen des Static-Dephasing-Regimes erf\"ullt sind, ist die Relaxationsrate vom Kugelradius unabh\"angig (siehe Gleichung (\ref{r2sternSD})). Dies ist nach Ungleichung (\ref{Ziener}) der Fall, wenn die Bedingung
\begin{equation}
R_{\text{S}}^2 \gg \frac{D}{\delta\omega} \frac{1}{\left( \frac{4}{9} - \frac{3}{8}\sqrt[3]{\eta} \right)}
\end{equation}
erf\"ullt ist.

\begin{figure}
\begin{center}
\includegraphics[width=12cm]{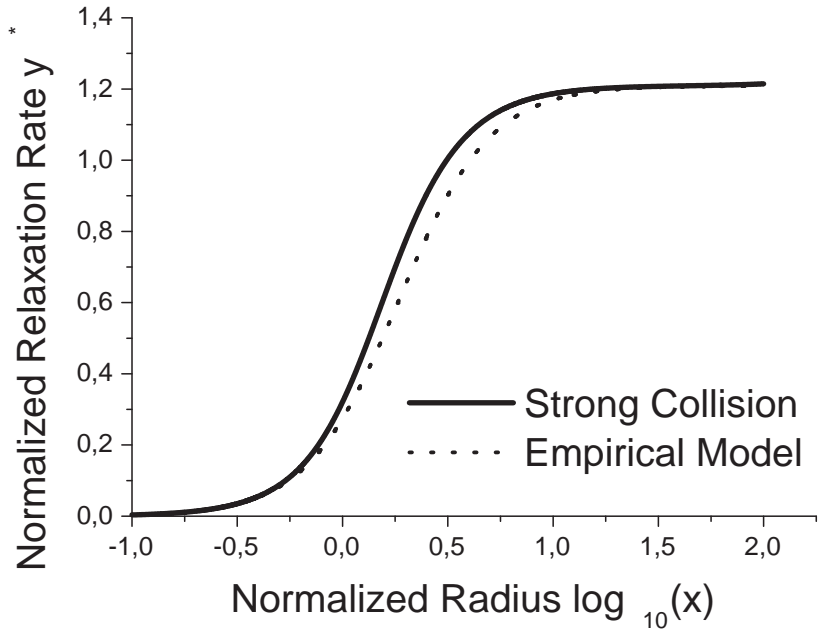}
\caption[Normierte Relaxationsrate]{\label{Fig:r2sternnorm}{\footnotesize Normierte Relaxationsrate. Die Abh\"angigkeit der normierten Relaxationsrate $ R_2^* $ vom normierten Kugelradius $ R_{\text{S}} $ ist dargestellt. Die durchgezogene Linie wurde nach Gleichung (\ref{MRT}) erhalten, die gepunktete Linie entspricht dem Interpolationsmodell (\ref{r2emp}) mit den Parametern $ c = \frac{16}{45} $ und
$ d = \frac{2 \pi}{3 \sqrt{3}} $.}}
\end{center}
\end{figure}

Um zu zeigen, dass die Ergebnisse des Strong-Collision-Modells \"uber den gesamten Dynamikbereich g\"ultig sind, werden diese Ergebnisse mit denen von Jensen und Chandra \cite{Jensen99} sowie mit denen von Yung \cite{Yung03} verglichen. Jensen und Chandra interpolierten den gesamten dynamischen Bereich durch einfache Addition der Relaxationszeiten des Static-Dephasing-Regimes (Gleichung (\ref{r2sternJ})) und des Motional-Narrowing-Regimes (Gleichung (\ref{r2sternMN2})), und erhielten
\begin{equation} \label{inter}
T_2^* \, = \, \frac{5}{4\,\eta\,\tau\,\delta\omega^2} \, + \,
\frac{1}{3\,k\,\eta\,\delta\omega} \,.
\end{equation}
Das gleiche Ergebnis wurde von Yung \cite{Yung03} gefunden, indem er eine empirische Interpolationsformel f\"ur die skalierte Relaxationsrate $ y^*=1/(\eta \,\delta\omega \, T_2^*) $ in Abh\"angigkeit vom skalierten Radius $x=\sqrt{\delta\omega/D}\,R_{\text{S}}$ fand:
\begin{equation} \label{r2emp}
y^{*} \, := \, \frac{c \, x^2}{1 \, + \, \frac{c}{d} \, x^2} \,,
\end{equation}
mit $c=16/45$ und $d=3k$. In Abbildung \ref{Fig:r2sternnorm} werden die Vorhersagen des Strong-Collision-Modells (Normierung des Plots in Abbildung \ref{Fig:r2sternrad}) mit Gleichung (\ref{r2emp}) verglichen.
Eine gute \"Ubereinstimmung der Ergebnisse des Strong-Collision-Modells und des empirischen Modells f\"ur alle Werte des Kugelradius $ R_{\text{S}} $ ist zu erkennen.

\begin{figure}
\begin{center}
\includegraphics[width=12cm]{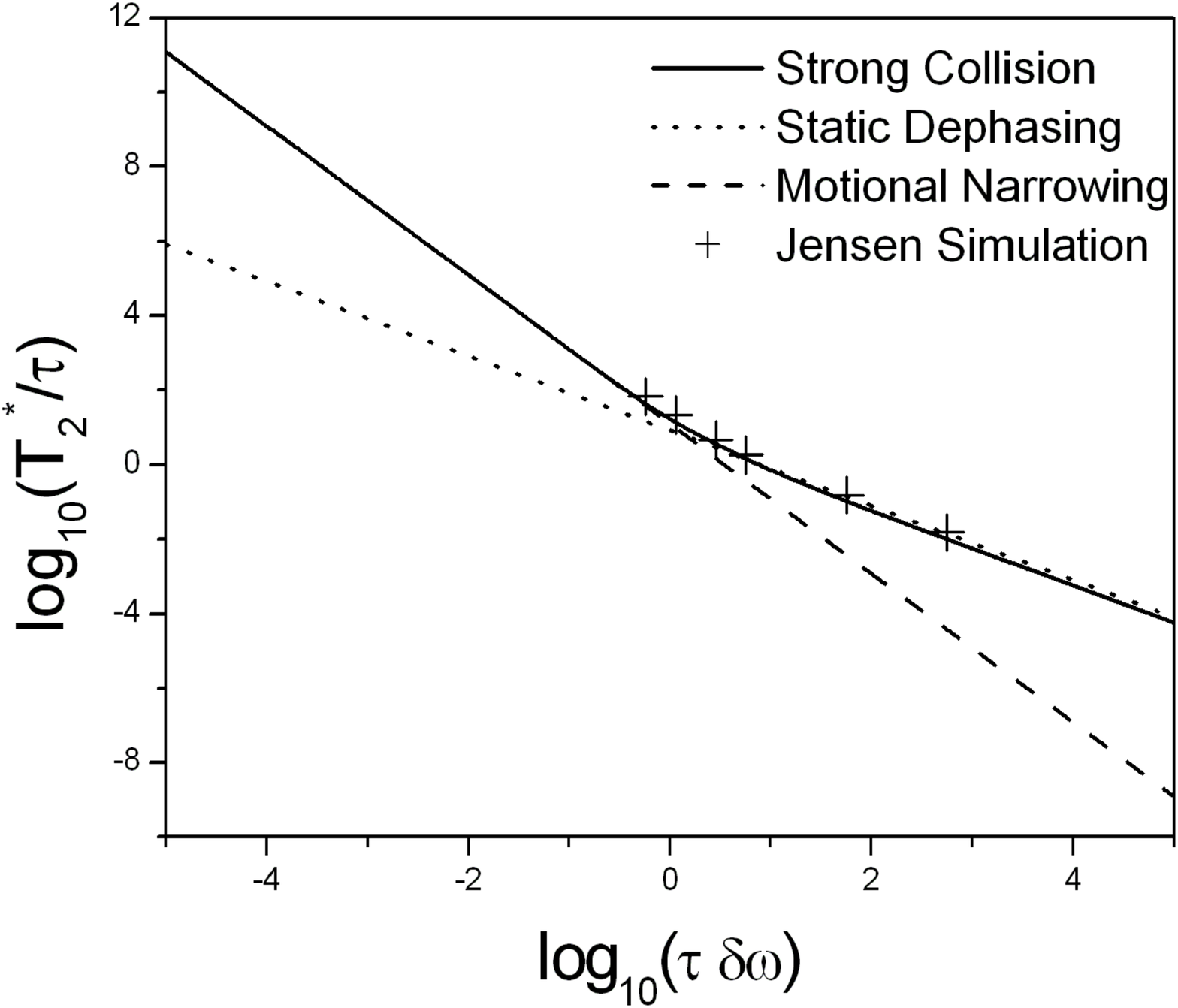}
\caption[Vergleich der Strong-Collision-Approximation mit Simulationen]{\label{Fig:r2sterndom}{\footnotesize Vergleich der Strong-Collision-Approximation mit Simulationen. Die Ergebnisse der Strong-Collision-Approximation nach Gleichung (\ref{MRT}) wurden mit den Simulationen von Jensen und Chandra f\"ur einen Volumenanteil $\eta = 0,1$. verglichen. Die Asymptoten entsprechen den Grenzf\"allen nach Gleichung (\ref{R2stern_mot_narr}) f\"ur das Motional-Narrowing-Regime und nach Gleichung (\ref{r2sternJ}) f\"ur das Static-Dephasing-Regime. Aus der Bedingung (\ref{Haacke}) ergibt sich der Wert $ \log_{10}(\tau \, \delta\omega) \gg 0,76 $ f\"ur die G\"ultigkeit des Static-Dephasing-Regimes.}}
\end{center}
\end{figure}

Um die Abh\"angigkeit der Relaxationsrate $R_2^*$ vom Frequenzshift $\delta\omega$ zu untersuchen, wurden in Abbildung \ref{Fig:r2sterndom} die Ergebnisse des Strong-Collision-Modells nach Gleichung (\ref{MRT}) mit der Interpolationsformel (\ref{inter}) und mit Monte-Carlo-Simulationen von Jensen und Chandra (siehe Abbildung 1 in \cite{Jensen99}) verglichen. Zwischen den Ergebnissen der Strong-Collision-N\"aherung, den Simulationen von Jensen und Chandra und der Interpolationsformel ist eine gute \"Ubereinstimmung zu erkennen. Zus\"atzlich kann der \"Ubergang zwischen den Diffusionsregimen identifiziert werden und der Wendepunkt (siehe Abbildung \ref{Fig:r2sterndom}) stimmt mit der Vorhersage von Yablonskiy und Haacke nach Ungleichung (\ref{Haacke}) \"uberein.

Die entscheidenden Parameter, welche die Relaxationsrate bestimmen, sind die Konzentration der USPIOs, der Radius des magnetischen Kernes, die Magnetisierungsdifferenz zwischen magnetischem Kern und umgebendem Gewebe sowie der Diffusionskoeffizient. Zwei Frequenzskalen bestimmen den urs\"achlichen Relaxationsmechanismus. Die dynamische Frequenzskala $ 1/\tau $ charakterisiert den stochastischen Prozess der Diffusion durch die Korrelationszeit. Die magnetische Frequenzskala, die durch die \"aquatoriale Frequenz $\delta\omega$ beschrieben wird, charakterisiert die St\"arke der Magnetfeldinhomogenit\"at aufgrund der Suszeptibilit\"atseffekte. Vergleicht man beide Frequenzen, k\"onnen wie in Abschnitt \ref{Subsec:Diffusionsregime} gezeigt, verschiedene Diffusionsregime unterschieden werden. Analytische Ausdr\"ucke f\"ur die Relaxationszeit $ T_2^* $ existierten bisher f\"ur jedes Diffusionsregime getrennt. Die Anwendbarkeit eines solchen Ausdrucks setzt die Kenntnis des zugrunde liegenden Diffusionsregimes voraus. Sind die Gewebeeigenschaften (d. h. der charakteristische Frequenzshift $ \delta\omega $ und die dynamische Frequenz $ 1 / \tau $) bekannt, kann entschieden werden, welche N\"aherung angewandt werden soll. Werden jedoch magnetisch markierte Zellen betrachtet, ist die Ausgangssituation genau umgekehrt. Die Messung der Relaxationszeit $ T_2^* $ soll ja erst die Information \"uber das zu untersuchende Gewebe liefern, d. h. man wei{\ss} nicht, in welchem Diffusionsregime man sich befindet und folglich ist auch nicht bekannt, welche N\"aherung benutzt werden soll, um die Konzentration der Zellen zu berechnen. Deshalb wurde eine quantitative Beschreibung der Relaxationszeit abgeleitet, die \"uber den gesamten Frequenzbereich g\"ultig ist und deshalb auch die Grenzf\"alle (Motional-Narrowing und Static-Dephasing) enth\"alt.

\begin{figure}
\begin{center}
\includegraphics[width=\textwidth]{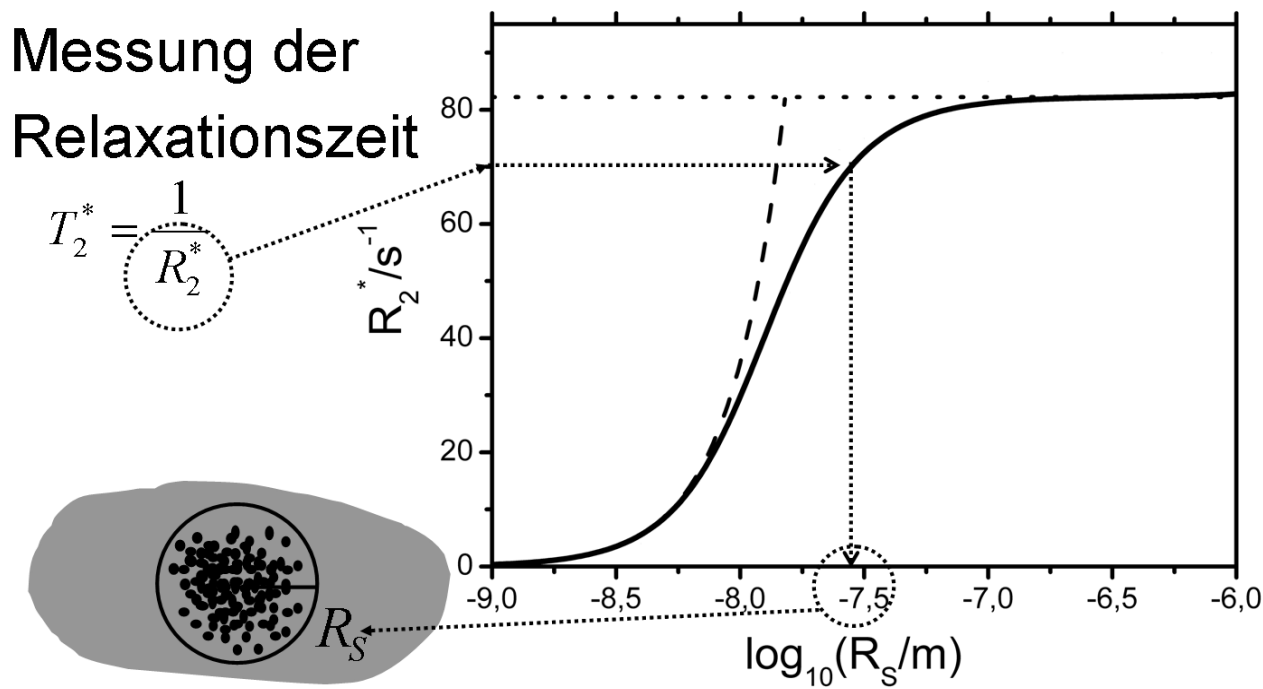}
\caption[Radiusbestimmung]{\label{Fig:Radiusbestimmung}{\footnotesize Radiusbestimmung. F\"ur die gleichen Parameter wie in Abbildung \ref{Fig:r2sternrad} wurde die Relaxationsrate in Abh\"angigkeit vom Kugelradius dargestellt. Die gemessene Relaxationsrate von $70 \, \text{s}^{-1}$ ergibt einen Kugelradius von $R_{\text{S}} \approx 10^{-7,5} \, \text{m} \approx 32 \, \text{nm}$.}}
\end{center}
\end{figure}

Mit den erhaltenen Ergebnissen k\"onnen jetzt aus der gemessenen Relaxationszeit $T_2^* = 1 / R_2^*$ Informationen \"uber die untersuchten Zellen gewonnen werden. Mit den bekannten Parametern (Volumenanteil $\eta$, Frequenzshift $\delta\omega$ und Diffusionskoeffizient $D$) kann die Relaxationsrate $R_2^*$ in Abh\"angigkeit vom Radius $R_{\text{S}}$ nach Gleichung (\ref{MRT}) ermittelt werden. Aus der gemessenen Relaxationsrate kann nun, wie in Abbildung \ref{Fig:Radiusbestimmung} veranschaulicht, der Radius ermittelt werden. Des Weiteren kann man auch erkennen, in welchem Diffusionsregime man sich befindet. Solange die N\"aherung nach Gleichung (\ref{R2stern_mot_narr}) gilt (in Abbildung \ref{Fig:Radiusbestimmung} als gestrichelte Linie dargestellt), befindet man sich im Motional-Narrowing-Grenzfall. Im Plateu-Bereich (in Abbildung \ref{Fig:Radiusbestimmung} als gepunktete Linie dargestellt) ist die Relaxationsrate vom Kugelradius unabh\"angig und man befindet sich im Static-Dephasing-Bereich. In dem in Abbildung \ref{Fig:Radiusbestimmung} skizzierten Beispiel ist man jedoch weder im Motional-Narrowing-Grenzfall noch im Static-Dephasing-Bereich.

\"Ahnlich kann man verfahren, wenn man beispielsweise die Konzentration bzw. die Dichte der markierten Zellen bestimmen will. Wenn man wei{\ss}, wie gro{\ss} die magnetischen Partikel im Inneren der Zelle sind, kann man die Abh\"angigkeit der Relaxationsrate vom Volumenanteil $\eta$ darstellen und dann aus der gemessenen Relaxationsrate den Volumenanteil ermitteln.

\chapter{\label{Kap:Skalierung}Skalierungsgesetze}
\pagestyle{headings}

\section{Skalierungen in der Kernspinresonanz}
Die transversalen Relaxationszeiten $T_2$ und $T_2^*$ sind fundamentale Gr\"o{\ss}en in der Kernspinresonanzbildgebung, besonders zur Charakterisierung von Geweben und deren Eigenschaften. Der Einfluss statischer Magnetfeldinhomogenit\"aten auf diese Relaxationszeiten ist von grunds\"atzlichem Interesse zum Verst\"andnis der Relaxationsprozesse innerhalb des Voxels. Die Relaxationszeiten $T_2$ und $T_2^*$ werden durch den Suszeptibilit\"atskontrast, durch \"au{\ss}ere Magnetfelder und durch die Diffusion im umgebenden Medium beeinflusst. Um Ver\"anderungen dieser Parameter und deren Auswirkung auf die Relaxationszeit beschreiben zu k\"onnen, ist es wichtig, das Skalierungsverhalten dieser Parameter zu verstehen. 

So stellt sich beispielsweise die Frage, wie sich Kontrastmittel verhalten, die einen st\"arkeren Suszeptibilit\"atssprung zum umgebenden Gewebe erzeugen. Reicht es aus, einfach die Feldst\"arke oder die Suszeptibilit\"atsdifferenz um einen bestimmten Faktor zu erh\"ohen, oder m\"ussen gleichzeitig auch die charakteristischen Abmessung des Teilchens ver\"andert werden, um die Relaxivit\"at zu erh\"ohen? Dabei ist es oft sinnvoll zu wissen, wie sich Gewebeparameter bei anderen Feldst\"arkewerten verhalten. Deshalb ist es wichtig, einen allgemeinen Zusammenhang zu finden, der die Ver\"anderungen der charakteristischen Abmessung und der \"au{\ss}eren Feldst\"arke mit der transversalen Relaxationszeit verbindet. Trotz dieser Tatsache gibt es in der Literatur nur wenige Arbeiten, die sich mit diesem Thema befassen. Weisskoff et al. \cite{Weisskoff94} diskutierten das Skalierungsverhalten der Relaxationsrate $R_2$ im Kontext mikroskopischer Suszeptibilit\"atsvariationen. Ausgehend von der Bloch-Torrey-Gleichung \cite{Torrey56} wurden in \cite{Weisskoff94} zwei spezielle Skalierungsgesetze erhalten und durch numerische Simulationen verifiziert.

Durch Anwenden der Strong-Collision-Approximation \cite{Bauer99,Bauer99PRL,Bauer99T2} kann eine rigorose Ableitung verallgemeinerter Skalierungsgesetze f\"ur die transversalen Relaxationszeiten $T_2$ und $T_2^*$ angeben werden. Zuerst wird ein einfaches Modell eines magnetisierten K\"orpers in einem Voxel betrachtet und dann wird die Strong-Collision-Approximation genutzt, um allgemeing\"ultige Skalierungsgesetze zu erhalten. Zur Illustration werden die Ergebnisse auf eine zylindrische Geometrie, die \"ublicherweise als Modell f\"ur eine Kapillare benutzt wird, angewandt.

\section{Modell des Voxels}
Wie in Abschnitt \ref{Subsec:Allgemeines Modell} beschrieben, wird ein K\"orper innerhalb eines Voxels betrachtet, der einen Suszeptibilit\"atsunterschied zum umgebenden Medium erzeugt (siehe Abbildung \ref{fig1}). Aus der Form des Gebietes $G$ ergibt sich die Geometriefunktion $f({\bf r})$, welche zur Berechnung der Korrelationszeit $\tau$ ben\"otigt wird. Einsetzen der lokalen Resonanzfrequenz (\ref{Eq2}) in den Ausdruck f\"ur die Korrelationszeit in Gleichung (\ref{Tauint}) ergibt
\begin{equation}
\label{Eq4}
\tau = \frac{1}{D}\,\frac{1}{\left\langle f^2({\bf r})\right\rangle V}\int_V\mathrm{d}^3{\bf r}\ f({\bf r})\left(-\,\frac{1}{\Delta}\right)f({\bf r})\, .
\end{equation}
Offensichtlich h\"angt diese Korrelationszeit nur von der Geometriefunktion $f({\bf r})$ des K\"orpers $G$ und dem Diffusionskoeffizienten $D$ ab. Nach Ausf\"uhren der Integration zeigt sich, dass die Korrelationszeit $\tau$ nur von einer charakteristischen L\"ange $L$, dem Diffusionskoeffizienten $D$ und dem Volumenanteil $\eta$ abh\"angt. Dies bedeutet, dass die Signaleigenschaften eines bestimmten Voxels vollst\"andig durch die Geometriefunktion $f({\bf r})$, den Diffusionskoeffizienten $D$ und der charakteristischen Frequenz $\delta\omega$ festgelegt sind. Mit der Korrelationszeit, die sich aus der Geometriefunktion $f({\bf r})$ des K\"orpers $G$ ergibt, kann das gesamte Problem auf die Untersuchung der Abh\"angigkeit $T^*_2 = T^*_2(\tau,\delta\omega)$ reduziert werden.

\section{Allgemeine Ergebnisse}
Um einen Zusammenhang zwischen der Relaxationszeit $T_2^*$, der Korrelationszeit $\tau$ und der charakteristischen Frequenz $\delta\omega$ zu finden, wird (wie in Abschnitt \ref{Sec:Diffusion} beschrieben) die Zeitentwicklung des Signals aus dem Voxel analysiert. Die Laplace-Transformierte $\hat{M}(s)$ des Magnetisierungs-Zeit-Verlaufs $M(t)$ l\"asst sich in der Strong-Collision-N\"aherung entsprechend Gleichung (\ref{Mdach}) durch die Laplace-Transformierte im Static-Dephasing-Regime $\hat{M}_0(s)$ ausdr\"ucken. Der Signal-Zeit-Verlauf im Static-Dephasing-Regime (Gleichung (\ref{SignalSD})) h\"angt nur von der lokalen Resonanzfrequenz und dem Dephasierungsvolumen ab.

Zur Bestimmung der Relaxationszeit $T^*_2$ wird die Mean-Relaxation-Time-Approximation genutzt \cite{Nadler85}. In dieser N\"aherung erh\"alt man:
\begin{equation}
\label{Eq11}
T^*_2 = \int_0^\infty\mathrm{d}t\ M(t) = \hat{M}(0)\, .
\end{equation}
Anwenden dieser Mean-Relaxation-Time-Approximation auf den allgemeinen Zusammenhang in Gleichung (\ref{Mdach}) liefert
\begin{equation}
T^*_2 = \frac{\hat{M}_0(\tau^{-1})}{1-\tau^{-1}\hat{M}_0(\tau^{-1})}\, .
\end{equation}
Die Laplace-Transformierte ergibt sich aus dem Signal-Zeit-Verlauf im Static-Dephasing-Regime (Gleichung (\ref{SignalSD})) zu
\begin{equation}
\hat{M}_0(s) = \int_V \frac{\text{d}^3{\bf r}}{s - \text{i} \omega({\bf r})} \,.
\end{equation}
Mit der in Gleichung (\ref{Eq2}) eingef\"uhrten Darstellung der lokalen Resonanzfrequenz in der Form $\omega({\bf r}) = \delta\omega f({\bf r})$ folgt die Abh\"angigkeit der Relaxationszeit $T_2^*$ von $\tau$ und $\delta\omega$:
\begin{equation}
\label{Eq12}
\frac{1}{T^*_2} = \frac{1}{\displaystyle{\frac{1}{V}\int_V\mathrm{d}^3{\bf r}\ \frac{1}{\tau^{-1}-\text{i}\delta\omega f({\bf r})}}}-\frac{1}{\tau}\, .
\end{equation}
Dieses Ergebnis impliziert direkt das Skalierungsgesetz
\begin{equation}
\label{Eq13}
T^*_2(\lambda\tau,\mu\delta\omega) = \frac{\lambda}{\mu}\, T^*_2(\mu\tau,\lambda\delta\omega)\, ,
\end{equation}
wobei die Skalierungsparameter $\lambda$ und $\mu$ beliebige reelle Zahlen sind. Da die Geometriefunktion zur Ableitung dieser Beziehung nicht n\"aher spezifiziert wurde, ist dieses Skalierungsgesetz f\"ur alle Objekte der gleichen Form g\"ultig. Die Parameter wie Gr\"o{\ss}e, Suszeptibilit\"at, \"au{\ss}eres Magnetfeld oder Diffusionskoeffizient des umgebenden Mediums k\"onnen beliebig gew\"ahlt werden. Diese Parameter bestimmen die zwei Skalierungsvariablen $\tau$ und $\delta\omega$ f\"ur eine vorgegebene Form des magnetischen K\"orpers.

Um ein \"ahnliches Ergebnis f\"ur die Spin-Echo-Relaxationszeit $T_2$ zu erhalten, wird die aus der Strong-Colission-N\"aherung herleitbare Beziehung
\begin{equation}
\label{Eq14}
T_2 = T^*_2 + \tau
\end{equation}
genutzt, die sowohl das Static-Dephasing-Regime als auch das Motional-Narrowing-Regime als Grenzfall enth\"alt \cite{Bauer99PRL,Bauer99T2}. Das resultierende Skalierungsgesetz kann in der Form
\begin{equation}
\label{Eq15} T_2(\lambda\tau,\mu\delta\omega) =
\frac{\lambda}{\mu}\, T_2(\mu\tau,\lambda\delta\omega)
\end{equation}
geschrieben werden. Die Gleichungen (\ref{Eq12}) und (\ref{Eq14}) wurden in der Strong-Colission-N\"aherung abgeleitet. Deshalb gelten die Skalierungsgesetze auch nur in dieser N\"aherung. Jedoch wurde in \cite{Bauer99} angegeben, dass die Vorhersagen dieses Modells auch weit \"uber die Annahmen der Strong-Colission-N\"aherung hinaus g\"ultig sind. Dies trifft dann auch f\"ur den G\"ultigkeitsbereich der Skalierungsgesetze zu. Entsprechend den Ausf\"uhrungen von Weisskoff et al. \cite{Weisskoff94} ist es m\"oglich, aus der Bloch-Torrey-Gleichung (\ref{BT}) die Relation
\begin{equation}
\label{eEq16}
T_2(L,D,\delta\omega) = T_2(\nu L,\nu^2 D,\delta\omega)
\end{equation}
herzuleiten, wobei $L$ eine charakteristische L\"ange des magnetisierten Objektes und $\nu$ eine beliebige reelle Konstante ist. Dieses Skalierungsgesetz ist abh\"angig von der charakteristischen Gr\"o{\ss}e $L$, dem Diffusionskoeffizienten $D$ und der charakteristischen Frequenz $\delta\omega$. Dies impliziert zusammen mit Gleichung (\ref{Eq15}), dass die Korrelationszeit eine Funktion von $L$ und $D$ ist. Im Skalierungsgesetz von Weisskoff werden $L$ und $D$ variiert, w\"ahrend $\delta\omega$ konstant bleibt. Dies entspricht in unserem Skalierungsgesetz nach Gleichung (\ref{Eq15}) dem Fall $\lambda = 1 = \mu$ und impliziert, dass auch $\tau$ konstant bleiben muss. Daraus kann gefolgert werden, dass $\tau$ eine Funktion des Verh\"altnisses von $L^2$ zu $D$ sein muss:
\begin{equation}
\label{eEq17}
\tau = \tau(L^2/D)\,.
\end{equation}
Um diese Abh\"angigkeit zu pr\"azisieren, kann die Koordinatentransformation ${\bf r} = L{\bf u}$ auf Gleichung (\ref{Eq4}) angewandt werden - was einer zentrischen Streckung entspricht -, wobei der Volumenanteil $\eta$ unver\"andert bleibt. Wird ber\"ucksichtigt, dass die Form der Geometriefunktion unter dieser Transformation variablenunabh\"angig ist, d. h. $f({\bf r}) = f({\bf u})$, ergibt sich f\"ur die Korrelationszeit $\tau = k(\eta) L^2/D$. Dies stimmt auch mit dem in Gleichung (\ref{tau1}) erhaltenen Ergebnis \"uberein. Ein analoger Zusammenhang wurden von Stables et al. postuliert (siehe Gleichung (15) in \cite{Stables98}). F\"ur eine vorgegebene Gestalt des magnetischen K\"orpers ergibt sich f\"ur die Abh\"angigkeit $k$ von $\eta$:
\begin{align}
\label{eeEq19}
k(\eta) = \frac{1}{\left\langle f^2({\bf u})\right\rangle V}\int_V\mathrm{d}^3{\bf u}\ f({\bf u})\left(-\,\frac{1}{\Delta}\right)f({\bf u})\, ,
\end{align}
wobei alle Gr\"o{\ss}en und Operatoren bez\"uglich der neuen Variable ${\bf u}$ angegeben sind. Die implizite Abh\"angigkeit von $\eta$ ist durch die Integrationsgebiete $G$ und $V$ in der Gleichung (\ref{Eq3}) bzw. in der Gleichung (\ref{Eq4}) gegeben.

\section{G\"ultigkeitsbereich der Strong-Collision-N\"aherung}
Mit dem allgemeinen Ausdruck f\"ur die Relaxationszeit nach Gleichung (\ref{Eq12}) kann nun ein mathematisches Kriterium f\"ur den G\"ultigkeitsbereich der Strong-Collision-N\"aherung angeben werden, welche im Fall $\tau<T_2^*$ g\"ultig ist. Wird der Zusammenhang aus Gleichung (\ref{MRT}) genutzt und in Gleichung (\ref{Eq12}) eingesetzt, ergibt sich das Kriterium
\begin{equation}
\label{Eq14a}
\mathrm{Re}\left[\left(\frac{1}{V}\int_V\mathrm{d}^3{\bf r}\ \frac{1}{1+\text{i}\tau\delta\omega f({\bf r})}\right)^{-1}\right] < 2 \ .
\end{equation}
Im Motional-Narrowing-Regime (in dem f\"ur die Relaxationszeit $T_2^{*}\propto 1/(\tau\delta\omega^2)$ gilt) f\"uhrt die G\"ultigkeitsbedingung f\"ur die Strong-Collision-N\"aherung $\tau<T_2^{*}$ zu der Ungleichung $\tau^2\delta\omega^2<1$. Im entgegengesetzten Grenzfall des Static-Dephasing-Regimes (in dem f\"ur die Relaxationszeit $T_2^{*} \propto 1/ \delta\omega$ gilt) f\"uhrt dieselbe Bedingung zu $\tau\delta\omega <1$. Daher kann Gleichung (\ref{Eq14a}) bis zum quadratischen Term in $\tau\delta\omega$ entwickelt werden und es ergibt sich
\begin{equation}
\label{Eq14b}
(\Delta f({\bf r}))^2 = \langle f({\bf r})^2\rangle - \langle f({\bf r})\rangle^2 <\frac{1}{2\tau^2\delta\omega^2}\ ,
\end{equation}
wobei der Erwartungswert einer Funktion in Gleichung (\ref{Var}) definiert ist. Die erhaltene Relation (\ref{Eq14b}) kann als mathematisches Kriterium benutzt werden, um den G\"ultigkeitsbereich der Strong-Collision-N\"aherung f\"ur eine bestimmte Form des untersuchten magnetischen K\"orpers festzulegen.

\section{Anwendung auf eine Kapillare}
Die oben erhaltenen allgemeing\"ultigen Skalierungsgesetze werden auf den konkreten Fall einer Kapillare angewandt. Die Abh\"angigkeit der transversalen Relaxationszeit von den Parametern des Kapillarsystems kann genutzt werden, um die Perfusionsreserve im Myokard zu bestimmen \cite{Wacker03}. Dazu wurden nach kombinierter Gabe von Vasodilatator und Kontrastmittel Messungen der transversalen Relaxationszeit durchgef\"uhrt. Das Konzept dieser Methode ist in der Arbeit \cite{Wacker03} erkl\"art. Sowohl die Konzentration des Vasodilatators als auch die Konzentration des Kontrastmittels beeinflussen die transversalen Relaxationszeiten. Der Vasodilatator beeinflusst den Kapillarradius $R_{\text{C}}$ und die Suszeptibilit\"atseffekte des Kontrastmittels ver\"andern den Parameter $\delta\omega$. Im Falle einer hohen Konzentration sowohl von Vasodilatator als auch von Kontrastmittel ist das Static-Dephasing-Regime das zugrunde liegende Diffusionsregime, w\"ahrend im entgegengesetzten Fall niedriger Konzentrationen von Vasodilatator und Kontrastmittel das Motional-Narrowing-Regime das zugrunde liegende Diffusionsregime ist. W\"ahrend das Static-Dephasing-Regime im Abschnitt \ref{Abschn.Zyl} ausf\"uhrlich untersucht wurde, k\"onnen die abgeleiteten Skalierungsgesetze benutzt werden, um allgemeine Ergebnisse zu erhalten. Diese Ergebnisse k\"onnen zur Beschreibung des Effektes von gleichzeitiger Gabe des Vasodilatators und des Kontrastmittels verwendet werden.

Zu diesem Zweck wird eine Kapillare mit dem Neigungswinkel $\theta$ in einem \"au{\ss}eren Magnetfeld $B_0$ betrachtet (siehe Abbildung \ref{krogh}). Die Suszeptibilit\"at im Inneren der Kapillare $\chi_i$ wird durch die Konzentration des Kontrastmittels bestimmt, w\"ahrend $\chi_e$ die Suszeptibilit\"at des umgebenden Mediums ist (siehe Abbildung \ref{krogh}). Der Radius der Kapillare $R_{\text{C}}$ kann direkt durch die Konzentration des Vasodilatators beeinflusst werden. Die lokale Larmor-Frequenz, die durch die Suszeptibilit\"atsdifferenz $\Delta \chi = \chi_i - \chi_e$ innerhalb des Voxels induziert wird, ist in Gleichung (\ref{om-Feld-Zyl}) gegeben.

\begin{figure}
\begin{center}
\includegraphics[width=15cm]{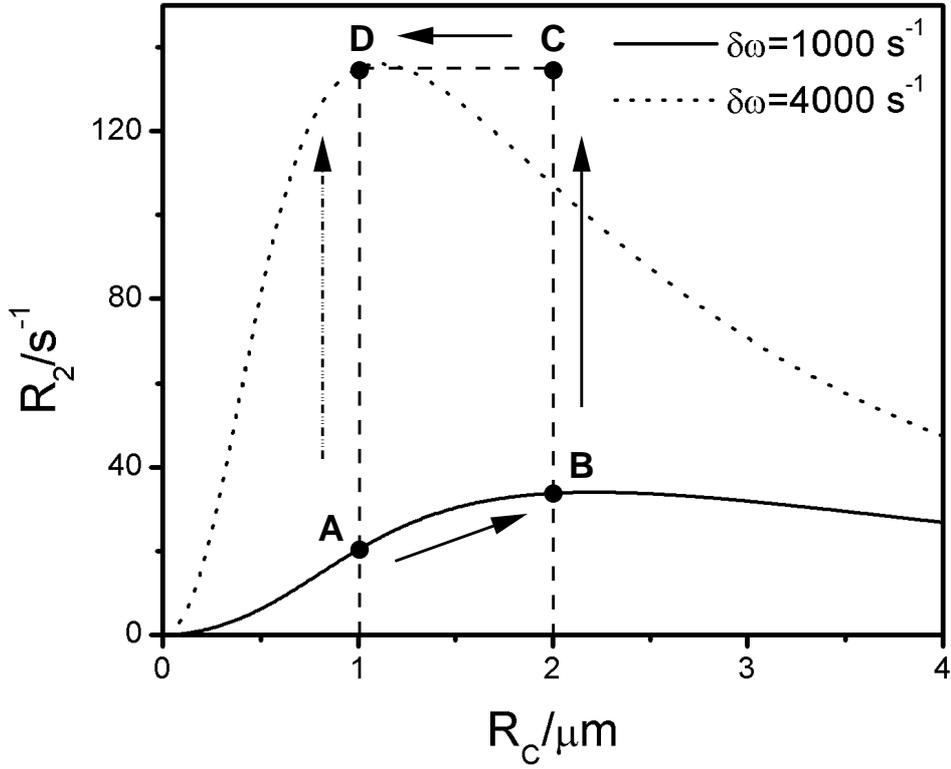}
\caption[Illustration des Skalierungsgesetzes]{{\footnotesize Illustration des Skalierungsgesetzes. Das Skalierungsgesetz $\lambda R_2(\sqrt{\lambda}R_{\text{C}},\delta\omega) = R_2(R_{\text{C}},\lambda\delta\omega)$ wird f\"ur den Parameter $\lambda=4$ erl\"autert. Die durchgezogene und die gestrichelte Kurve wurden aus Gleichung (\ref{Eq19}) mit den Werten $D=1\, \mu\mathrm{m}^2\mathrm{ms}^{-1}$ und $\eta = 0,05$ erhalten.
Eine Verdoppelung des Kapillarradius (von Punkt A zu Punkt B) und anschlie{\ss}ende Multiplikation der entsprechenden Relaxationsrate mit dem Skalierungsparameter $\lambda=4$ (von Punkt B zu Punkt C) ergibt den gleichen Wert f\"ur die Relaxationsrate wie eine Multiplikation des urspr\"unglichen Wertes der Relaxivit\"at bei der charakteristischen Frequenz $\delta\omega$ mit dem Skalierungsparameter $\lambda=4$ unter Beibehaltung des urspr\"unglichen Radius (von Punkt A zu Punkt D).}}
\label{fig4}
\end{center}
\end{figure}

Die Korrelationszeit der um die Kapillare diffundierenden Spins wurde in Abschnitt \ref{Subsec:Cylinders} berechnet und ist in den Gleichungen (\ref{taurad}) und (\ref{tau}) f\"ur strahlende bzw. reflektierende Randbedingungen angegeben. Der Zusammenhang zwischen der transversalen Relaxationszeit und der geometrischen Anordnung der Kapillare wurde bereits in verschiedenen Arbeiten untersucht \cite{Bauer99,Kiselev99,Bauer99T2,Kiselev02}:
\begin{equation}
\label{Eq19}
\frac{1}{\tau R_2} = 1+
\frac{1+\eta}{\left[\sqrt{1+(\eta\tau\,\delta\omega)^2}-1\right]+
\eta\left[\sqrt{1+(\tau\,\delta\omega)^2}-1\right]}\, .
\end{equation}
Mit den beiden Gleichungen (\ref{tau1}) und (\ref{Eq19}) ist es nun m\"oglich, die Suszeptibilit\"atsdifferenz $\Delta\chi$, den Kapillarradius $R_{\text{C}}$ und die Relaxationszeiten $T_2$ und $T_2^*$ miteinander zu verkn\"upfen. Um die Anwendbarkeit der Skalierungsgesetze zu illustrieren, wurde der Einfluss der Erweiterung des Kapillarradius $R_{\text{C}}$ auf die Relaxationsrate $R_2 =1/T_2$ f\"ur verschiedene Suszeptibilit\"atsdifferenzen untersucht. Mit dem allgemeinen Skalierungsgesetz nach Gleichung (\ref{Eq15}) und dem Ausdruck f\"ur die Korrelationszeit nach Gleichung (\ref{tau1}), ergibt sich im Fall $\mu=1$ folgende Form des Skalierungsgesetzes f\"ur die transversale Relaxationszeit: $\lambda\, R_2(\sqrt{\lambda}R_{\text{C}},\delta\omega) = R_2(R_{\text{C}},\lambda\delta\omega)$. Dieser Spezialfall des allgemeinen Skalierungsverhaltens stimmt mit dem Ergebnis von Weisskoff et al. (Gleichung (9) in \cite{Weisskoff94}) \"uberein. Eine Anwendung dieses Skalierungsgesetzes ist in Abbildung \ref{fig4} dargestellt. 

\section{Weitere Beispiele zu den Skalierungsgesetzen}
Die oben angegebenen allgemeinen Skalierungsgesetze in den Gleichungen (\ref{Eq13}) und (\ref{Eq15}) k\"onnen noch weiter vereinfacht werden, um zus\"atzliche Anwendungsm\"oglichkeiten aufzuzeigen. Dazu wird auf der linken Seite von Gleichung (\ref{Eq13}) die Substitution $\tau^{'} = \lambda \tau $ und $\delta\omega^{'} = \mu \delta\omega $ vorgenommen. Auf der rechten Seite von Gleichung (\ref{Eq13}) wird dann $\tau$ durch $\tau = \tau^{'} / \lambda$ und $\delta\omega$ durch $\delta\omega = \delta\omega^{'} / \mu$ ersetzt. Damit ergibt sich
\begin{align}
T^*_2(\tau^{'},\delta\omega^{'}) = \frac{\lambda}{\mu} \, T^*_2 \left( \frac{\mu}{\lambda} \tau^{'}, \frac{\lambda}{\mu}\delta\omega^{'} \right) \,.
\end{align}
Jetzt k\"onnen $\tau^{'}$ und $\delta\omega^{'}$ wieder in $\tau$ und $\delta\omega$ umbenannt werden. Es zeigt sich, dass nur noch der Quotient $\alpha = \lambda/\mu$ in die Betrachtungen eingeht. Damit k\"onnen die Skalierungsgesetze in der Form
\begin{align}
T^*_2(\tau,\delta\omega) & = \alpha \, T^*_2 \left( \frac{1}{\alpha} \tau, \alpha \delta\omega \right)\;\;\;\; \text{und} \\[2ex]
T_2(\tau,\delta\omega) & = \alpha \, T_2 \left( \frac{1}{\alpha} \tau, \alpha \delta\omega \right)
\end{align}
geschrieben werden. Diese Form der Skalierungsgesetze erlaubt es nun, die Parameter, welche die Relaxationszeiten beeinflussen, durch nur einen einzigen Skalierungsfaktor $\alpha$ zu beschreiben.

Als Beispiel wird ein beliebig geformtes Kontrastmittelteilchen betrachtet. Dies kann entweder ein kugelf\"ormiger USPIO oder ein zylinderf\"ormiges Nanor\"ohrchen sein. Wird statt der Relaxationszeit die Relaxationsrate $R_2^{*} = 1 / T_2^{*}$ betrachtet, so ergibt sich die Beziehung
\begin{align} \label{R2stern-skal}
\alpha R^*_2(\tau,\delta\omega) = R^*_2 \left( \frac{1}{\alpha} \tau, \alpha \delta\omega \right) \,.
\end{align}
Dieser Zusammenhang ist nun leicht interpretierbar. Ein Kontrastmittelteichen erzeugt aufgrund des Suszeptibilit\"atsunterschiedes zur Umgebung (charakterisiert durch den Frequenzshift $\delta\omega$) und unter Ber\"ucksichtigung der Diffusion um das Kontrastmittelteilchen (charakterisiert durch die Korrelationszeit $\tau$) die Relaxationsrate $R_2^{*}(\tau,\delta\omega)$. Will man nun diese Relaxationsrate um den Faktor $\alpha$ erh\"ohen (siehe linke Seite der Gleichung (\ref{R2stern-skal})), reicht es nicht aus, nur den Frequenzshift um den Faktor $\alpha$ auf den Wert $\alpha \delta\omega$ zu vergr\"o{\ss}ern, es muss auch gleichzeitig die Korrelationszeit um den Faktor $\alpha$ auf den Wert $\tau / \alpha$ verkleinert werden. Da die Korrelationszeit proportional zum Quadrat eines charakteristischen Durchmessers des Kontrastmittelteilchens ist ($\tau \propto R^2$, wobei $R$ der Radius eines USPIOs oder eines Nanor\"ohrchens ist), muss also auch die Abmessung des Kontrastmittelteichens um den Faktor $1/\sqrt{\alpha}$ verkleinert werden.

\chapter{Zusammenfassung}
Das Dephasierungsverhalten und die daraus resultierende Relaxation der Magnetisierung sind Grundlage aller auf der Kernspinresonanz basierenden bildgebenden Verfahren. Das erhaltene Signal der pr\"azedierenden Protonen wird wesentlich von den Eigenschaften des untersuchten Gewebes bestimmt. Insbesondere die durch magnetisierte Stoffe wie z. B. desoxygeniertes Blut (BOLD-Effekt) oder magnetische Nanopartikel erzeugten Suszeptibilit\"atsspr\"unge gewinnen zunehmend Bedeutung in der biomedizinischen Bildgebung. In der vorliegenden Arbeit wurden die Einfl\"usse von Feldinhomogenit\"aten auf das NMR-Signal untersucht.

Nach einer kurzen Einleitung in Kapitel \ref{Kap.Einleitung}, die einen \"Uberblick zum Inhalt der Arbeit gibt, folgt in Kapitel \ref{Kap.Grundlagen} die Vorstellung der f\"ur die folgenden Untersuchungen n\"otigen Grundlagen, sowohl aus physikalischer und mathematischer, als auch aus biologischer Sicht. Ein allgemeines Modell der Magnetfeldinhomogenit\"aten wird entwickelt und die lokale Resonanzfrequenz um Zylinder und Kugeln wird angegeben. Des Weiteren wird die f\"ur die nachfolgende Beschreibung der Diffusionseffekte um magnetisierte K\"orper benutzte \glqq Strong-Collision-N\"aherung\grqq\ erkl\"art.

Wie in den physikalischen Grundlagen dargestellt wurde, kann die Diffusion in einem inhomogenen Magnetfeld als stochastischer Prozess im Sinne einer Brownschen Molekularbewegung aufgefasst werden. Eine wesentliche Gr\"o{\ss}e zur Beschreibung dieses Prozesses ist die Korrelationszeit. In Kapitel \ref{Kap.Korr} wurde diese Korrelationszeit der Diffusionsbewegung um einen magnetisierten K\"orper untersucht. Ein Gro{\ss}teil dieser Ergebnisse ist in unserer Arbeit \cite{Ziener06a} ver\"offentlicht. F\"ur die im Folgenden interessierenden Spezialf\"alle in Form von Zylindern und Kugeln konnte ein Zusammenhang zwischen der Korrelationszeit, dem Diffusionskoeffizienten, dem Volumenanteil und dem Zylinder- bzw. Kugelradius hergestellt werden. Des Weiteren wurde die Oberfl\"achenbeschaffenheit des magnetisierten K\"orpers ber\"ucksichtigt. Im Falle einer impermeablen Oberfl\"ache wurden reflektierende Randbedingungen angenommen, d. h. sobald ein Spin die Oberfl\"ache ber\"uhrt, wird die Trajektorie durch eine symmetrische Trajektorie fortgesetzt. K\"onnen die Wassermolek\"ule die Oberfl\"ache des magnetisierten K\"orpers jedoch durchdringen, m\"ussen strahlende Randbedingungen angenommen werden. Diese k\"onnen im Sinne eines aus der Physiologie bekannten Oberfl\"achenpermeabilit\"atsproduktes ber\"ucksichtigt werden.

Die St\"arke der Suszeptibilit\"atseffekte wird durch die charakteristische Frequenz $\delta\omega \propto \Delta \chi B_0$ beschrieben, wobei $\Delta \chi$ die Suszeptibilit\"atsdifferenz zum umgebenden Medium ist. Diese sogenannte statische Frequenzskala kann mit der dynamischen Frequenzskala, die als das Inverse der Korrelationszeit $1/\tau$ definiert ist, verglichen werden. Dadurch ist eine Einteilung der Diffusionsregime m\"oglich. Wenn die durch die Suszeptibilit\"atsdifferenz hervorgerufenen Effekte gr\"o{\ss}er sind als der Einfluss der Diffusion, d. h. $\delta\omega \gg 1/\tau$, ist das Static-Dephasing-Regime das zugrunde liegende Diffusionsregime. Im umgekehrten Grenzfall $\delta\omega \ll 1/\tau$ herrscht das Motional-Narrowing-Regime. Mit den Ergebnissen aus Kapitel \ref{Kap.Korr} kann nun aus den charakteristischen Parametern des Gewebes (Diffusionskoeffizient, Durchmesser des magnetisierten K\"orpers, Volumenanteil, Suszeptibilit\"atsdifferenz) ermittelt werden, welches Diffusionsregime das zugrunde liegende ist. Damit kann entschieden werden, welche N\"aherungsmethode zur Beschreibung der Relaxation angewandt werden muss.

F\"ur den Fall des Motional-Narrowing-Regimes, der z. B. zur Beschreibung der Relaxationseffekte von injizierbaren MR-Kontrastmitteln angewandt wird, kann die Relaxationszeit leicht aus der Korrelationszeit und dem Erwartungswert der lokalen Frequenz um ein Kontrastmittelteilchen entsprechend der Gleichung $1/T_2 = \tau \langle \omega^2(\mathrm{r}) \rangle$ ermittelt werden. F\"ur niedrige Kontrastmittelkonzentrationen ergibt sich wie erwartet ein linearer Zusammenhang zwischen Relaxationsrate und Konzentration des Kontrastmittels.

Zur Charakterisierung biologischer Gewebe, die Feldinhomogenit\"aten enthalten, eignen sich nicht nur die Relaxationszeiten, sondern auch die Frequenzverteilung, die von den Feldinhomogenit\"aten erzeugt wird. Diese Frequenzverteilungen werden in Kapitel \ref{Kap:Frequenz} untersucht. Mit dieser Methode ist es m\"oglich, die NMR-Signalentstehung im magnetisch inhomogenen Gewebe \"uber den gesamten Dynamikbereich, also vom Motional-Narrowing-Regime zum Static-Dephasing-Regime zu beschreiben. Dazu wird die Frequenzverteilung innerhalb eines Voxels beschrieben. Zuerst wurde das einfach zu beschreibende Static-Dephasing-Regime untersucht, in dem die Diffusion vernachl\"assigt werden kann. Durch Anwenden des aus der statistischen Physik bekannten Formalismus der Zustandsdichten konnte die Frequenzverteilung um ein zylinderf\"ormiges Objekt analytisch ermittelt werden. Die Ergebnisse sind in unserer Arbeit \cite{Ziener05MAGMA} ver\"offentlicht. Diese analytischen Werte konnten durch Simulationen und durch numerische Rechnungen best\"atigt werden. Ein Phantom zur Messung der exakten Form der Frequenzverteilung wurde entwickelt und auch die experimentellen Ergebnisse stimmen mit den analytischen Resultaten \"uberein. An diesem Beispiel der Frequenzverteilung um eine Kapillare konnte sehr gut gezeigt werden, wie die vier Grundlagen, auf denen der Prozess der physikalischen Erkenntnisgewinnung basiert -- n\"amlich analytische Ergebnisse, Computersimulationen, numerische Rechnungen und die experimentelle Best\"atigung -- genutzt wurden, um saubere und reproduzierbare Ergebnisse zu erhalten. Aus dieser intensiven Bearbeitung -- sowohl von theoretischer als auch experimenteller Seite -- ergaben sich weitere interessante Fragestellungen, wie z. B. der Einfluss der Diffusionseffekte des umgebenden Mediums auf die Frequenzverteilung. Um diese Fragestellung ersch\"opfend zu beantworten, wurde die im Grundlagenabschnitt beschriebene Strong-Collision-Approximation genutzt. Durch Verallgemeinerung der Ergebnisse des Static-Dephasing-Regimes konnte ein Formalismus zur Beschreibung der Frequenzverteilung, der \"uber den gesamten Dynamikbereich g\"ultig ist, abgeleitet werden. Die erhaltenen Ergebnisse stimmen mit den aus der Linienformtheorie nach Kubo erhaltenen Resultaten \"uberein. Diese Ergebnisse wurden in der Arbeit \cite{ZienerPRE} ver\"offentlicht.

Mit der in Kapitel \ref{Kap.Korr} untersuchten Korrelationszeit konnte die Frequenzverteilung um zylinderf\"ormige und kugelf\"ormige Objekte \"uber den gesamten Dynamikbereich beschrieben werden. Mit diesen Ergebnissen k\"onnen nun nicht nur Gradientenechoexperimente, sondern auch weiterentwickelte Pulssequenzen, wie z. B. SSFP-Sequenzen, beschrieben werden.

Die Charakterisierung von Geweben durch Relaxationszeiten setzt die Annahme eines exponentiellen Magnetisierungszerfalles voraus. Die bisherigen Verfahren zur Beschreibung der Dephasierung in Bildgebungsexperimenten beruhen auf dem Anderson-Weiss-Modell. Ausgangspunkt dieses Modells ist die Annahme einer gau{\ss}verteilten St\"arke der St\"orfelder bzw. St\"orfrequenzen, die durch Magnetfeldinhomogenit\"aten induziert werden. Entscheidend f\"ur den Magnetisierungszerfall ist jedoch der Phasenwinkel. Problematisch ist, dass Anderson und Weiss als bewiesen annahmen, es sei die Gau{\ss}verteilung des Phasenwinkels automatisch bei einer Gau{\ss}verteilung der St\"orfelder gegeben. 

Eine exakte Untersuchung des Zusammenhangs zwischen Frequenz- und Phasenverteilung wurde in Kapitel \ref{Chap:Dephas} vorgenommen und in unserer Arbeit \cite{Bauer05} ver\"offentlicht. Dabei wurden auch erstmals die exakten Kriterien f\"ur die Anwendbarkeit des Anderson-Weiss-Modell herausgearbeitet. Zentraler Untersuchungsgegenstand ist die \"Ubergangswahrscheinlichkeit eines diffundierenden Spins von der St\"orfrequenz $\omega_1$ zur Frequenz $\omega_2$ innerhalb des Zeitintervalls $\Delta t$, auch als Greensche Funktion $G(\omega_2, \omega_1, \Delta t)$ bezeichnet. Diese Betrachtungsweise kn\"upft direkt an die bereits von Einstein eingef\"uhrte stochastische Beschreibung der Diffusion an.

Im Rahmen dieser Untersuchungen konnte gezeigt werden, dass eine gau{\ss}f\"ormige Verteilung der Phasenwinkel \"aquivalent zu einer gau{\ss}f\"ormigen \"Ubergangswahrscheinlichkeit der Resonanzfrequenzen ist, jedoch folgt aus einer gau{\ss}f\"ormigen Frequenzverteilung nicht eine gau{\ss}f\"ormige Phasenwinkelverteilung. Dies bedeutet, dass die Annahme einer Gau{\ss}schen Dephasierung der Phasenwinkel die \"Ubergangsdynamik zwischen lokalen Frequenzen auf eine sehr spezielle Klasse von stochastischen Prozessen beschr\"ankt. Nur wenn die \"Ubergangsdynamik durch einen Ornstein-Uhlenbeck-Prozess beschrieben werden kann, also s\"amtliche Charakteristika eines station\"aren Gau{\ss}schen Prozesses aufweist, ist das Anderson-Weiss-Modell der Dephasierung anwendbar. W\"ahrend Anderson und Weiss als einzige Voraussetzung f\"ur die Anwendbarkeit ihres Modells eine gau{\ss}f\"ormige Gleichgewichtsverteilung in den St\"orfeldfrequenzen nannten, konnte gezeigt werden, dass diese Voraussetzung nicht ausreichend ist, sondern dass noch weitere Anforderungen an die Stochastik des \"Ubergangsprozesses zu stellen sind.

Zur praktischen Anwendung auf klinische Fragestellungen reichen jedoch die transversalen Relaxationszeiten zur Charakterisierung von Geweben aus. Dazu wurde in Kapitel \ref{Kap:Relaxation} die transversale Relaxationszeit magnetisch markierter Zellen untersucht. Die Ergebnisse wurden in unserer Arbeit \cite{Ziener05MRM} ver\"offentlicht. Zur Ableitung der Ergebnisse wurde angenommen, dass die Kontrastmittelteilchen (USPIOs) von den Zellen phagozytiert werden und sich zu einem Kern innerhalb der Zelle vereinigen. Dieser Kern wird von einer Wasserh\"ulle umgeben, in der die Dephasierung stattfindet. In Abh\"angigkeit von dem Volumenanteil der magnetischen Teilchen (Kontrastmittelkonzentration), dem Diffusionskoeffizienten des umgebenden Mediums und dem Radius der magnetischen Teilchen konnte die Relaxationszeit $T_2^*$ bestimmt werden.

Einfache Skalierungsgesetze sind n\"utzliche Werkzeuge, um Effekte zu verstehen, die durch die Ver\"anderung verschiedener Parameter auftreten, welche das NMR-Experiment beeinflussen. In Kapitel \ref{Kap:Skalierung} wird das allgemeine Skalierungsverhalten der transversalen Relaxationszeiten diskutiert. Die Ergebnisse wurden in unserer Arbeit \cite{Ziener06b} ver\"offentlicht. Dabei wird der Dephasierungsprozess um einen beliebig geformten K\"orper betrachtet, die in einem Voxel eingebettet ist. Das Signal von diesem Voxel kann als exponentieller Abfall mit einer transversalen Relaxationszeit angen\"ahert werden. Diese transversale Relaxationszeit wird durch verschiedene Parameter beeinflusst: durch die Suszeptibilit\"atsdifferenz zwischen magnetischen K\"orper und umgebenden Medium, durch die \"au{\ss}ere Feldst\"arke, durch den Diffusionskoeffizienten des umgebenden Mediums sowie durch eine charakteristische L\"ange des magnetischen K\"orpers. Mit den Skalierungsgesetzen kann nun beschrieben werden, wie sich beispielsweise eine Vervielfachung der \"au{\ss}eren Magnetfeldst\"arke, eine Vervielfachung des Diffusionskoeffizienten (durch Erw\"armung des umgebenden Mediums) oder eine Vervielfachung der charakteristischen Abmessungen (\"Ubergang vom Tierexperiment zur klinischen Anwendung) auf die transversalen Relaxationszeiten auswirken. Die Skalierungsgesetze werden am Beispiel einer blutgef\"ullten Kapillare unter dem gleichzeitigen Einfluss von Kontrastmittel und Vasodilatator illustriert.

\chapter*{Brief Summary}
\pagestyle{plain}
\addcontentsline{toc}{chapter}{Brief Summary}
The properties of dephasing and the resulting relaxation of the magnetization are the basic principle on which all magnetic resonance imaging methods are based. The signal obtained from the gyrating spins is essentially determined by the properties of the considered tissue. Especially the susceptibility differences caused by magnetized materials (for example, deoxygenated blood, BOLD-effect) or magnetic nanoparticles are becoming more important for biomedical imaging. In the present work, the influence of such field inhomogeneities on the NMR-signal is analyzed.

A short introduction in Chapter \ref{Kap.Einleitung} gives a survey of the content of the work. In Chapter \ref{Kap.Grundlagen}, the physical and mathematical as well as biological fundamentals which are necessary for the following analysis are presented. A general model of the magnetic field inhomogeneities is developed and the local resonance frequency around a cylinder and a sphere is given. Furthermore, the strong collision approximation, which is used for the subsequent description of the diffusion effects around magnetized objects, is explained.

As illustrated in the section about the physical basics, the diffusion in an inhomogeneous magnetic field can be considered a stochastic process analogous to Brownian motion. An essential parameter to describe this process is the correlation time. In Chapter \ref{Kap.Korr}, this correlation time of the diffusion around a magnetized object is analyzed. Further results have been published in the work \cite{Ziener06a}. For the below-mentioned special cases of cylinders and spheres, it was possible to give a relation between the correlation time, diffusion coefficient, volume fraction, and radius. Furthermore, the surface structure of the magnetized object was considered. In the case of an impermeable surface, reflective boundary conditions were assumed, i.e. if a spin reaches the surface, the trajectory of this spin is continued by a symmetric one. If water molecules can penetrate the surface of the magnetized object, radiative boundary conditions have to be assumed. These boundary conditions can be considered in the same sense of the surface-permeability-product, which is well known from physiology.

The intensity of the susceptibility effects is described by the characteristic frequency $\delta\omega \propto \Delta \chi B_0$, where $\Delta \chi$ is the susceptibility difference of the surrounding medium. This so-called static frequency scale can be compared with the dynamic frequency scale, which is defined as the inverse of the correlation time. Thus, it is possible to classify the diffusion regimes. If the effects caused by the susceptibility difference are greater than the influence of the diffusion (i.e., $\delta\omega \gg 1/\tau$) the underlying diffusion regime is the static dephasing regime. In the opposite limiting case $\delta\omega \ll 1/\tau$ the motional-narrowing regime dominates. With the results of Chapter \ref{Kap.Korr} in hand, it is possible to determine the underlying diffusion regime from the characteristic parameters of the tissue (diffusion coefficient, diameter of the magnetized object, volume fraction, susceptibility difference). It is then possible to decide which approximation has to be applied to describe the relaxation effects.

In the motional narrowing regime, which is applied to describe the relaxation effects of MR contrast agents, the relaxation time can be easily calculated from the correlation time and the expectation value of the local frequency around a contrast agent particle according to the equation $1/T_2 = \tau \langle \omega^2(\mathrm{r}) \rangle$. Small values of the concentration of the contrast agent yield, as expected, a linear relationship between the relaxation rate and the concentration of the contrast agent.

Not only the relaxation time, but also the frequency distribution, which is caused by the field inhomogeneities, is applicable to characterizing biological tissues containing field inhomogeneities. These frequency distributions are analyzed in Chapter \ref{Kap:Frequenz}. Using this method, it is possible to describe the NMR signal formation in magnetic inhomogeneous tissue over the whole dynamic range from the motional narrowing regime to the static dephasing regime. For this reason, the frequency distribution inside a voxel is analyzed. First, the static dephasing regime, in which diffusion effects can be neglected, was analyzed. The formalism of the density of states, which is well known from statistical physics, could be applied to calculate the frequency distribution around a cylindrical object. The results are published in the work \cite{Ziener05MAGMA}. These analytical results could be verified by simulations and numerical analysis. To measure the exact form of the frequency distribution, an MRI phantom was developed. The experimental values were in good agreement with the analytical results. In the example of the frequency distribution around a capillary, it was demonstrated very well how the four fundamentals on which the gain of knowledge in physics is based - analytical analysis, computational simulation, numerical calculation, and the experimental verification - were utilized to obtain exact and reproducible results. This intensive treatment raised further new and interesting problems, such as the influence of the diffusion effects of the surrounding medium on the frequency distribution. To consider this issue, the strong collision approximation, described in the chapter on basic physical principles, was utilized. Generalizing the results of the static dephasing regime, it was possible to deduce a formalism valid over the whole dynamic range to describe the frequency distribution. The obtained results are in agreement with the results obtained from Kubos line shape theory. These results are published in the work \cite{ZienerPRE}.

With the correlation time studied in Chapter \ref{Kap.Korr} in hand, it was possible to describe the frequency distribution around cylindrical and spherical objects over the whole dynamic range. These results can be used to describe more sophisticated pulse sequences (for example, SSFP-sequences).

The characterization of tissues by relaxation times presumes an exponential decay of the magnetization. Until now, the methods used to describe the dephasing in magnetic resonance imaging experiments have been based on the Anderson-Weiss-Model. The starting point of this model is the assumption of a Gaussian distribution of the fields and frequencies induced by the magnetic field inhomogeneities. However, the phase angle is essential for the time evolution of the magnetization. The problem is that Anderson and Weiss assumed that a Gaussian distribution of the phase angle necessarily follows from a Gaussian distribution of the resonance frequencies. 

An exact analysis of the relation between the distribution of the resonance frequencies and the distribution of the phase angle is performed in Chapter \ref{Chap:Dephas} of this work and published in \cite{Bauer05}. First, exact criteria for the applicability of the Anderson-Weiss-Model are achieved. The object of investigation is the transition probability of a diffusing spin from the resonance frequency $\omega_1$ to the resonance frequency $\omega_2$ during the time interval $\Delta t$, which is also denoted as Green's function $G(\omega_2, \omega_1, \Delta t)$ This approach ties in directly with the description of the diffusion as a stochastic process, as established by Einstein.

In the framework of this analysis, it could be shown that a Gaussian distribution of the phase angle is equivalent to a Gaussian transition probability of the resonance frequencies, but a Gaussian distribution of the frequencies does not imply a Gaussian distribution of the phase angles. This means that the assumption of a Gaussian dephasing of the phase angles restricts the transition probability between local frequencies to a very limited class of stochastic processes. Only if the transition dynamics can be described by a Ornstein-Uhlenbeck-process, which exhibits all characteristics of a stationary Gaussian process, is the Anderson-Weiss-model applicable. While Anderson and Weiss assumed that only a Gaussian distribution of the frequencies is necessary for the applicability of their model, it could be shown that this assumption is not sufficient. Rather, further assumptions about the stochastic of the transition process have to be fulfilled.

However, for practical application in medical problems, the transverse relaxation times suffice to characterize tissues. Therefore, in Chapter \ref{Kap:Relaxation} the transverse relaxation time of magnetically labeled cells is analyzed. The results are published in the work \cite{Ziener05MRM}. To deduce the results, it was assumed that the contrast agents (USPIOs) are phagocytosed and agglomerate to a magnetic core inside the cell. This core is surrounded by a shell of water in which the dephasing occurs. It was possible to determine the relaxation time $T_2^*$ as a function of the volume fraction of the magnetic core (concentration of the contrast agent), diffusion coefficient of the surrounding medium and the radius of the magnetic core.

Simple scaling laws are a useful tool to understand effects which arise from the variation of different parameters influencing the NMR-experiment. In Chapter \ref{Kap:Skalierung}, the general scaling behavior of the transverse relaxation times is discussed. The results are published in the work \cite{Ziener06b}. To achieve this, the diffusion process around an arbitrary object embedded in a voxel is considered. The signal of this voxel can be approximated as an exponential decay with a transverse relaxation time. This transverse relaxation time is influenced by many parameters: the susceptibility difference between the magnetized object and the surrounding medium, the strength of the external magnetic field, the diffusion coefficient of the surrounding medium, as well as a characteristic length of the magnetized object. With the scaling laws in hand, it is possible to describe the influence of a multiplication of the strength of the external magnetic field, a multiplication of the diffusion coefficient (by heating of the surrounding medium), or a multiplication of the characteristic diameter (change from animal experiment to the measurement of humans) on the transverse relaxation time. The scaling laws are visualized by the example of a blood filled capillary under the simultaneous influence of contrast agent and vasodilatator.

\chapter*{Publikationsliste}
\addcontentsline{toc}{chapter}{Publikationsliste}
\pagestyle{plain}
\vspace{-0.8cm}
\section*{Ver\"offentlichungen als Erstautor}
\begin{enumerate}
\item {\bf C. H. Ziener}, S. Glutsch, and F. Bechstedt. {\it RKKY interaction in semiconductors:   Effects of magnetic field and screening}, Phys. Rev. B 70, 075205 (2004).
\item {\bf C. H. Ziener}, W. R. Bauer, and P. M. Jakob. {\it Transverse Relaxation of Cells Labeled with Magnetic Nanoparticles}, Magn. Reson. Med. 54, 702-706 (2005).
\item {\bf C. H. Ziener}, W. R. Bauer, and P. M. Jakob. {\it Frequency distribution and signal formation around a vessel}, Magn. Reson. Mater. Phy. 18, 225-230 (2005).
\item {\bf C. H. Ziener}, W. R. Bauer, G. Melkus, T. Weber, V. Herold, P. M. Jakob. {\it Structure-specific magnetic field inhomogeneities and its effect on the correlation time}, Magn. Reson. Imaging 24, 1341-1347 (2006).
\item {\bf C. H. Ziener}, T. Kampf, G. Melkus, P. M. Jakob, W. R. Bauer. {\it Scaling Laws for Transverse Relaxation Times}, J. Magn. Reson. 184, 169-175 (2007).
\item {\bf C. H. Ziener}, T. Kampf, W. R. Bauer, P. M. Jakob, S. Glutsch, F. Bechstedt. {\it Quantum Beats in Semiconductors}, International Journal of Modern Physics B 21, Nos. 8-9, 1621-1625 (2007).
\item{\bf C. H. Ziener}, T. Kampf, G. Melkus, V. Herold, T. Weber, G. Reents, P. M. Jakob, W. R. Bauer. {\it Local frequency density of states around field inhomogeneities in magnetic resonance imaging: Effects of diffusion}, Phys. Rev. E 76, 031915 (2007).
\item {\bf C. H. Ziener}, T. Kampf, V. Herold, P. M. Jakob, W. R. Bauer, W. Nadler. {\it Frequency autocorrelation function of stochastically fluctuating fields caused by specific magnetic field inhomogeneities}, J. Chem. Phys. 129, 014507 (2008).
\end{enumerate}
\section*{Ver\"offentlichung als Zweitautor}
\begin{enumerate}
\setcounter{enumi}{8}
\item W. R. Bauer, {\bf C. H. Ziener}, and P. M. Jakob. {\it Non-Gaussian spin dephasing}, Phys. Rev. A 71, 053412 (2005)
\end{enumerate}
\section*{Sonstige Ver\"offentlichung}
\begin{enumerate}
\setcounter{enumi}{9}
\item G. Klug, T. Kampf, {\bf C. H. Ziener}, M. Parczyk, E. Bauer, V. Herold, E. Rommel, P. M. Jakob, W. R. Bauer. {\it Murine atherosclerotic plaque imaging with the USPIO Ferumoxtran-10}, Frontiers in Biosci., im Druck.
\end{enumerate}

\newpage

\section*{Diplomarbeit}
\begin{itemize}
\item {\bf C. H. Ziener}. {\it Spinquantenschwebungen in semimagnetischen Halbleitern}, Jena (2003).
\end{itemize}

\section*{Wissenschaftspreis}
\begin{itemize}
\item {\bf C. H. Ziener}. {\it Frequency distribution in a vascular network}, Young Investigator Award der ESMRMB, zweiter Preis, Basel (2005).
\end{itemize}

\section*{Vortr\"age}
\begin{enumerate}
\item {\bf C. H. Ziener}, W. R. Bauer, and P. M. Jakob. {\it Frequency distribution in a vascular network}, ESMRMB, Vortrag 148 (Basel 2005).
\item {\bf C. H. Ziener}, W. R. Bauer, P. M. Jakob. {\it Skalierungsgesetze f\"ur transversale Relaxationszeiten}, 8. Jahrestreffen der Deutschen Sektion der ISMRM (M\"unster 2005).
\item {\bf C. H. Ziener}, T. Kampf, G. Melkus, W. R. Bauer, and P. M. Jakob. {\it SSFP signal analysis}, ESMRMB, Vortrag 42 (Warschau 2006).
\item {\bf C. H. Ziener}, T. Kampf, G. Melkus, W. R. Bauer, P. M. Jakob. {\it Diffusionsabh\"angige Frequenzverteilungen}, 9. Jahrestreffen der Deutschen Sektion der ISMRM (Jena 2006).
\item {\bf C. H. Ziener}, T. Weber, W. R. Bauer, and P. M. Jakob. {\it Quantification of the Spinal Cord Axon Diameter using an Extension of the PGSE Sequence}, Proc. Int. Soc. Magn. Reson. Med. 2007:2290 (Berlin 2007).
\item {\bf C. H. Ziener}, T. Kampf, W. R. Bauer, and P. M. Jakob. {\it Magnetic resonance imaging of magnetically labelled cells}, Fellows Meeting 2007 der Ernst-Schering-Foundation (Berlin 2007).
\end{enumerate}

\section*{Eingeladener Vortrag}
\begin{itemize}
\item {\bf C. H. Ziener}. {\it From Microscopic Field Inhomogeneities to a Macroscopic MR-Signal}, Bayer Schering Pharma Symposium \glqq Keeping Track of Innovation\grqq\  anl\"asslich des Joint Annual Meeting ISMRM-ESMRMB (Berlin 2007).
\end{itemize}

\section*{Poster}
\begin{enumerate}
\item {\bf C. H. Ziener}, W. R. Bauer, P. M. Jakob. {\it Relaxationsverhalten magnetisch markierter Zellen}, 7. Jahrestreffen der Deutschen Sektion der ISMRM (Mainz 2004).
\item {\bf C. H. Ziener}, W. R. Bauer, and P. M. Jakob. {\it Transverse Relaxation of Cells Labeled with Magnetic Nanoparticles}, Proc. Int. Soc. Magn. Reson. Med. 2005:2611 (Miami 2005).
\newpage
\item {\bf C. H. Ziener}, T. Kampf, G. Melkus, R. Kharrazian, M. Choli, W. R. Bauer, C. Faber, P. M. Jakob. {\it SSFP Signal Formed by a Lorentzian Frequency Distribution}, International Symposium on Biomedical Magnetic Resonance Imaging and Spectroscopy at Very High Fields, Poster 14 (W\"urzburg 2006).
\item {\bf C. H. Ziener}, T. Kampf, S. Glutsch, W. R. Bauer, P. M. Jakob, F. Bechstedt. {\it Quantum Beates in Magnetic Semiconductors}, 17th International Conference on High Magnetic Fields in Semiconductor Physics (HMF), Poster HMF$\_5\_5$ (W\"urzburg 2006).
\item T. Kampf, {\bf C. H. Ziener}, G. Melkus, A. Purea, M. Parczyk, W. R. Bauer, P. M. Jakob. {\it USPIO-Modelle im Vergleich}, 9. Jahrestreffen der Deutschen Sektion der ISMRM (Jena 2006).
\item T. Kampf, {\bf C. H. Ziener}, P. M. Jakob, W. R. Bauer. {\it Dependence of the frequency distribution on the orientation of the voxel}, Molekulare Bildgebung 07, Poster 1 (Kiel 2007).
\item {\bf C. H. Ziener}, T. Kampf, W. R. Bauer, P. M. Jakob. {\it Diffusion Dependent Frequency Distribution}, 9th International Conference on Magnetic Resonance Microscopy, Poster 42 (Aachen 2007).
\item T. Kampf, {\bf C. H. Ziener}, X. Helluy, P. M. Jakob, W. R. Bauer. {\it Computation of inter and intra voxel diffusion using MC-simulations in frequency and spatial domain: a comparison}, 9th International Conference on Magnetic Resonance Microscopy, Poster 34 (Aachen 2007).
\item {\bf C. Ziener}, V. Herold, G. Klug, M. Parczyk, E. Rommel, P. Jakob, W. R. Bauer. {\it Nichtinvasive in vivo Messung der regionalen Pulswellengeschwindigkeit mittels hochaufl\"osender MRI}, 74. Jahrestagung der
Deutschen Gesellschaft f\"ur Kardiologie – Herz - und Kreislaufforschung e.V., Poster 1491 (Mannheim 2008).
\item V. Herold, M. Parczyk, {\bf C. Ziener}, G. Klug, E. Rommel, W. R. Bauer, P. Jakob {\it In vivo Magnetresonanzbildgebung zur Messung der lokalen Pulswellengeschwindigkeit an der Maus bei 17,6 Tesla}, 74. Jahrestagung der Deutschen Gesellschaft f\"ur Kardiologie – Herz - und Kreislaufforschung e.V., Poster 841 (Mannheim 2008).
\item {\bf C. H. Ziener}, V. Herold, M. Parczyk, G. Klug, T. Kampf, E. Rommel, P. Jakob, W. Bauer. {\it In-vivo-Messung der regionalen und lokalen Pulswellengeschwindigkeit in der Aorta der Maus mittels MR-Bildgebung bei 17,6 Tesla}, 114. Kongress der Deutschen Gesellschaft f\"ur Innere Medizin, Poster 245 (Wiesbaden 2008).
\item T. C. Basse-Luesebrink, T. Kampf, {\bf C. H. Ziener}, G. Klug, W. R. Bauer, P. M. Jakob, and D. Haddad. {\it Evaluation of sensitivity increase by T1 and T2 contrast agents in 19F MRI of PF15C}, Proc. Int. Soc. Magn. Reson. Med. 2008:1655 (Toronto 2008).
\item V. Herold, G. Klug, M. Parczyk, {\bf C. Ziener}, T. Weber, S. Sarkar, W. R. Bauer, E. Rommel, and P. M. Jakob. {\it In vivo measurement of local pulse-wave velocity in mice with MRI at 17.6 T}, Proc. Int. Soc. Magn. Reson. Med. 2008:907 (Toronto 2008).
\item {\bf C. H. Ziener}, T. Kampf, V. Herold, P. M. Jakob, W. R. Bauer and W. Nadler. {\it Temporal correlation function around spheres and cylinders}, 9th International Bologna Conference Magnetic Resonance in Porous Media (MRPM9), Poster 101 (Cambridge MA, USA 2008).
\end{enumerate}

\chapter*{Lebenslauf}
\addcontentsline{toc}{chapter}{Lebenslauf}

\section*{Pers\"onliche Daten}
\begin{tabular}[t]{p{0.25\textwidth}p{0.8\textwidth}}
Vor- und Zuname & \underline{Christian} Herbert Ziener\\
Geburtsdatum &18.12.1978\\
Geburtsort &Weimar\\
Staatsangeh\"origkeit &deutsch\\
Familienstand &ledig\\
Adresse &dienstlich:\hspace{3.7cm}privat:\\
        &Experimentelle Physik 5\hspace{1.1cm}Straubm\"uhlweg SWH2\\
        &Am Hubland\hspace{3.15cm}Zimmer 211\\
        &97074 W\"urzburg\hspace{2.55cm}97078 W\"urzburg\\
Telefon &0931-888\,4957\hspace{3.05cm}0931-203\,88262\\
E-mail &{\tt ziener@physik.uni-wuerzburg.de}
\end{tabular}

\vspace*{2.5cm}
\section*{Werdegang}
\begin{tabular}[t]{p{0.25\textwidth}p{0.69\textwidth}}
1985-1991 & Polytechnische Oberschule \glqq Friedrich Le\ss ner\grqq\  in Blankenhain\\
{}\\
1991-1993 & Geschwister-Scholl-Gymnasium in Bad Berka\\
{}\\
1993-1997 & Carl-Zeiss-Gymnasium in Jena, Spezialschule mathematisch-naturwis\-senschaftlich-technischer Richtung, Abschluss: Abitur, Durchschnittsnote: 1,1; Internat der Spezialschule in Jena\\
{}\\
1997-1998 & Wehrdienst bei der Bundeswehr in Eschweiler und in Gotha\\
{}\\
1998-2003 & Studium der Physik an der Friedrich-Schiller-Universit\"at in Jena, Diplomarbeit am Institut f\"ur Festk\"orpertheorie und Theoretische Optik. Abschluss: Diplomphysiker, Durchschnittsnote: 1,0 (mit Auszeichnung), Nebenfach: Funktionalanalysis\\
\end{tabular}

\newpage
\thispagestyle{empty}
\begin{tabular}[t]{p{0.25\textwidth}p{0.69\textwidth}}
seit 2004 & Promotionsstudium am Lehrstuhl f\"ur Experimentelle Physik 5 der Julius-Maximilians-Universit\"at W\"urzburg\\
{}\\
seit WS 2004 & Studium der Humanmedizin an der Julius-Maximilians-Universit\"at W\"urzburg\\
{}\\
2005 -2006 & Stipendiat der Schering-Stiftung\\
{}\\
08/2006 & Erster Abschnitt der \"Arztlichen Pr\"ufung, Durchschnittsnote: 1,5 (sehr gut), Wahlpflichtfach: Magnetresonanzverfahren in der kardiologischen Grundlagenforschung und in klinischer Anwendung\\
{}\\
seit 2007 & Stipendiat des Berufsverbandes Deutscher Internisten
\end{tabular}

\vspace{3cm}

W\"urzburg, 04. September 2008\\
\vspace{2cm}

Christian H. Ziener

\chapter*{Eidesstattliche Versicherung}
\addcontentsline{toc}{chapter}{Eidesstattliche Versicherung}
gem\"a{\ss} \S 5 Absatz 1 Satz 4 und Absatz 2 Satz 2 der Promotionsordnung der Fakult\"at f\"ur Physik und Astronomie der Bayerischen Julius-Maximilians-Universit\"at W\"urzburg.

Hiermit versichere ich an Eides statt, dass ich, Christian Herbert Ziener, geboren am 18.12.1978 in Weimar, die Dissertation selbst\"andig angefertigt habe. Ich habe keine anderen Hilfsmittel als die in der Arbeit angegebenen benutzt. Alle Ausf\"uhrungen, die w\"ortlich oder sinngem\"a{\ss} \"ubernommen wurden, sind als solche gekennzeichnet. Die Dissertation wurde bisher weder vollst\"andig noch teilweise einer anderen Hochschule mit dem Ziel, einen akademischen Grad zu erwerben, vorgelegt.

Am 24. 09. 2003 wurde mir von der Friedrich-Schiller-Universit\"at Jena der akademische Grad \glqq Diplomphysiker\grqq\ verliehen. Weitere akademische Grade habe ich weder erworben, noch versucht zu erwerben. Mir wurde kein akademischer Grad entzogen. Es wurde kein strafrechtliches Ermittlungsverfahren oder ein Disziplinarverfahren gegen mich eingeleitet.

Am 31. 08. 2006 habe ich den Ersten Abschnitt der \"Arztlichen Pr\"ufung bestanden. Den Zweiten Abschnitt der \"Arztlichen Pr\"ufung werde ich voraussichtlich im Jahre 2010 absolvieren.
\\

\vspace{3cm}

W\"urzburg, 04. September 2008\\
\vspace{2cm}

Christian H. Ziener

\begin{thebibliography}{200}
\addcontentsline{toc}{chapter}{Literaturverzeichnis}

\bibitem{Lauterbur72} Lauterbur PC. Measurements of local nuclear magnetic-resonance relaxation-times. Bull Am Phys Soc 1973;18:86.

\bibitem{Lauterbur73} Lauterbur PC. Image formation by induced local interactions - Examples employing nuclear magnetic resonance. Nature 1973;242:190-191.

\bibitem{Ogawa90} Ogawa S, Lee TM, Kay AR, Tank DW. Brain magnetic resonance imaging with contrast dependent on blood oxygenation. Proc Natl Acad Sci USA 1990;87:9868-9872.

\bibitem{Pauling36} Pauling L, Coryell CD. The magnetic properties and structure of hemoglobin, oxyhemoglobin and carbonmonoxyhemoglobin. Proc Natl Acad Sci USA 1936;22:210-216.

\bibitem{Wacker03} Wacker CM, Hartlep AW, Pfleger S, Schad LR, Ertl G, Bauer WR. Susceptibility-sensitive magnetic resonance imaging detects human myocardium supplied by a stenotic coronary artery without a contrast agent. J Am Coll Cardiol 2003;41:834-840.

\bibitem{Yablonskiy94} Yablonskiy DA, Haacke EM. Theory of NMR signal behavior in magnetically inhomogeneous tissues: the static dephasing regime. \MRM{1994}{32}{749-763}.

\bibitem{Callaghan} Callaghan PT. Principles of Nuclear Magnetic Resonance Microscopy. Clarendon Press, Oxford, 1991.

\bibitem{Bauer99} Bauer WR, Nadler W, Bock M, Schad LR, Wacker C, Hartlep A, Ertl G. Theory of the BOLD effect in the capillary region: an analytical approach for the determination of T2 in the capillary network of myocardium. \MRM{1999}{41}{51-62}.

\bibitem{Bauer99PRL} Bauer WR, Nadler W, Bock M, Schad LR, Wacker C, Hartlep A, Ertl G. Theory of coherent and incoherent nuclear spin dephasing in the heart. \PRL{1999}{83}{4215-4218}.

\bibitem{Kiselev99} Kiselev VG, Posse S. Analytical model of susceptibility-induced MR signal dephasing: effect of diffusion in a microvascular network. \MRM{1999}{41}{499-509}.

\bibitem{Sukstanskii02} Sukstanskii AL, Yablonskiy DA. Effects of restricted diffusion on MR signal formation. \JMR{2002}{157}{92-105}.

\bibitem{Sukstanskii03} Sukstanskii AL, Yablonskiy DA. Gaussian approximation in the theory of MR signal formation in the presence of structure-specific magnetic field inhomogeneities. \JMR{2003}{163}{236-247}.

\bibitem{Sukstanskii04} Sukstanskii AL, Yablonskiy DA. Gaussian approximation in the theory of MR signal formation in the presence of structure-specific magnetic field inhomogeneities. Effects of impermeable susceptibility inclusions. \JMR{2004}{167}{56-67}.

\bibitem{Anderson53} Anderson PW, Weiss PR. Exchange narrowing in paramagnetic resonance. Rev Mod Phys 1953;25:269-276.

\bibitem{Kennan94} Kennan RP, Zhong J, Gore JC. Intravascular susceptibility contrast mechanisms in tissues. \MRM{1994}{31}{9-21}.

\bibitem{Haase86} Haase A, Frahm J, Matthaei D, Hanicke W, Merboldt K. FLASH imaging: rapid NMR imaging using low flip-angle pulses. \JMR{1986}{67}{258-266}.

\bibitem{Frahm87} Frahm J, H\"anicke W, and Merboldt KD. Transverse coherence in rapid FLASH NMR imaging. \JMR{1987}{72}{307-314}.

\bibitem{Wood87} Wood ML, Silver M, Runge VM. Optimization of Spoiler Gradients in FLASH MRI. \MRI{1987}{5}{455-463}.

\bibitem{Crawley88} Crawley AP, Wood ML, Henkelman RM. Elimination of transverse coherences in FLASH MRI. \MRM{1988}{8}{248-260}.

\bibitem{Ernst66} Ernst RR and Anderson WA. Application of Fourier transform spectroscopy to magnetic resonance. Rev Sci Instrum 1966;37:93-102.

\bibitem{Haacke99} Haacke EM, Brown RW, Thompson MR and Venkatesan R. Magnetic Resonance Imaging: Physical Principles and Sequence Design. John Wiley, New York, 1999.

\bibitem{Hinshaw76} Hinshaw WS. Image formation by magnetic resonance: the sensitive point method. J Appl Phys 1976;47:3709-3721.

\bibitem{Freeman71} Freeman R, Hill HDW. Phase and intensity anomalies in Fourier transform NMR. \JMR{1971}{4}{366-383}.

\bibitem{Gyngell88} Gyngell ML. The Steady-State Signals in Short-Repetition-Time Sequences. \JMR{1989}{81}{474-483}.

\bibitem{Scheffler03} Scheffler K, Hennig J. Is TrueFISP a gradient-echo or a spin-echo sequence? \MRM{2003}{49}{395-397}.

\bibitem{Jackson04} Jackson JD. Classical Electrodynamics. 3rd ed. John Wiley and Sons Ltd. New York 2004.

\bibitem{Salomir03} Salomir R, de Senneville BD, Moonen CTW. A fast calculation method for magnetic field inhomogeneity due to an arbitrary distribution of bulk susceptibility. Concepts Magn Reson B 2003;9B:26-34.

\bibitem{Landau} Landau LD, Lifshitz EM. Course of Theoretical Physics, Vol.\ 2, 2nd.\ ed.\ Pergamon, Oxford, 1999.

\bibitem{Krogh19} Krogh A. The number and the distribution of capillaries in muscle with the calculation of the oxygen pressure necessary for supplying the tissue. J Physiol (Lond) 1919;52:409-415.

\bibitem{Reichenbach01} Reichenbach JR, Haacke EM. High-resolution BOLD venographic imaging: a window into brain function. NMR Biomed 2001;14:453-467.

\bibitem{Torrey56} Torrey HC. Bloch Equations with Diffusion Terms. \PR{1956}{104}{563-565}.

\bibitem{Nadler85} Nadler W, Schulten K. Generalized moment expansion for Brownian relaxation processes. \JCP{1985}{82}{151-160}.

\bibitem{Stables98} Stables LA, Kennan RP, Gore JC. Asymmetric spin-echo imaging of magnetically inhomogeneous systems: theory, experiment, and numerical studies. \MRM{1998}{40}{432-442}.

\bibitem{Cowan97} Cowan B. Nuclear Magnetic Resonance and Relaxation. University Press, Cambridge, 1997.

\bibitem{Bauer05} Bauer WR, Ziener CH, and Jakob PM. Non-Gaussian spin dephasing. \PRA{2005}{71}{053412}.

\bibitem {Spees01} Spees WM, Yablonskiy DA, Oswood MC, Ackerman JJ. Water proton MR properties of human blood at 1.5 Tesla: magnetic susceptibility, T(1), T(2), T*(2), and non-Lorentzian signal behavior. \MRM{2001}{45}{533-542}.

\bibitem{Abragam89} Abragam A. Principles of Nuclear Magnetism. Oxford University Press, New York, 1989.

\bibitem{Jensen00} Jensen JH, Chandra R. NMR relaxation in tissues with weak magnetic inhomogeneities. \MRM{2000}{44}{144-156}.

\bibitem{Bauer92} Bauer WR, Schulten K. Theory of contrast agents in magnetic resonance imaging: coupling of spin relaxation and transport. \MRM{1992}{26}{16-39}.

\bibitem{Szabo80} Szabo A, Schulten K, Schulten Z. First passage time approach to diffusion controlled reactions. \JCP{1980}{72}{4350-4357}.

\bibitem{Dattagupta74} Dattagupta S, Blume M. Stochastic theory of line shape. I. Nonsecular effects in the strong-collision model. \PRB{1974}{10}{4540-4550}.

\bibitem{Dattagupta76} Dattagupta S, Blume M. Stochastic theory of spin relaxation in liquids. \PRA{1976}{14}{480-494}.

\bibitem{Lynden-Bell71} Lynden-Bell RM. A density matrix formulation of the theory of magnetic resonance spectra in slowly reorienting systems. Mol Phys 1971;22:837-851.

\bibitem{Ziener06a} Ziener CH, Bauer WR, Melkus G, Weber T, Herold V, Jakob PM. Structure-specific magnetic field inhomogeneities and its effect on the correlation time. \MRI{2006}{24}{1341-1347}.

\bibitem{Ziener06b} Ziener CH, Kampf T, Melkus G, Jakob PM, Bauer WR. Scaling laws for transverse relaxation times. \JMR{2007}{184}{169-175}.

\bibitem{Yung03} Yung KT. Empirical models of transverse relaxation for spherical magnetic perturbers. \MRI{2003}{21}{451-463}.

\bibitem{Carr58} Carr HY. Steady-State Free Precession in Nuclear Magnetic Resonance. \PR{1958}{112}{1693-1701}.

\bibitem{Lebel06} Lebel RM, Menon RS, Bowen CV. Relaxometry model of strong dipolar perturbers for balanced-SSFP: application to quantification of SPIO loaded cells. \MRM{2006}{55}{583-591}.

\bibitem{Cheng01} Cheng YC, Haacke EM, Yu YJ. An exact form for the magnetic field density of states for a dipole. \MRI{2001}{19}{1017-1023}.

\bibitem{Bakker94} Bakker CJG, Bhagwandien R, Moerland MA, Ramos LMP. Simulation of susceptibility artifacts in 2D and 3D Fourier transform spin-echo and gradient-echo magnetic resonance imaging. \MRI{1994}{12}{767-774}.

\bibitem{Ziener05MAGMA} Ziener CH, Bauer WR, Jakob PM. Frequency distribution and signal formation around a vessel. Magn Reson Mater Phy 2005;18:225-230.

\bibitem{Landau5} Landau LD, Lifshitz EM. Course of Theoretical Physics, Vol.\ 5, 2nd.\ ed. Pergamon, Oxford, 1999.

\bibitem{Bowen02} Bowen CV, Zhang X, Saab G, Gareau PJ, Rutt BK. Application of the static dephasing regime theory to superparamagnetic iron-oxide loaded cells. \MRM{2002}{48}{52-61}.

\bibitem{Bassingwaighte74} Bassingthwaighte JB, Yipintsoi T, Harvey RB. Microvasculature of the dog left ventricular myocardium. Microvasc Res 1974;7:229-249.

\bibitem{Bauer99T2} Bauer WR, Nadler W, Bock M, Schad LR, Wacker C, Hartlep A, Ertl G. The relationship between T2* and T2 in myocardium. \MRM{1999}{42}{1004-1010}.

\bibitem{Moiny92} Moiny F, Gillis, P, Roch A, Muller RN. Transverse relaxation of superparamagnetic contrast agents: a numerical analysis. Book of Abstracts: Eleventh Annual Meeting of the Society of Magnetic Resonance in Medicine 1992;2:1431.

\bibitem{Brooks01} Brooks RA, Moiny F, Gillis P. On T2-shortening by weakly magnetized particles: the chemical exchange model. \MRM{2001}{45}{1014-1020}.

\bibitem{Baklanov04} Baklanov DV, Demuinck ED, Thompson CA, Pearlman JD. Novel double contrast MRI technique for intramyocardial detection of percutaneously transplanted autologous cells. \MRM{2004}{52}{1438-1442}.

\bibitem{Stoller91} Stoller SD, Happer W, Dyson FJ. Transverse spin relaxation in inhomogeneous magnetic fields. \PRA{1991}{44}{7459}.

\bibitem{Puetz91} P\"utz B, Barsky D, Schulten K. Edge enhancement by diffusion: Microscopic magnetic resonance imaging of an ultra-thin glass capillary. \CPL{1991}{183}{391-396}.

\bibitem{Puetz92} P\"utz B, Barsky D, Schulten K. Edge enhancement by diffusion in microscopic magnetic resonance imaging. \JMR{1992}{97}{27-53}.

\bibitem{Barsky92} Barsky D, P\"utz B, Schulten K. Diffusional edge enhancement observed by NMR in thin glass capillaries. \CPL{1992}{200}{88-96}.

\bibitem{Kubo62} Kubo R. Fluctuations, Relaxation and Resonance in Magnetic Systems. Oliver \& Boyd, Edinburgh, 1962.

\bibitem{Kubo63} Kubo R. Stochastic Liouville Equations. \JMP{1963}{4}{174-183}.

\bibitem{Kiselev98} Kiselev VG, Posse S. Analytical theory of susceptibility induced NMR signal dephasing in a cerebrovascular network. \PRL{1998}{81}{5696-5699}.

\bibitem{Bauer02} Bauer WR, Nadler W. Spin dephasing in the strong collision approximation. \PRE{2002}{65}{066123}.

\bibitem{Norris00} Norris DG. The effects of microscopic tissue parameters on the diffusion weighted magnetic resonance imaging experiment. NMR Biomed 2001;14:77-93.

\bibitem{ZienerPRE} Ziener CH, Kampf T, Melkus G, Herold V, Weber T, Reents G, Jakob PM, Bauer WR. Local frequency density of states around field inhomogeneities in magnetic resonance imaging: effects of diffusion. \PRE{2007}{76}{031915}

\bibitem{Hahn50} Hahn EL. Spin Echoes. \PR{1950}{80}{580-594}.

\bibitem{Seppenwoolde05} Seppenwoolde JH, van Zijtveld M, Bakker CJ. Spectral characterization of local magnetic field inhomogeneities. Phys Med Biol 2005;50:361-372.

\bibitem{Case87} Case TA, Durney CH, Ailion DC, Cutillo AG, Morris AH. A mathematical model of diamagnetic line broadening in lung tissue and similar heterogeneous systems: calculations and measurements. \JMR{1987}{73}{304-314}.

\bibitem{Durney89} Durney CH, Bertolina JA, Ailion DC, Christman R, Cutillo AG, Morris AH, Hashemi S. Calculation and interpretation of inhomogeneous line broadening in models of lungs and other heterogeneous structures. \JMR{1989}{85}{554-570}.

\bibitem{Bertolina91} Bertolina JA, Durney CH, Ailion DC, Cutillo AG, Morris AH, Goodrich KC. Experimental verification of inhomogeneous line-broadening calculations in lung models and other inhomogenous structures. \JMR{1992}{99}{161-169}.

\bibitem{Cutillo96} Cutillo AG. Application of Magnetic Resonance to the Study of Lung. Futura Publishing, Armonk, New York, 1996.

\bibitem{Oberhettinger72} Oberhettinger F. Tables of Bessel Transforms. Springer Verlag, Berlin, Heidelberg, New York, 1972.

\bibitem{Gradstein81} Gradstein IS and Ryshik IM. Summen-, Produkt- und Integraltafeln/ Tables of Series, Products, and Integrals. Verlag Harry Deutsch, Thun, Frankfurt/Main, 1981.

\bibitem{Zimmerman56} Zimmerman JR, Foster MR. Standardization of N.M.R. high resolution spectra. \JPC{1957}{61}{282-289}.

\bibitem{Scheffler01} Scheffler K, Seifritz E, Bilecen D, Venkatesan R, Hennig J, Deimling M, Haacke EM. Detection of BOLD changes by means of a frequency-sensitive trueFISP technique: preliminary results. NMR Biomed 2001;14:490-496.

\bibitem{Glutsch04} Glutsch S. Excitons in Low-Dimensional Semiconductors: Theory, Numerical Methods, Applications. Springer, Berlin, 2004.

\bibitem{Bottomley84} Bottomley PA. Selective volume method for performing localized NMR spectroscopy. United States Patent 1984:4 480 228.

\bibitem{Bottomley87} Bottomley PA. Spatial Localization in NMR Spectroscopy in Vivo. Ann N Y Acad Sci 1987;508:333-348.

\bibitem{Weissleder97} Weissleder R, Cheng H, Bogdanova A, Bogdanova A Jr. Magnetically labeled cells can be detected by MR imaging. \JMRI{1997}{7}{258-263}.

\bibitem{Stroh05} Stroh A, Faber C, Neuberger T, Lorenz P, Sieland K, Jakob PM, Webb A, Pilgrimm H, Schober R, Pohl EE, Zimmer C. In vivo detection limits of magnetically labeled embryonic stem cells in the rat brain using high-field (17.6 T) magnetic resonance imaging. Neuroimage 2005;24:635-645.

\bibitem{Ziener05MRM} Ziener CH, Bauer WR, Jakob PM. Transverse relaxation of cells labeled with magnetic nanoparticles. \MRM{2005}{54}{702-706}.

\bibitem{Gillis87}Gillis P, Koenig SH. Transverse relaxation of solvent protons induced by magnetized spheres: application to ferritin, erythrocytes, and magnetite. \MRM{1987}{5}{323-345}.

\bibitem{Jensen99} Jensen JH, Chandra R. Transverse Relaxation Time Field Dependence for Tissues with Microscopic Magnetic Susceptibility Variations. Proc Int Soc Magn Reson Med 1999:656.

\bibitem{Muller91} Muller RN, Gillis P, Moiny F, Roch A. Transverse relaxivity of particulate MRI contrast media: from theories to experiments. \MRM{1991}{22}{178-182}.

\bibitem{Weisskoff94} Weisskoff RM, Zuo CS, Boxerman JL, Rosen BR. Microscopic susceptibility variation and transverse relaxation: theory and experiment. \MRM{1994}{31}{601-610}.

\bibitem{Kiselev02} Kiselev VG, Novikov DS. Transverse NMR relaxation as a probe of mesoscopic structure. \PRL{2002}{89}{278101}.

\end{thebibliography}
\end{document}